\documentclass[twocolumn,times,twocolappendix]{aastex701}
\newcommand{\igr}{IGR 17062-6143}
\newcommand{\sax}{SAX J1712.6-3739}
\newcommand{\swift}{Swift J1734.5-3027}
\newcommand{\fouru}{4U 1850-087}
\newcommand{\xrt}{\textit{Swift}-XRT}

\shorttitle{Investigating Rapid Fluctuations in Type I X-ray Bursts}
\shortauthors{Ballantyne et al.}
\begin{document}

\title{Do Accretion Disks Get Bent Out of Shape? Investigating the Origin of Rapid Fluctuations in the Tails of Long Thermonuclear X-ray Bursts}

\author[orcid=0000-0001-8128-6976]{D.R. Ballantyne}
\affiliation{Center for Relativistic Astrophysics, School of Physics, Georgia Institute of Technology, 837 State Street, Atlanta, GA USA 30332}
\email[show]{david.ballantyne@physics.gatech.edu}  

\author[orcid=0000-0002-0092-3548]{N. Degenaar} 
\affiliation{Anton Pannekoek Institute for Astronomy, University of Amsterdam, Science Park 904, 1098 XH, Amsterdam, the Netherlands}
\email{N.D.Degenaar@uva.nl}

%\author[]{others...}
%\affiliation{somewhere}
%\email{fakeemail3@google.com}

\begin{abstract}
Thermonuclear X-ray bursts from the surfaces of neutron stars
affect the surrounding accretion flow, revealing details
of the underlying physical processes influencing accretion physics. In
this context, we perform a spectral analysis of 5 long X-ray bursts
that exhibit temporary rapid flux variations during the tails
of their light curves. In all cases, spectra extracted from before,
during and after the time of fluctuations show evidence for
relativistic, ionized reflection with the blackbody from the neutron
star often hidden from view. Both the properties of the reflecting
region and the observed fraction of the blackbody vary as the
fluctuations start and stop. These results are compared to a sample of
5 similar bursts without fluctuations in their light curves, and we
find that spectra from 4 of the bursts in the control sample are
best described by a simple absorbed blackbody. The bursts with fluctuations
are longer and more energetic than those without fluctuations, in
agreement with the prediction that radiatively-driven warps are impacting
the accretion disks of the bursts with fluctuations. This result would also
imply that these accretion disks must have large viscosity parameters. The lack of reflection in bursts from the control sample may result from lower accretion rates, as these disks would have
smaller surface densities and would expand and become Compton-thin due
to heating from the burst. High-throughput, time-resolved spectral
analysis of X-ray bursts that undergo warping would give insight into
the strength of accretion disk viscosity.  
\end{abstract}



\section{Introduction}
\label{sect:intro}
X-ray bursts are explosions on the surfaces of accreting neutron stars (NSs) in low-mass X-ray binaries (LMXBs) caused by an unstable nuclear runaway in the accreted material \citep[e.g.,][]{lvt93,gk21}. The resulting hot stellar surface glows as a blackbody in X-rays as it cools and matter is replenished. Observationally, the light curves of X-ray bursts are characterized by a rapid rise to a peak flux that can approach or even exceed the local Eddington limit of the star, followed by an exponential decay \citep[e.g.,][]{gall08,minbar20}. The decay times of the majority of bursts (often referred to as Type I bursts) are $\sim10$--$20$~s, but, on occasion, longer bursts with decay times of $\sim100$~s and $\sim1000$~s (referred to as intermediate duration bursts or superbursts, respectively) are observed \citep[e.g.,][]{sb03,cum04}. 

X-ray bursts release a significant amount of radiative energy into their environment ($\sim 10^{39-41}$~ergs; e.g., \citealt{degenaar18}) which then interacts with the surrounding accretion flow, potentially leading to a variety of structural changes to the accretion disk \citep{be05,degenaar18}. Observational and theoretical studies have found evidence that the flood of radiation produced during bursts significantly heats and ionizes the accretion disk \citep[e.g.,][]{bs04,keek14b,keek17,fbb19,speicher22}, cools and potentially collapses the corona \citep[e.g.,][]{mc03,ji14,fbmw18,sanchez20,speicher20}, erodes the inner edge of the accretion disk through Poynting-Robertson drag \citep[e.g.,][]{walker89,worpel15,fbb19,speicher23}, and may even warp the disk via a radiative instability \citep{ball23}. There is also evidence that bursts may impact the production of jets in LMXBs \citep{russell24}. As X-ray bursts occur at regular intervals in most LMXBs, they provide an unique opportunity to study the behavior of accretion disks under extreme conditions through a repeatable experiment. 

With well over 100 known bursting sources in the Galaxy\footnote{https://www.sronpersonalpages.nl/$\sim$jeanz/bursterlist.html} \citep[e.g.,][]{int19} and over 7000 unique bursts detected by X-ray instruments \citep[e.g.,][]{minbar20}, it is possible to find unusual phenomena in burst data that may provide novel new insights into accretion physics that would be otherwise impossible through the study of steady accretion-powered sources. One example of this is the sudden onset and disappearance of rapid fluctuations in the flux of a small number of intermediate duration bursts \citep[e.g.,][]{igb11,degenaar13,barriere15,int19}. The fluctuations are typically achromatic above $2$~keV, go both above and below the normal decay of the flux with timescales that vary between $\sim 1$~s to $\sim 1$~min, and have amplitudes of $\sim 70$\%. The fluctuations always appear during the decay phase of the burst and last for 10s to 100s of seconds before stopping \citep[e.g.,][]{degenaar18}. None of the systems exhibiting fluctuations are known to be eclipsing or observed to be `dippers' \citep{int19}, indicating that this effect is not a result of viewing these sources at a high inclination angle. All the bursts that show fluctuations are highly luminous photospheric radius expansion events, where a wind may be driven off the surface of the NS \citep[e.g.,][]{lvt93,herrera20,guich21}, and have decay times longer than typical Type I bursts. However, there are many long photospheric radius expansion bursts that do not show fluctuations in their light curves \citep{int19}. This collection of properties has eluded an explanation, but the timescales of the fluctuations strongly argue that the cause of the fluctuations must result from an interaction between the burst and the accretion disk. 

In a proof-of-concept study, \citet{ball23} found that, depending on the strength of the disk viscosity, X-ray bursts could cause short-lived warping of the surrounding accretion disk due to a radiation-driven instability. For a given viscosity, the condition for triggering the warping instability is more easily reached for both longer and more luminous X-ray bursts, similar to the group of luminous intermediate duration bursts that show fluctuations. In addition, evolution calculations of disk warps in the linear regime exhibited rapid changes in the warp amplitude on timescales similar to the observed fluctuations, and \citet{ball23} was able to identify this as the radial diffusion timescale through the inner part of the disk. As the warping instability depends so sensitively on the disk viscosity, the fluctuations seen in the decay of intermediate duration bursts are potential probes of the viscous processes in their accretion disks.

It is therefore of interest to better understand the properties of the accretion disk during the observed fluctuations to determine if they can be connected to the disk warping instability. Earlier analyses of the fluctuations characterized the spectra only with blackbodies and were unable to infer changes to the accretion environment \citep[e.g.,][]{degenaar13,int19}. Here, we compile a sample of 5 intermediate duration bursts observed by \xrt\ that exhibited rapid fluctuations during their decay, and perform a detailed spectral analysis at the times before, during and following the fluctuations with the goal of searching for evidence of changes in the accretion structure. To isolate the impact of the fluctuations, we perform a similar analysis on a control sample of 5 other intermediate duration bursts without fluctuations. Section~\ref{sect:data} describes the sample selection, data reduction and extraction of spectral products. In Sect.~\ref{sect:igr} we provide a detailed description of the spectral analysis of the 2012 intermediate duration burst from \igr, and the results from the other four sources with fluctuations are presented in Sect.~\ref{sect:other}. The analysis of the control sample of bursts without fluctuations is found in Sect.~\ref{sect:control}. The results are discussed in Sect.~\ref{sect:discuss} and we present our conclusions in Sect.~\ref{sect:concl}.

\section{Sample Selection and Data Reduction} 
\label{sect:data}

\subsection{Source and data sample} 
\label{subsect:sample}
In the literature there a total of 12 long bursts, from 10 different sources, reported to show distinct fluctuations during their burst tails (\citealt{int19}, their Table~3). These were detected with different satellites and instruments (\textit{Ginga}, \textit{RXTE}, \textit{INTEGRAL}, \textit{BeppoSAX}, \textit{Swift}, \textit{NuSTAR}), with the largest sample detected by a single telescope being five bursts seen with \textit{Swift}. We thus focus on these five \textit{Swift} bursts to allow for a consistent and comparative analysis. The soft X-ray coverage provided by \textit{Swift} is also sensitive to X-ray reflection signals from the accretion disk \citep[e.g.,][]{garcia22,speicher22} which will be important in attempting to measure changes to the disk structure. As a control sample we take five long bursts without fluctuations but with comparable duration and \textit{Swift} coverage, also drawn from the work of \citet{int19}. Table~\ref{tab:sample} lists the two samples of bursts, along with information on the \textit{Swift} observations.

\begin{deluxetable*}{ccccccccc}[t!]
\tabletypesize{}
%\tablewidth{0} 
\tablecaption{Source sample and \xrt\ observations. \label{tab:sample}}
\tablehead{
\colhead{Source} & \colhead{Date} & \colhead{ObsID} & \colhead{Start time} & $t_{5\%}$ & \colhead{$t_{\mathrm{pre/1}}$} & \colhead{$t_{\mathrm{fluc/2}}$} & \colhead{$t_{\mathrm{post/3}}$} & \colhead{$N_{\mathrm{H,Gal}}$} \\
\colhead{} & \colhead{} & \colhead{} & \colhead{(MET)} & \colhead{(s)} & \colhead{(s)} & \colhead{(s)} & \colhead{(s)} & \colhead{($10^{22}$~cm$^{-2}$)}
} % Note
\startdata
\multicolumn{9}{c}{Bursts with fluctuations}   \\
\hline
IGR J17062--6143 & 2012-06-25 & 00525148000 & 362357117.4089714 & 1300 & 224 & 540 & 107 & 0.102 \\ 
% swxwt0to2s6_20010101v014.rmf
SAX J1712.6--3739 & 2011-09-26 & 00504101000 & 338760817.204697 & 1050 & 93 & 521 & 90 & 1.13 \\ 
%s wxwt0to2s6_20110101v015.rmf
% NOTE: pre- and post-modulation data are short
& 2014-08-18 & 00609879000 & 430075313.185269 & $>5530$ & 115 & 1098 & - &  1.13 \\ 
% swxwt0to2s6_20131212v015.rmf 
% NOTE: no data available after the fluctuation phase 
Swift J1734.5--3027 & 2013-09-01 & 00569022000 & 399719734.788132 & 600 & - & 193 & 1478 & 0.674 \\ 
% swxwt0to2s6_20130101v015.rmf
% NOTE: no data available before the fluctuation phase
4U 1850--087 & 2014-03-10 & 00591237000 & 416178802.780347 & 1400 & 359 & 262 & 162 & 0.251 \\ 
% swxwt0to2s6_20131212v015.rmf
\hline
\multicolumn{9}{c}{Bursts without fluctuations}   \\
\hline
4U 1246--58 & 2006-08-11 & 00223918000 & 176957038.373 & 290 & 101 & 150 & 496 & 0.335 \\
% swxwt0to2s0_20010101v012.rmf
XTE J1701--407 & 2008-07-17 & 00317205000 & 237993244.948 & 190 & 100 & 148 & 500 & 1.33 \\
% swxwt0to2s6_20010101v015.rmf
SAX J1712.6--3739 & 2010-07-01 & 00426405000 & 299687992.944 & 140 & 79 & 141 & 345 & 1.13\\
% swxwt0to2s6_20090101v015.rmf
XTE J1810--189 & 2011-06-19 & 00455640000 & 330137029.071 & 750 & 100 & 150 & 440 & 1.61 \\
% swxwt0to2s6_20110101v015.rmf
SAX J1806.5--2215 & 2017-04-01 & 00745022000 & 512765126.844 & 550 & 100 & 151 & 457 & 1.0 \\
% swxwt0to2s6_20131212v015.rmf
\hline
\enddata
\tablecomments{The start time of the \xrt\ exposures is given in Mission Elapsed Time (MET). The burst duration is measured by $t_{5\%}$ which is the length of time that the burst flux is greater than $5\%$ of the peak flux \citep{int19}. For the five bursts that exhibit modulations, the duration of the intervals before, during and after the fluctuations are indicated as $t_{\mathrm{pre}}$, $t_{\mathrm{fluc}}$ and $t_{\mathrm{post}}$, respectively (Fig.~\ref{fig:five_lc_mod}). Dashes indicate that the specified interval is not available. The three intervals analyzed in the control sample have durations of $t_1$, $t_2$ and $t_3$ (Fig.~\ref{fig:five_lc_nomod}). The final column lists the Galactic column density towards each source, $N_{\mathrm{H,Gal}}$ \citep{hi4pi16}.}
\end{deluxetable*}

\subsection{\textit{Swift} data reduction} 
\label{subsect:reduction}
All \textit{Swift} bursts were initially detected around their peak with the wide-field Burst Alert Telescope (BAT). These BAT detections triggered automatic followed-up, typically within $\sim$minutes, with the pointed X-Ray Telescope (XRT), providing coverage of the X-ray burst tail. Our analysis thus focuses on the XRT data, which were all obtained in Windowed Timing (WT) mode.

The \xrt\ data were obtained from the HEASARC archive and the extraction of data products (light curves and spectra) was performed using the \textsc{heasoft} suite (v.6.23). Raw data were first processed using the \textsc{xrtpipeline} using the default quality cuts. Light curves and spectra were subsequently obtained using XSelect. For all sources we used a circular source extraction region, centered on the pixel with the highest count rate, with a radius of 50 arcseconds. To extract background photons, we used a circular region of the same size located towards the edge of the WT readout column. Response matrix files corresponding to the observing epochs were obtained from the  CALDB and arf files were generated using \textsc{xrtmkarf} based on the observation-specific exposure maps. All spectra were grouped using \textsc{grappha} to contain a minimum of 20 photons per bin. 
% NOTE ND: I also extracted average burst spectra, but I don't think you used them in the analysis, so I didn't mention them. Is that correct? Otherwise I should mention them.
% DRB: Yes, I did not ultimately use them in the analysis so it is OK not to mention it.

\subsection{Selection of burst intervals} 
\label{subsect:lc}
\begin{figure*}[t!]
\begin{center}
\includegraphics[width=0.45\textwidth]{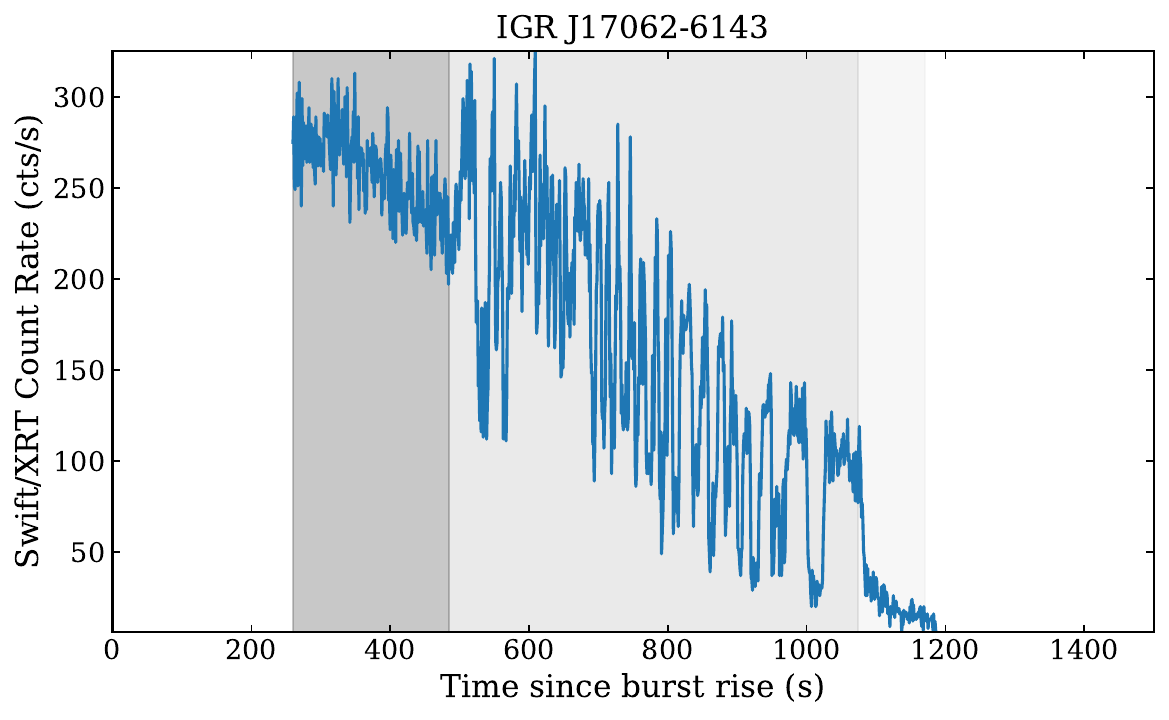}
\includegraphics[width=0.45\textwidth]{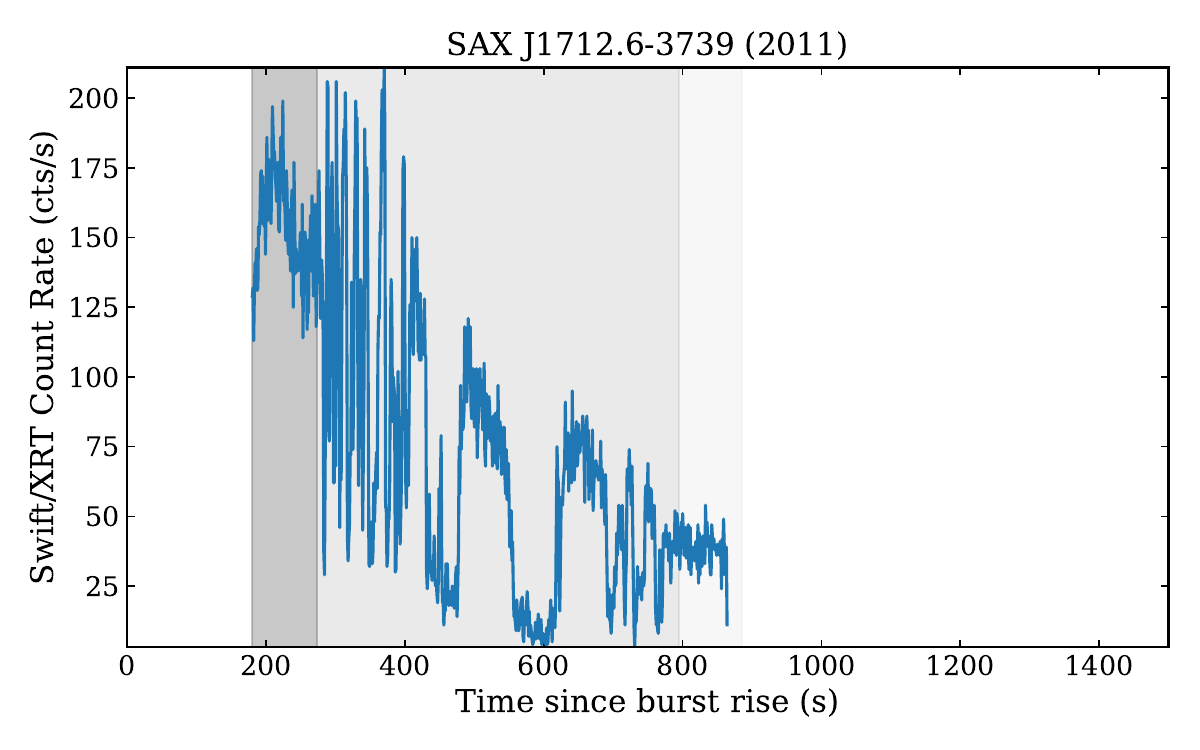}
\includegraphics[width=0.45\textwidth]{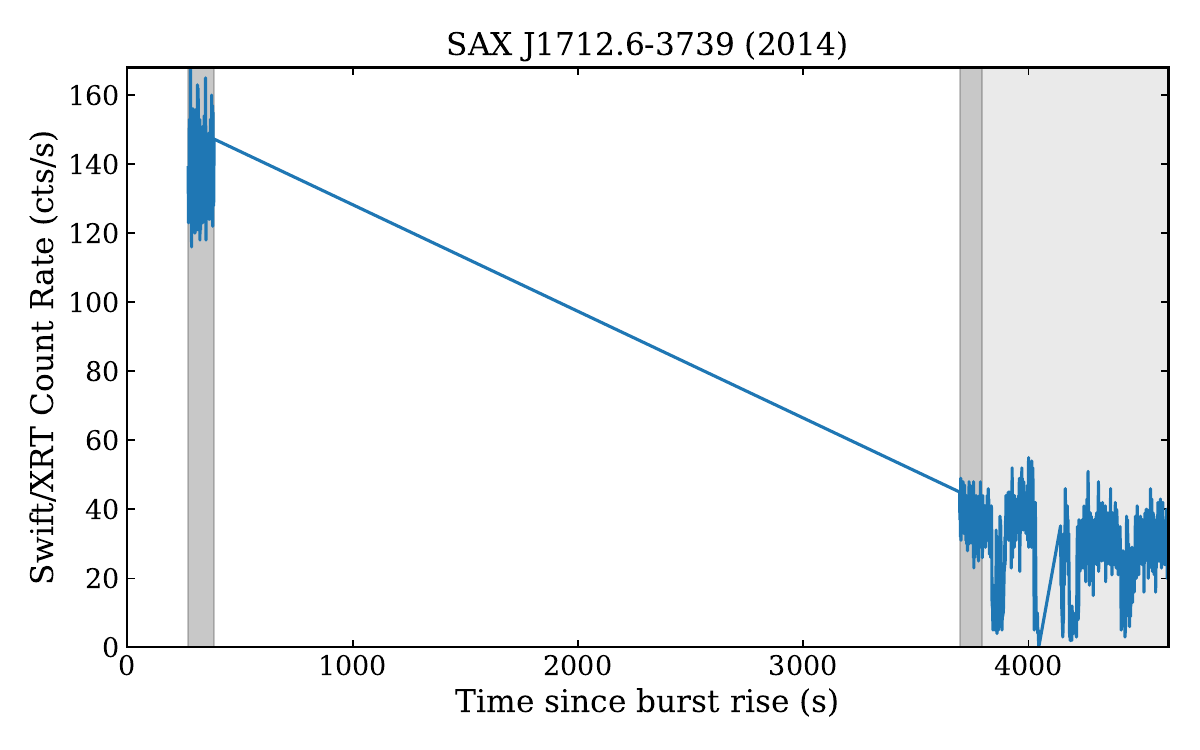}
\includegraphics[width=0.45\textwidth]{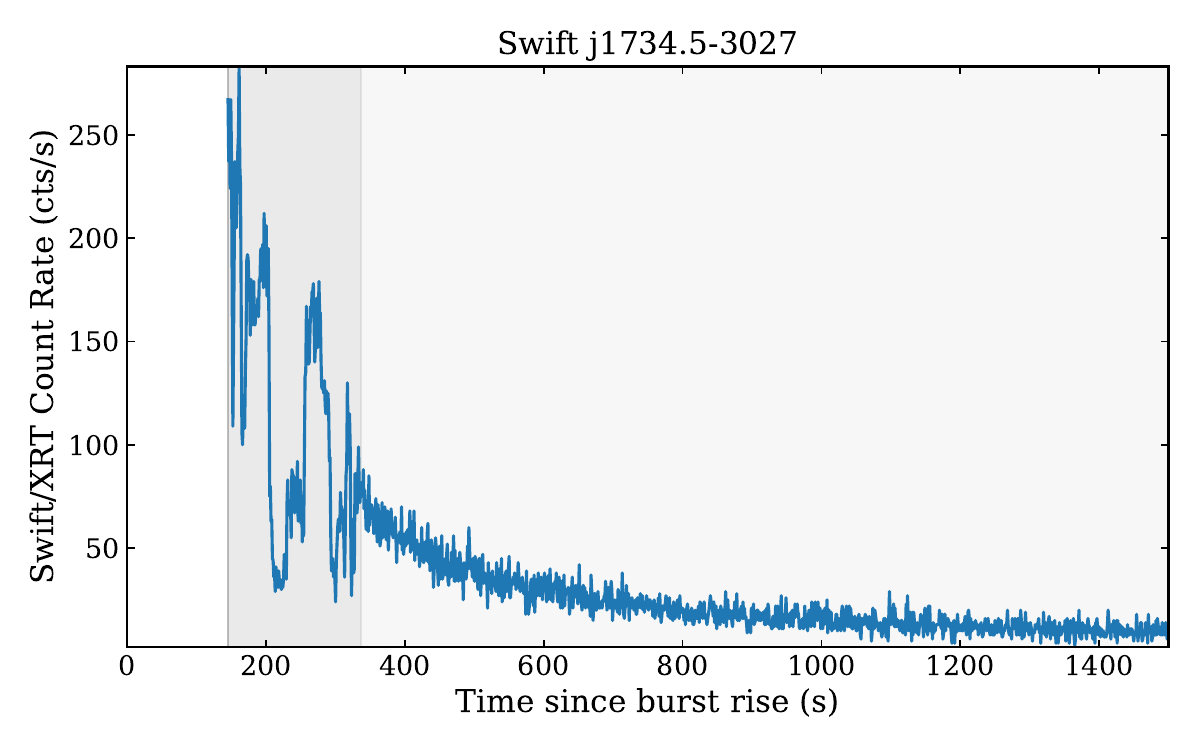}
\includegraphics[width=0.45\textwidth]{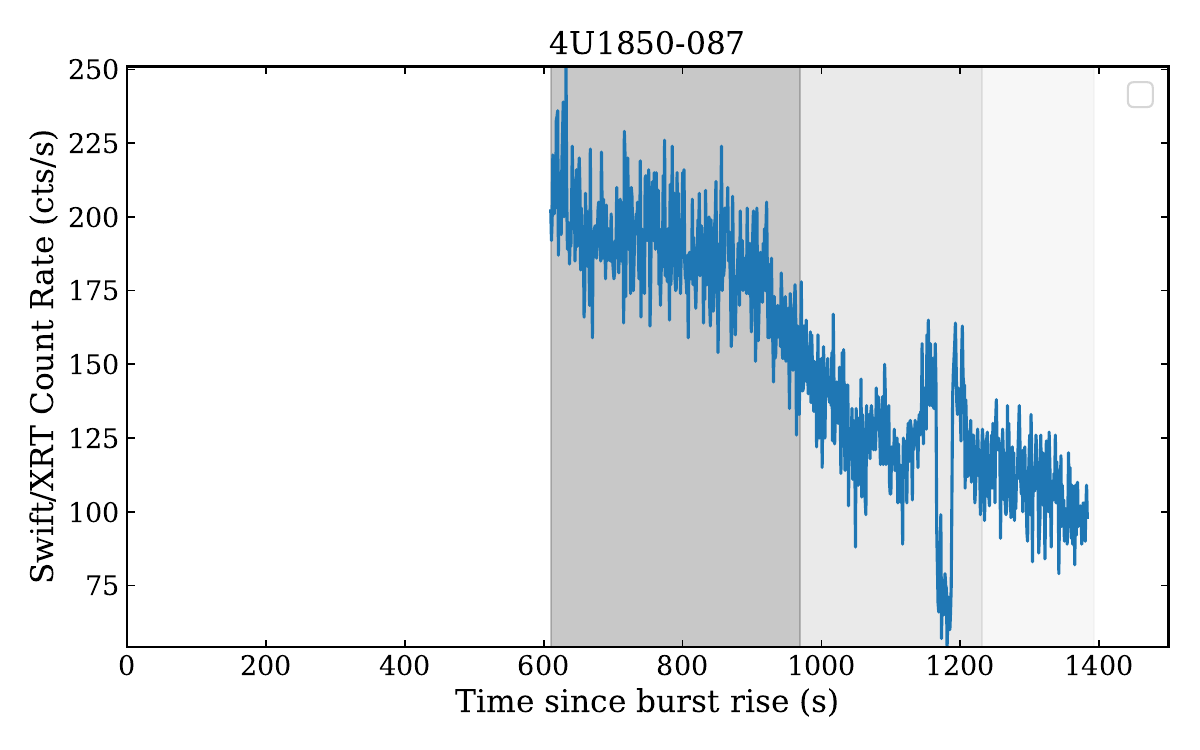}
\end{center}
\caption{\xrt\ light curves of the five bursts with fluctuations chosen for analysis. The time resolution is 1 s and $t = 0$ corresponds to the onset of the burst rise found by \citet{int19}. Colored regions indicate the different time intervals used for spectral extraction, where dark, medium and light gray specify the intervals before (PRE), during (FLUC), and after the fluctuations (POST), respectively. All light curves are plotted on the same time scale of 0-1500~s for a direct comparison between the different bursts, except for the 2014 burst of SAX J1712.6-3739, for which the data with the fluctuations are obtained at much later times.}
\label{fig:five_lc_mod}
\end{figure*}

\begin{figure*}[t!]
\begin{center}
\includegraphics[width=0.45\textwidth]{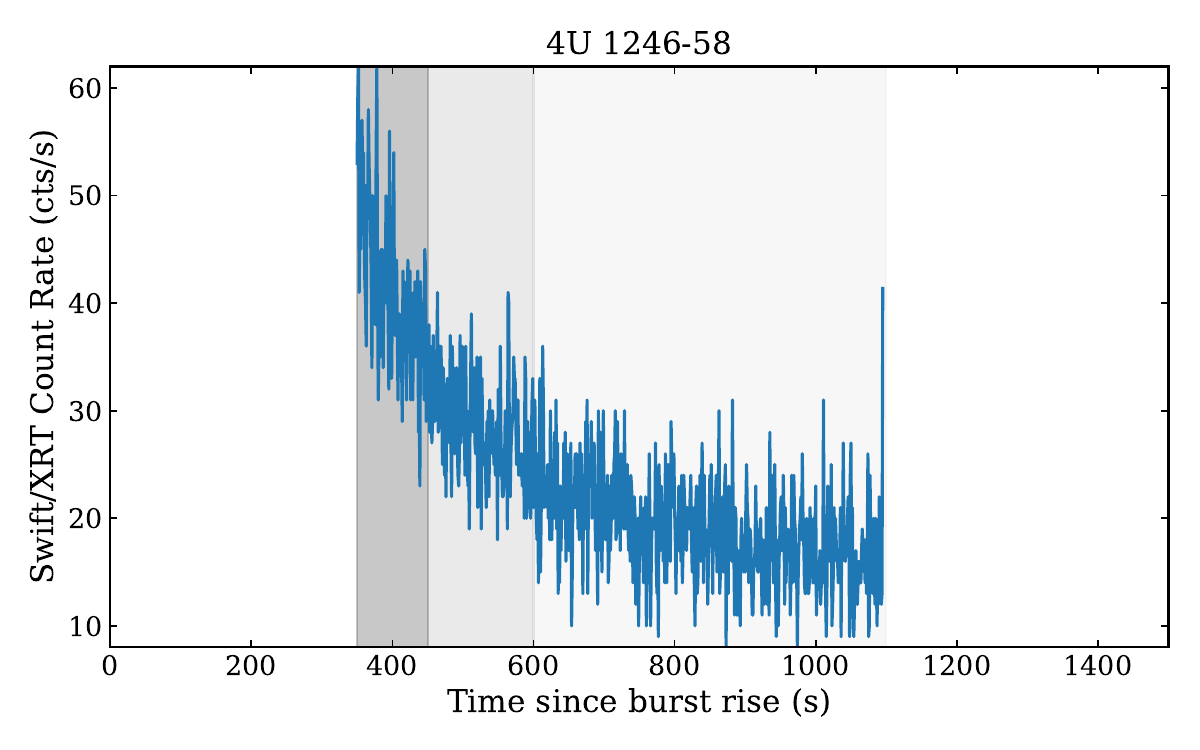}
\includegraphics[width=0.45\textwidth]{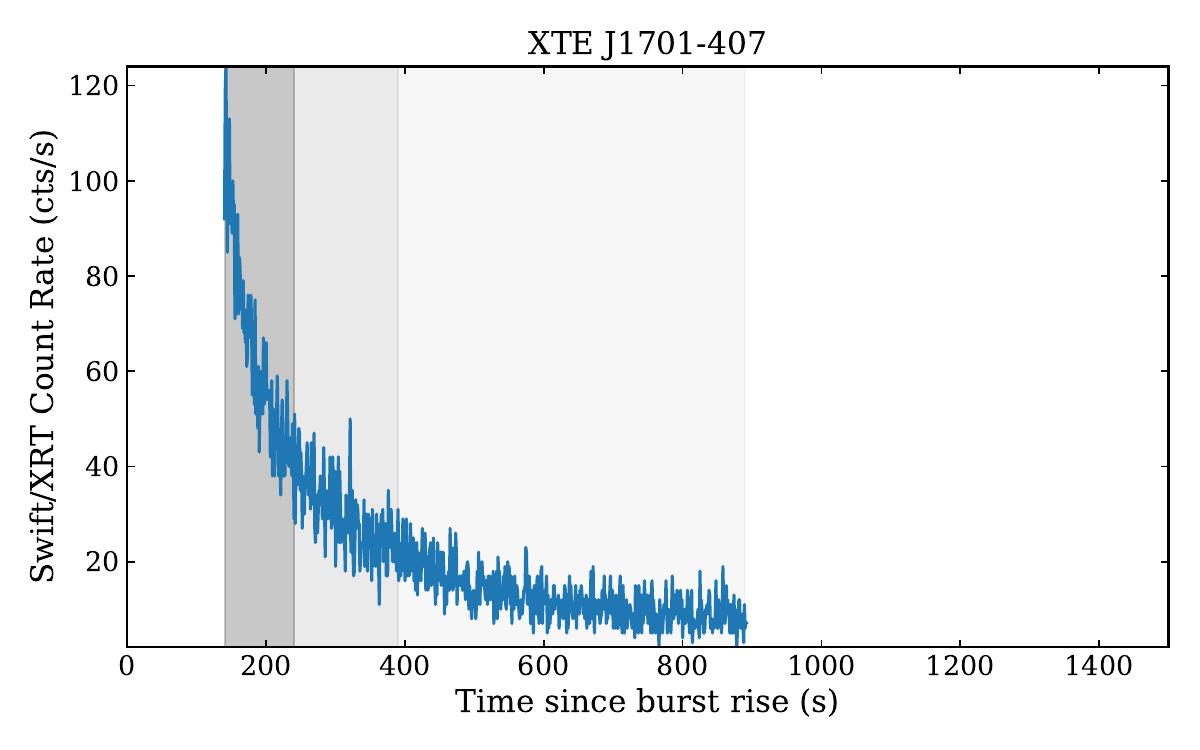}
\includegraphics[width=0.45\textwidth]{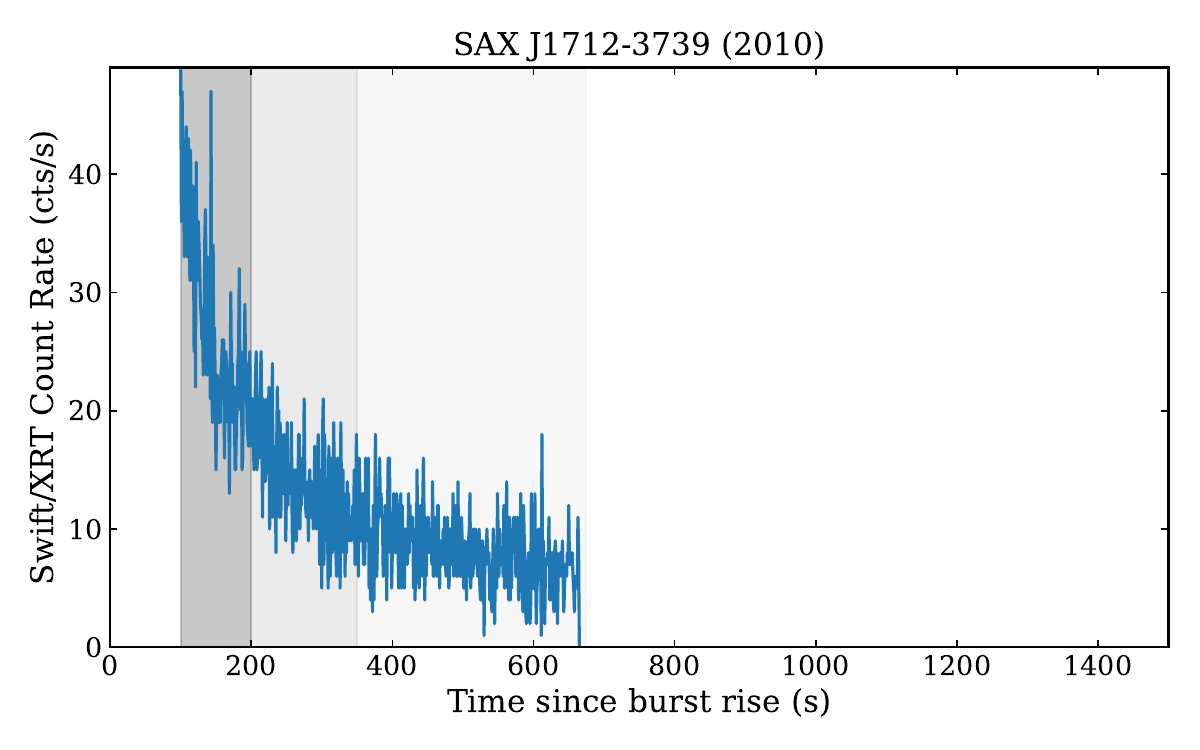}
\includegraphics[width=0.45\textwidth]{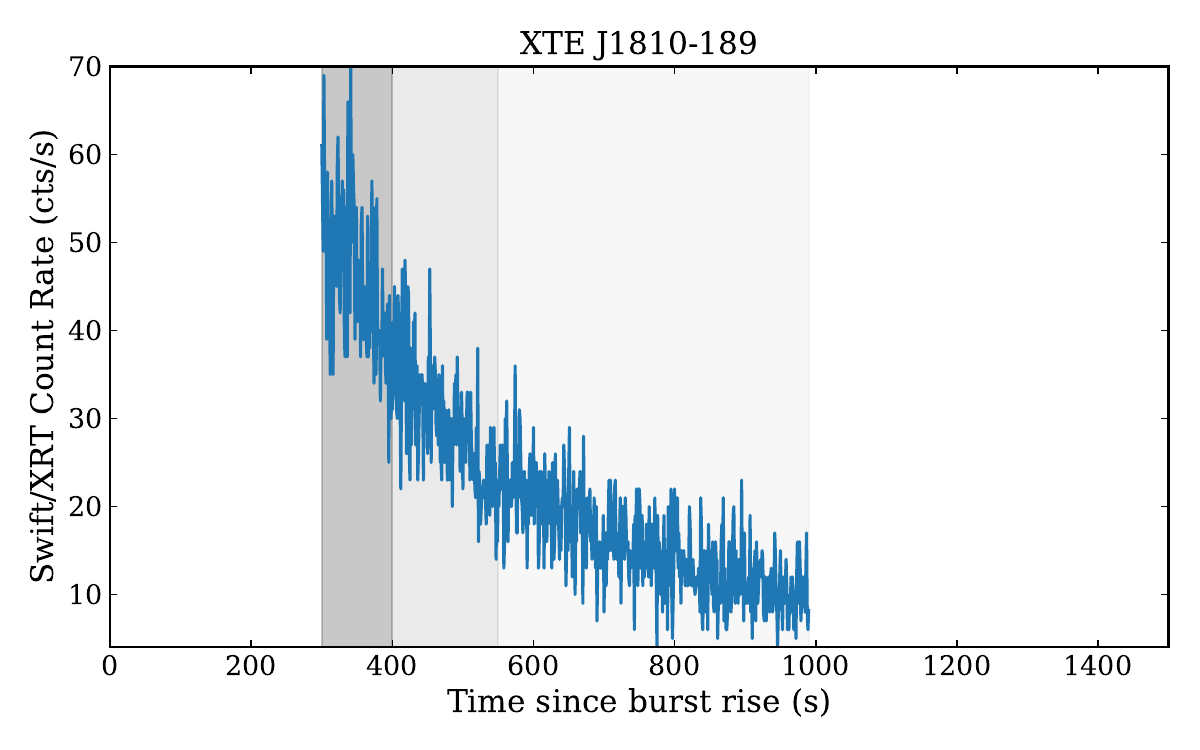}
\includegraphics[width=0.45\textwidth]{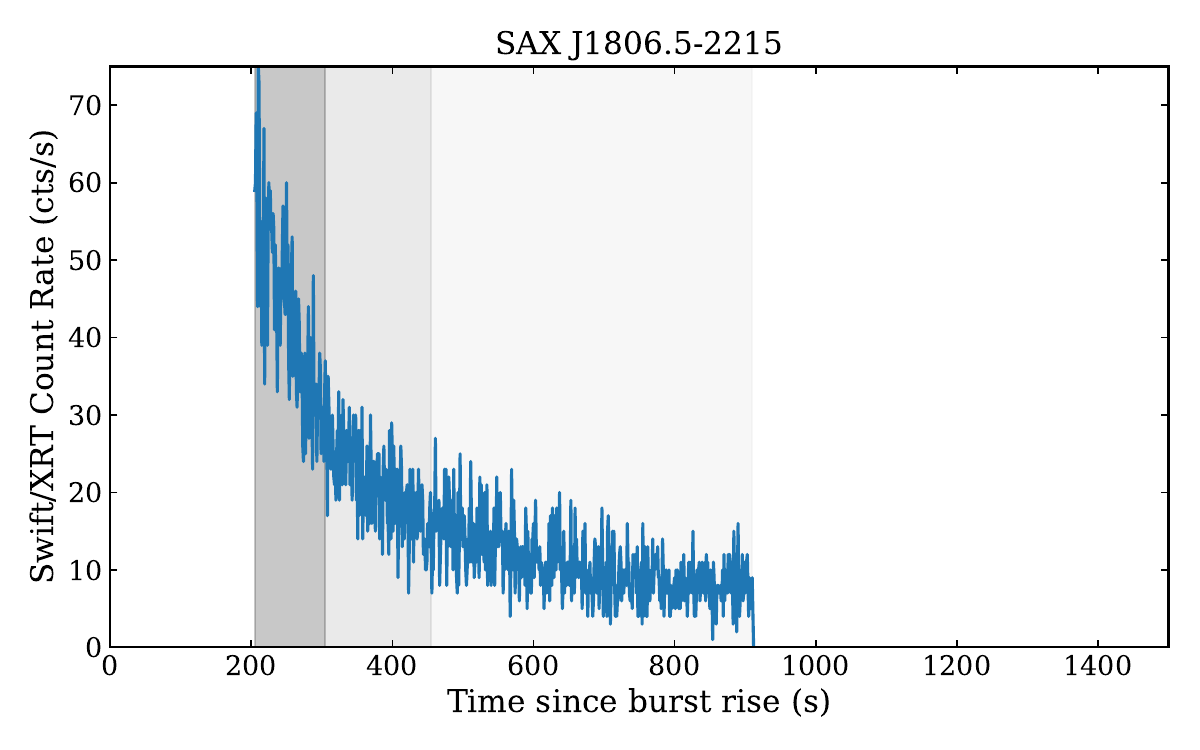}
\end{center}
\caption{\xrt\ light curves (1-s resolution) of the five bursts without fluctuations, which comprises our control sample (lower half of Table~\ref{tab:sample}). The layout of the light curves is the same as Fig.~\ref{fig:five_lc_mod}, with the three colored regions defining the intervals in each burst extracted for spectral analysis. The three intervals are referred to as Intervals 1, 2, and 3, starting at the beginning of each observation. The duration of the first two intervals ($t_1$ and $t_2$) are chosen to approximately bisect the decay of each burst, with the third interval (lasting a time $t_3$) reaching the end of each light curve. The duration of each interval is listed in Table~\ref{tab:sample}.}
\label{fig:five_lc_nomod}
\end{figure*}

The burst light curves for our samples are shown in Fig.~\ref{fig:five_lc_mod} (fluctuations) and Fig.~\ref{fig:five_lc_nomod} (no fluctuations/control sample). Previous spectral analysis of bursts with fluctuations by \citet{degenaar13} and \citet{int19} showed that the spectral shape of the 'high' and 'low' times during the fluctuations was largely unchanged, particularly above 2~keV. Therefore, for the sources in our sample with fluctuations, we extracted a single spectrum covering the fluctuation phase as well as spectra covering times before and after the fluctuations. For the control sample, we extracted spectra for different intervals along the burst tail to perform a qualitatively similar analysis as done for the bursts with fluctuations. In this case, the first two intervals are chosen to roughly bisect the decay of the light curve, with the final interval reaching to the end of the burst tail. The intervals used for the time-resolved spectral analysis are indicated by the colored regions in Fig.~\ref{fig:five_lc_mod} and Fig.~\ref{fig:five_lc_nomod} and are listed in Table~\ref{tab:sample}. Details on the light curve properties of the fluctuations, such as the total duration, amplitude and repetition time, are reported in \citet{int19}. We here focus on the spectral properties.

%\begin{figure*}
%\centering
%\includegraphics[width=\textwidth]{five_panel_lightcurve.pdf}
%\caption{
%Swift/XRT light curves of the five bursts with fluctuations analysed in this work. Colored regions indicate the different time intervals used for spectral extraction, where xx, xx and xx concern the intervals before, during, and after the fluctations, respectively.
%}
%\label{fig:LC_mod}
%\end{figure*}

\subsection{Spectral analysis} 
\label{subsect:spec}
Following extraction, there are 13 spectra from the sample with fluctuations and 15 spectra from the control sample. Spectral analysis is performed using \textsc{xspec} v.12.15.1 \citep{arn96}, and, following \citet{int19}, data below $0.7$~keV are ignored for all fits. As the spectra were extracted from intervals with very different count rates, the highest energy bin that was above background often changed from spectrum to spectrum. Table~\ref{tab:spectra} provides the energy range that is analyzed and the corresponding total number of counts in each spectrum.

\begin{deluxetable*}{cccc}[t!]
\tabletypesize{}
%\tablewidth{0} 
\tablecaption{Details of the spectra analyzed in each interval. \label{tab:spectra}}
\tablehead{
\colhead{Source} & \colhead{Interval} & \colhead{Energy Rangy} & \colhead{Counts} \\
\colhead{} & \colhead{} & \colhead{(keV)} & \colhead{}
} % Note
\startdata
\multicolumn{4}{c}{Bursts with fluctuations}   \\
\hline
IGR J17062--6143 & PRE & 0.7--9.5 & 19387 \\ 
 & FLUC & 0.7--9.5 & 39367 \\
 & POST & 0.7--7 & 2025 \\
SAX J1712.6--3739 (2011) & PRE & 0.7--9.5 & 13849 \\ 
 & FLUC & 0.7--10 & 34793 \\
 & POST & 0.7--8.9 & 3381 \\
SAX J1712.6--3739 (2014) & PRE & 0.7--10 & 14940 \\ 
 & FLUC & 0.7--10 & 28306 \\
Swift J1734.5--3027 & FLUC & 0.7--10 & 21944 \\ 
 & POST & 0.7--9 & 28792 \\
4U 1850--087 & PRE & 0.7--10 & 64894 \\ 
 & FLUC & 0.7--10 & 32100 \\
 & POST & 0.7--9.5 & 17127 \\
\hline
\multicolumn{4}{c}{Bursts without fluctuations}   \\
\hline
4U 1246--58 & 1 & 0.7--7.5 & 4062 \\
 & 2 & 0.7--7 & 4098 \\
 & 3 & 0.7--7.2 & 8885 \\
XTE J1701--407 & 1 & 0.7--9.5 & 6249 \\
 & 2 & 0.7--8.9 & 4256 \\
 & 3 & 0.7--8.2 & 6031 \\
SAX J1712.6--3739 (2010) & 1 & 0.7--8.5 & 2190 \\
 & 2 & 0.7--7.2 & 2130 \\
 & 3 & 0.7--7.5 & 2791 \\
XTE J1810--189 & 1 & 0.7--10 & 4508 \\
 & 2 & 0.7--9.5 & 4381 \\
 & 3 & 0.7--8.2 & 6569 \\
SAX J1806.5--2215 & 1 & 0.7--9 & 3869 \\
 & 2 & 0.7--9 & 2931 \\
 & 3 & 0.7--9 & 4501 \\
 \hline
\enddata
%\tablecomments{}
\end{deluxetable*}

Following standard X-ray burst modeling practices \citep[e.g.,][]{int19}, each spectrum is fit with an initial model of an absorbed blackbody spectrum, where the \texttt{TBabs} absorption model is used with 'wilm' abundances \citep{tbabs00} and the column density is fixed at the Galactic value (Table~\ref{tab:sample}). The blackbody describes the burst emission from the NS. If this model is unsuccessful then the absorbing column density is thawed and allowed to vary. 

If significant residuals still remain then a \texttt{xillverNS} reflection spectrum is added to the model \citep{garcia22} and the column density is reset to the Galactic value. The \texttt{xillverNS} spectrum describes the reprocessed emission from a constant density slab illuminated by a blackbody. Reflection of the bright NS emission by the surrounding accretion disk is an expected outcome during X-ray bursts \citep{speicher22} and has been observed in several bursts \citep[e.g.,][]{keek17,guver22}. If the disk structure is changing during the fluctuations seen in Fig.~\ref{fig:five_lc_mod} --- due to a radiative warping instability or other causes --- then evidence for this could be found by searching for changes in the reflection spectrum during the fluctuation phase. 

The blackbody temperature in the \texttt{xillverNS} model is set to be equal to the one from the NS. We fix the density of the slab to be $10^{19}$~cm$^{-3}$, the largest value allowed by \texttt{xillverNS}, and the inclination to $30^{\circ}$. The ionization parameter $\xi$ and iron abundance $A_{\mathrm{Fe}}$ are allowed to vary, and \texttt{refl\_frac=-1} in order to produce only the reflection spectrum. In \textsc{xspec} vernacular, the model is \texttt{TBabs*(bbodyrad+xillverNS)}. The normalizations of the illuminating blackbody and the \texttt{xillverNS} model are determined by `cflux' models that compute the $0.1$--$10$~keV fluxes of each component. The absorbing column density is allowed to vary if the initial \texttt{xillverNS} fit is unsatisfactory. 

The above procedure generally yields an acceptable fit to most of the spectra we consider. For spectra that require reflection, we also test if the relativistically blurred version of \texttt{xillverNS}, called \texttt{relxillNS}, is a better description of the data. This may occur if the reprocessed radiation originates in the regions of the accretion disk closest to the NS. In \texttt{relxillNS}, the spin parameter of the NS is fixed at $0.2$ \citep{cook94,ss98,speicher23,neuweiler26}, the reflecting zone extends from the innermost stable circular orbit to $400$ gravitational radii ($r_g=GM/c^2$, where $M$ is the mass of the NS), and the emissivity index is fixed at $3.0$ \citep{wilkins18}. As there are no good constraints on the viewing angle into any of these sources, the inclination angle within \texttt{relxillNS} is initially frozen at 30~degrees, but we also check if a freed inclination improves the fit. 

We find that a few of the analyzed burst intervals require additional emission components in the model in order to account for residuals or features that were not sufficiently described by this baseline model. These situations are discussed further in Sect.~\ref{sect:igr} and Appendix~\ref{app:fluct} during the presentation of the individual sources. Error-bars are the $90$\% confidence levels computed for a single parameter of interest using the `error' command in \textsc{xspec}. 

\section{Fluctuations in \igr}
\label{sect:igr}
It is instructive to initially focus on the results from fitting the three intervals from the 2011 burst of \igr\ (Fig.~\ref{fig:five_lc_mod}; top left). The source has the lowest Galactic column density of the sample (Table~\ref{tab:sample}) and has the largest number of counts in any of the 5 FLUC intervals considered. Thus, \igr\ may have the highest likelihood of showing evidence of accretion disk changes during the flux modulation phase. In addition, \igr\ is known to exhibit an emission line at $\approx 1$~keV, and the origin of this line is investigated in some detail.

\begin{figure*}[p!]
\begin{center}
\includegraphics[width=0.45\textwidth]{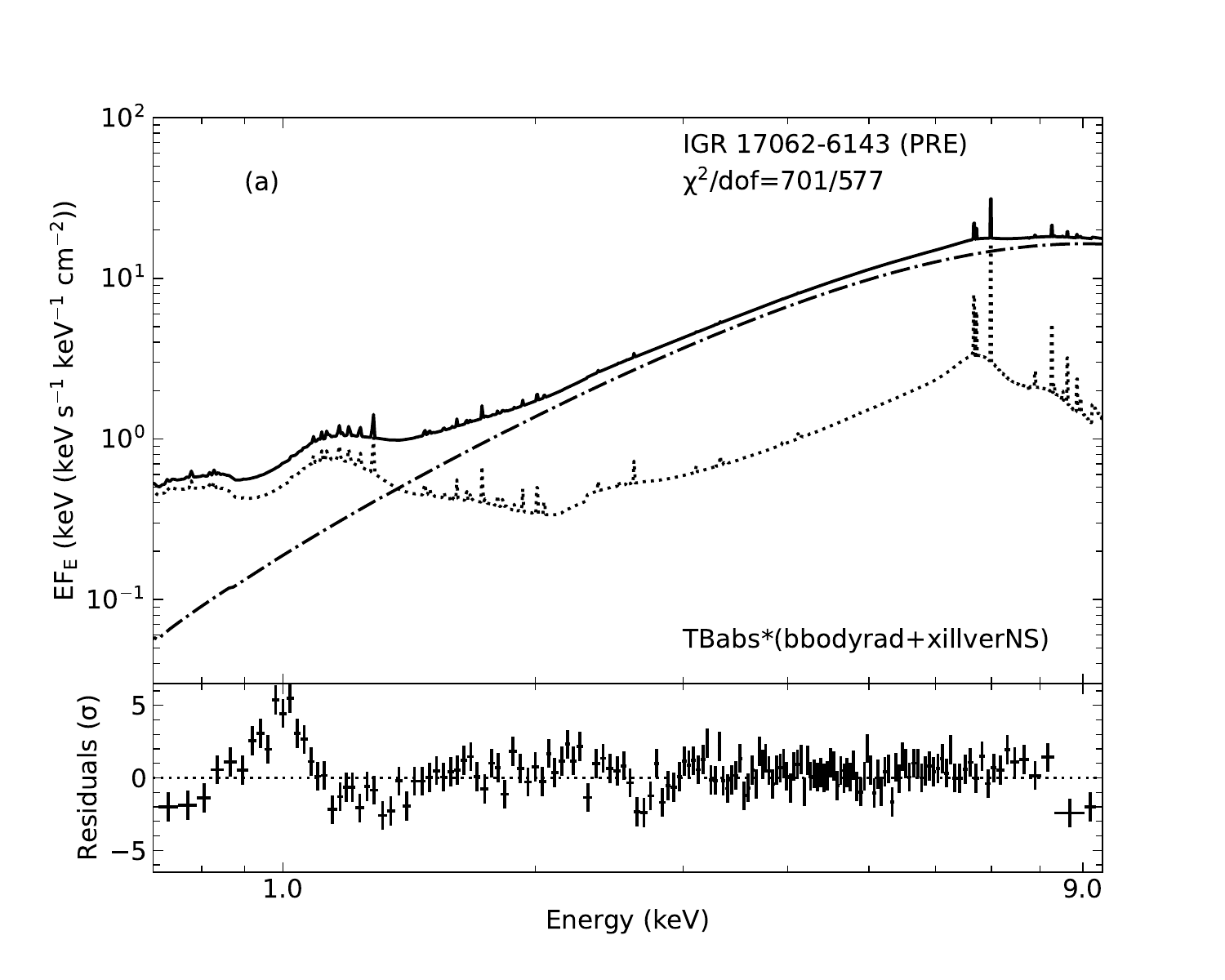}
\includegraphics[width=0.45\textwidth]{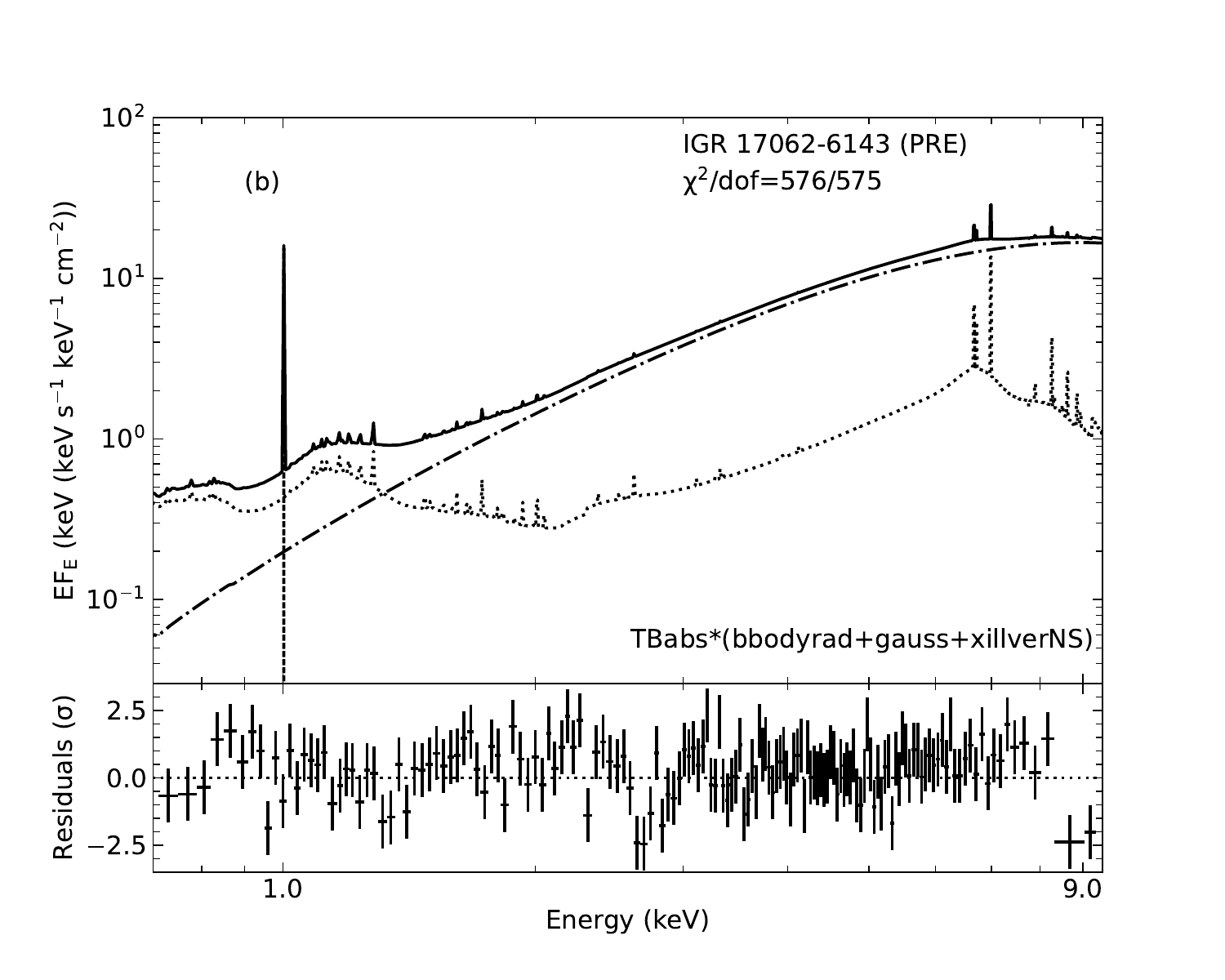}
\includegraphics[width=0.45\textwidth]{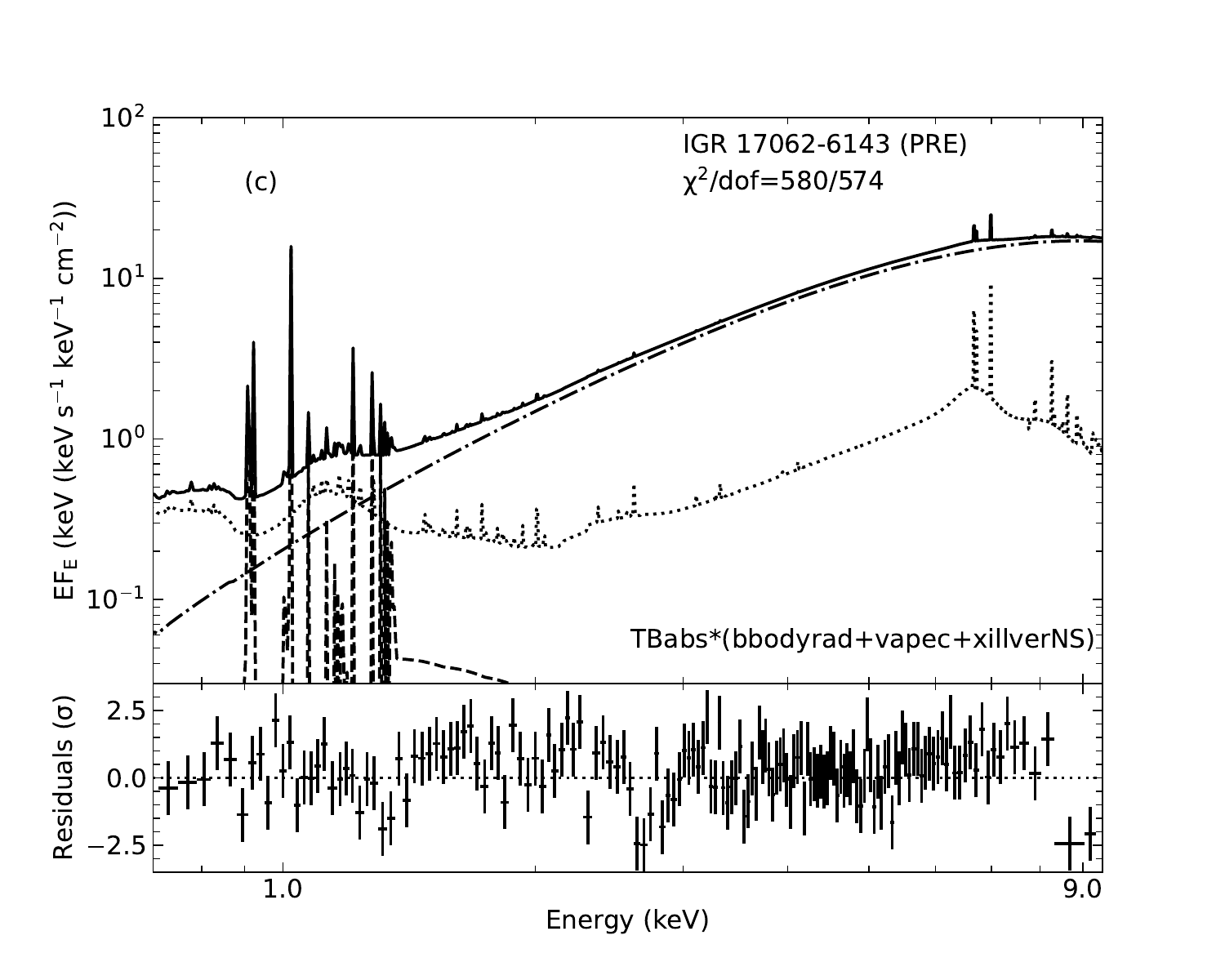}
\includegraphics[width=0.45\textwidth]{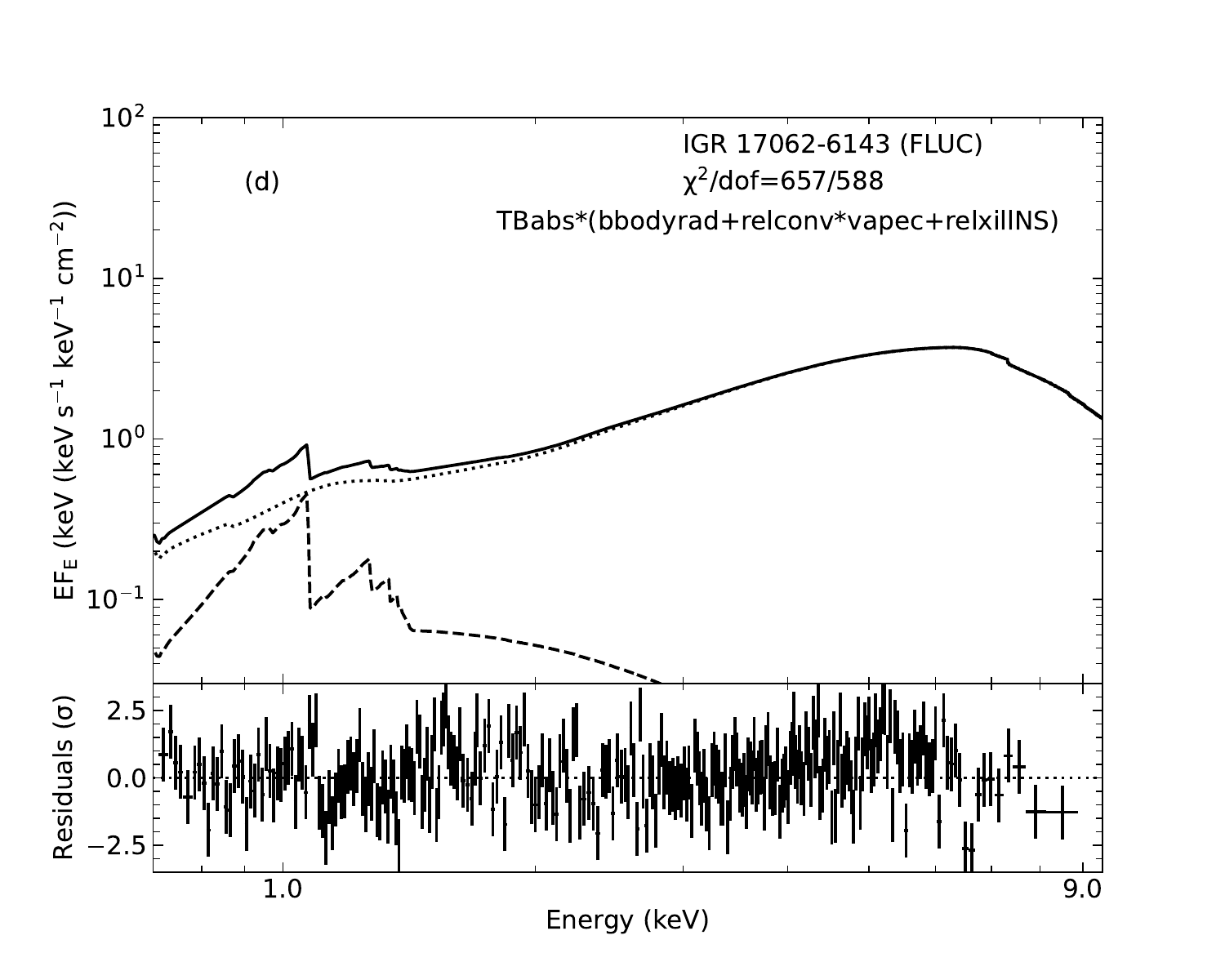}
\includegraphics[width=0.45\textwidth]{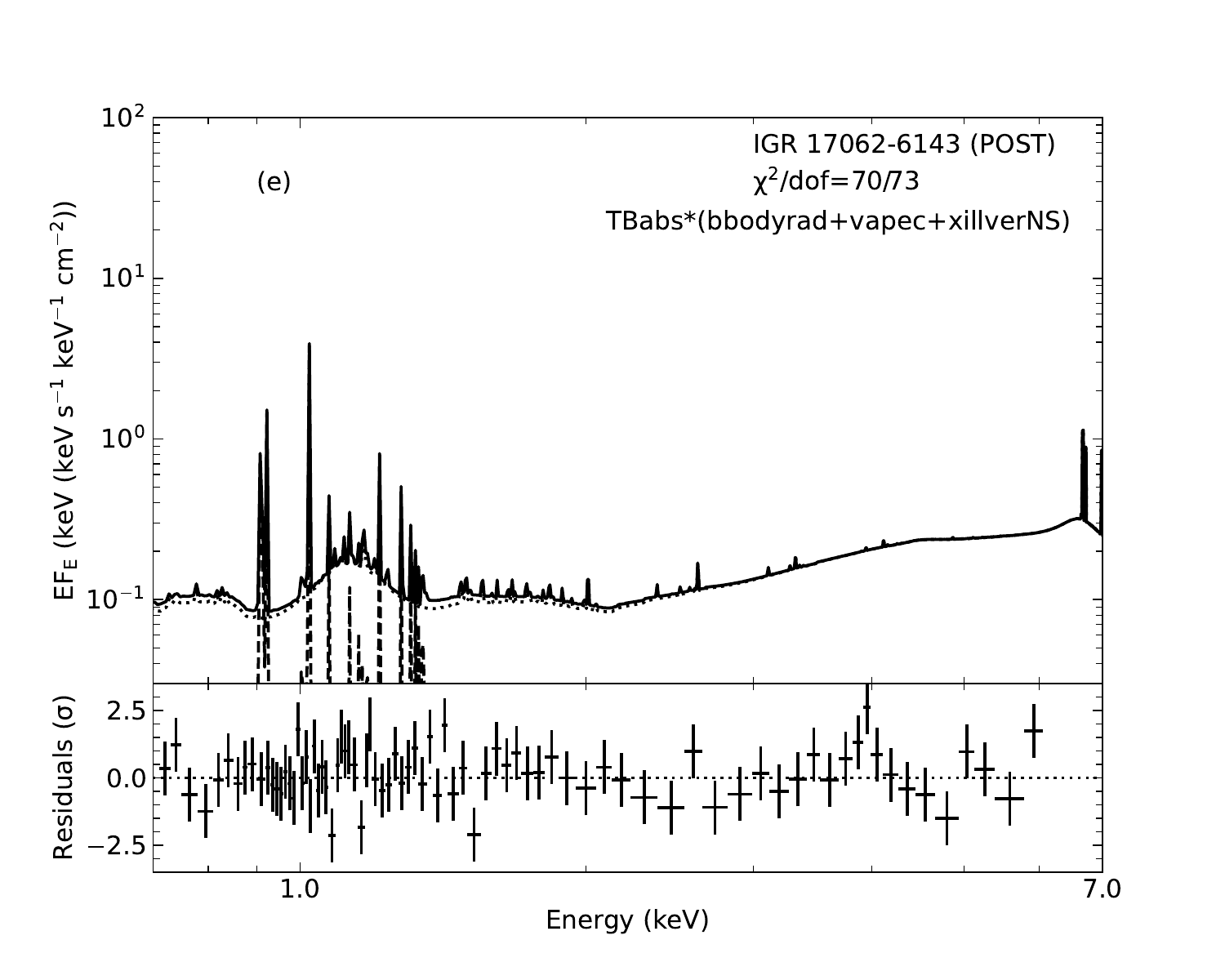}
\end{center}
\caption{Plots of the best-fit models and residuals (in units of $\sigma$) for the PRE, FLUC and POST intervals from the \igr\ burst (Fig.~\ref{fig:five_lc_mod} and Table~\ref{tab:sample}). The solid line in the upper-part of each plot shows the final model (plotted in $EF_E$ units) including the burst blackbody (dot-dashed line) and the \texttt{xillver} or \texttt{relxill} component (dotted line). Panels (a), (b) and (c) show the development of the final model describing the PRE interval from the \igr\ burst including a \texttt{vapec} component (dashed line). Panel (d) plots the results from the FLUC interval during the rapid modulations where the spectrum is dominated by relativistically blurred emission. Panel (e) shows the result from the POST interval, indicating that the spectrum remains dominated by ionized reflection. Table~\ref{tab:igrfits} lists the parameters for the final PRE, FLUC and POST intervals.}
\label{fig:igrfits}
\end{figure*}

In their initial analysis of this burst, \citet{degenaar13} found that the spectrum exhibited a soft excess when fit with a single absorbed blackbody. A soft excess is a natural outcome of reflection from the surrounding accretion disk \citep[e.g.,][]{ball04,garcia22,speicher22}, so we begin with modeling the PRE interval of \igr\ with a model that includes a \texttt{xillverNS} reflection component. As seen in Figure~\ref{fig:igrfits}(a), this model provides a good fit to most of the spectrum but leaves a significant line-like residual at $\approx 1$~keV that was also highlighted by \citet{degenaar13}\footnote{\citet{degenaar13} also find weak absorption features in the spectra of \igr\ at energies $> 7.5$~keV. These features have negligible impact on the quality of the broadband fits of interest here.}. Adding a narrow ($\sigma=0$) Gaussian line with central energy $1.00\pm0.01$~keV yields an excellent fit to the spectrum ($\chi^2$/dof$=576/575$; Fig.~\ref{fig:igrfits}(b)). The equivalent width of the emission line is $122$~eV. Neither replacing the \texttt{xillver} model with the relativistic version (\texttt{relxillNS}), nor allowing the absorbing column density to vary, meaningfully improves the fit from this point.

A $1$~keV emission line has been observed in the spectra of several other X-ray bursts \citep[e.g.,][]{bult19,strohmayer19,jaisawal25}, including other bursts from \igr\ \citep[e.g.,][]{keek17,bult21}. This line has been attributed to either the Fe-L complex or Ne \textsc{x} Ly$\alpha$ in a reflection spectrum. However, the \texttt{xillverNS} models includes both of these emission features and are unable to account for the $1$~keV line in the PRE \igr\ spectrum. Indeed, the \texttt{xillverNS} models shown in Fig.~\ref{fig:igrfits}(a,b) both have iron abundances maxed out at $10\times$ Solar and cannot fit the emission line. Unfortunately, it is not possible to vary the abundances of other metals in the \texttt{xillverNS} models.

The soft excess produced by X-ray reflection is a result of a bremsstrahlung spectrum emitted by the hot surface of the irradiated gas and has a strong dependence on the disk density \citep[e.g.,][]{ball04}. The gas density of accretion disks around neutron stars is expected to be $>10^{20-21}$~cm$^{-3}$ \citep{ss73}, above the maximum density allowed by \texttt{xillverNS} ($10^{19}$~cm$^{-3}$). Therefore, there is reason to expect that \texttt{xillverNS} is underpredicting the soft excess in \igr. To account for this, and to better constrain the origin of the $1$~keV emission line, a \texttt{vapec} spectrum \citep{atomdb12} is added to the \igr\ PRE model. \texttt{vapec} provides the spectrum from a collisionally-ionized plasma which consists of emission lines superimposed on a bremsstrahlung continuum with temperature $kT_{\mathrm{vapec}}$. The relative line strengths predicted by \texttt{vapec} will not be correct for the photoionized gas from the surface of the accretion disk. However, the line energies will be correct, and the energy resolution of the \xrt\ spectra is not sufficient to meaningfully distinguish the two scenarios. Therefore, adding the \texttt{vapec} model will approximate the impact of any missing soft excess emission and provide a mechanism to identify the $1$~keV emission line. Indeed, Fig.~\ref{fig:igrfits}(c) shows that replacing the Gaussian line with a \texttt{vapec} spectrum results in a good fit to the \igr\ PRE interval. By individually varying each metal in the \texttt{vapec} model, it is found that Ne is the only element needed in the spectrum and it satisfactorily describes the $1$~keV emission feature. An overabundance of Ne is consistent with \igr\ being an ultracompact X-ray binary \citep{jd03,hernIGR19,strohmayer19}. 

\begin{deluxetable*}{cccccccccc}[t!]
\tabletypesize{}
%\tablewidth{0} 
\tablecaption{Best-fit parameters found from the PRE, FLUC and POST intervals of the \igr\ intermediate duration burst.\label{tab:igrfits}}
\tablehead{
\colhead{Interval} & \colhead{Flux} & \colhead{BB Fraction} & \colhead{$\Delta N_H$} & \colhead{$\log \xi$} & \colhead{$kT_{\mathrm{BB}}$} & \colhead{$A_{\mathrm{Fe}}$} & \colhead{$kT_{\mathrm{vapec}}$} & \colhead{$A_{\mathrm{Ne}}$} & \colhead{$\chi^2$/dof} \\
 & \colhead{(erg cm$^{-2}$ s$^{-1}$)} & & \colhead{($10^{22}$~cm$^{-2}$)} & & \colhead{(keV)} & & \colhead{(keV)} & & 
} 
\startdata
PRE & $2.8\times 10^{-8}$ & $0.89$ & 0 & $3.30^{+0.15}_{-0.20}$ & $2.27^{+0.07}_{-0.06}$ & $7.70^{+2.3p}_{-3.06}$ & $0.61^{+0.11}_{-0.07}$ & $>1.8$ & $580/574$\\
FLUC & $7.1\times 10^{-9}$ & $<0.05$ & $+(0.15\pm0.05)$ & $3.81^{+0.07}_{-0.09}$ & $1.87^{+0.05}_{-0.04}$ & $10^{+0p}_{-1.5}$ & $0.86^{+0.39}_{-0.21}$ & $>1.3$ & $657/588$ \\
POST & $7.7\times 10^{-10}$ & $<0.17$ & $0$ & $3.20^{+0.17}_{-0.12}$ & $1.41^{+0.09}_{-0.1}$ & $10^{+0p}_{-2.87}$ & $0.51^{+0.08}_{-0.06}$ & $>1.2$ & $70/73$\\
\enddata
\tablecomments{The PRE and POST intervals are fit with the model \texttt{TBabs*(bbodyrad+vapec+xillverNS)}, while the FLUC spectrum is fit with the model \texttt{TBabs*(bbodyrad+relconv*vapec+relxillNS)}. The best-fit models and the residuals to the fits are shown in Figure~\ref{fig:igrfits}. The observed $0.1$--$10$~keV flux is listed in the 2nd column, followed by the fraction of the total flux contributed directly by the primary blackbody from the NS. $\Delta N_{H}$ is the change in column density from the Galactic column (see Table~\ref{tab:sample}). The key parameters of the reflection model are the ionization parameter ($\log \xi$), the temperature of the irradiating blackbody ($kT_{\mathrm{BB}}$), and the iron abundance in Solar units ($A_{\mathrm{Fe}}$). The \texttt{vapec} model is characterized by its temperature ($kT_{\mathrm{vapec}}$) and elemental abundances, but only Ne is found to have a non-zero abundance ($A_{\mathrm{Ne}}$) in the three intervals. The upper-limits to the blackbody fraction and the lower-limits to $A_{\mathrm{Ne}}$ are both calculated using a $\Delta \chi^2=+2.706$ criterion from the best-fit. A `p' in an error-bar indicates the parameter pegged at its upper or lower bound. } 
\end{deluxetable*}

With the \texttt{vapec} model providing the final element of a self-consistent fit to the PRE interval of \igr, we find that the spectrum of the burst before the modulations begin is dominated by the blackbody from the surface of the NS, comprising $\approx 90$\% of the total flux (Table~\ref{tab:igrfits}). The remaining flux, including the soft excess, is provided by a non-relativistic ionized ($\log \xi=3.3$) reflection spectrum, described by the combination of the \texttt{xillverNS} and \texttt{vapec} models. The non-relativistic fit suggests that the reflecting region is located some distance from the NS. To estimate the distance, we replace \texttt{xillverNS} with \texttt{relxillNS} and apply the relativistic blurring model \texttt{relconv} to the \texttt{vapec} spectrum. Allowing the inner radius to vary gives a distance of $\approx 163$~$r_g$ ($\chi^2/$dof$=579/573$) showing that the reflection spectrum is primarily produced at a radius of at least $\approx 3.4\times 10^7$~cm along the disk (assuming a $1.4$~M$_{\odot}$ NS). For reference, \cite{hernIGR19} model the full multi-wavelength spectral energy distribution of the accretion disk and measure the outer disk radius in this system as $2.2\times 10^{10}$~cm. In addition, the PRE spectrum does not require any neutral absorption in excess of the Galactic column.  

Applying a spectral model with reflection and a \texttt{vapec} component to the FLUC interval (when the source undergoes rapid modulations in flux) leads to several significant changes in the characterization of the spectrum (Fig.~\ref{fig:igrfits}(d) and Table~\ref{tab:igrfits}). First, the blackbody emission from the NS vanishes (the upper-limit is just $5$\% of the total flux) and the spectrum from \igr\ becomes entirely reflection dominated. Second, the reflection spectrum in the FLUC phase has a higher ionization parameter than in the earlier PRE phase ($\log \xi=3.81$ compared to $\log \xi = 3.3$), despite a drop in flux of $\approx 4$. In addition, the FLUC reflection spectrum is best described by relativistic reflection arising from the inner regions of the accretion disk. Employing the non-relativistic \texttt{xillverNS} model, as was done with the PRE spectrum, increases $\chi^2$ by $14$ for the same number of degrees of freedom. Allowing the inner radius of the relativistic reflector to vary does not improve the fit but constrains the inner edge to be $\la 12$~$r_g$. Lastly, we find an increase in the absorbing column density of $\approx 1.5\times 10^{21}$~cm$^{-2}$ during the FLUC interval.

The \igr\ spectrum changes again during the short POST interval following the fluctuations (Fig.~\ref{fig:igrfits}(e) and Table~\ref{tab:igrfits}). The total flux from the source has dropped by an order of magnitude compared to the FLUC interval, but the blackbody from the NS remains hidden from view (the upper-limit is 17\% of the flux) with the spectrum returning to being well described by non-relativistic ionized reflection ($\log \xi=3.2$). Relativistically blurred reflection results in a worse fit ($\Delta \chi^2=+16$), and, after allowing the inner radius to vary, is unable to constrain an inner radius. As with the PRE interval, no additional column density above the Galactic value is needed in the fit.

Overall, the results shown in Fig.~\ref{fig:igrfits} and Table~\ref{tab:igrfits} provide compelling evidence for a change in the accretion disk geometry during the time when rapid fluctuations are seen in the burst light curve. During this period, the direct blackbody emission from the NS surface is no longer visible, reflection from the inner-most regions of the disk becomes dominant and there is an excess of absorption along the line-of-sight. After the fluctuations stop, the column density is consistent with the Galactic value and reflection from the inner disk is no longer visible. Before considering possible explanations for these observations, we investigate whether the other 4 bursts in our sample that exhibit fluctuations (top half of Table~\ref{tab:sample} and Fig.~\ref{fig:five_lc_mod}) show similar changes in their spectra during the flux variations.

\section{Spectral Properties of Bursts with Fluctuations} 
\label{sect:other}
Details of the spectral fits to the other four bursts that exhibit fluctuations are collected in Appendix~\ref{app:fluct}. The analysis follows the same steps as described above for \igr\, but a 1~keV emission line was not found in any of the remaining 10 spectra. Therefore, if needed, a simple \texttt{bremss} model is used to model the soft excess. Unlike \igr, a strong relativistic reflection component is required to fit the spectra in all the intervals from the other four sources. Tables and figures describing the best fit models for the four sources are found in Appendix~\ref{app:fluct}. 

\begin{figure*}[t!]
\begin{center}
\includegraphics[width=0.49\textwidth]{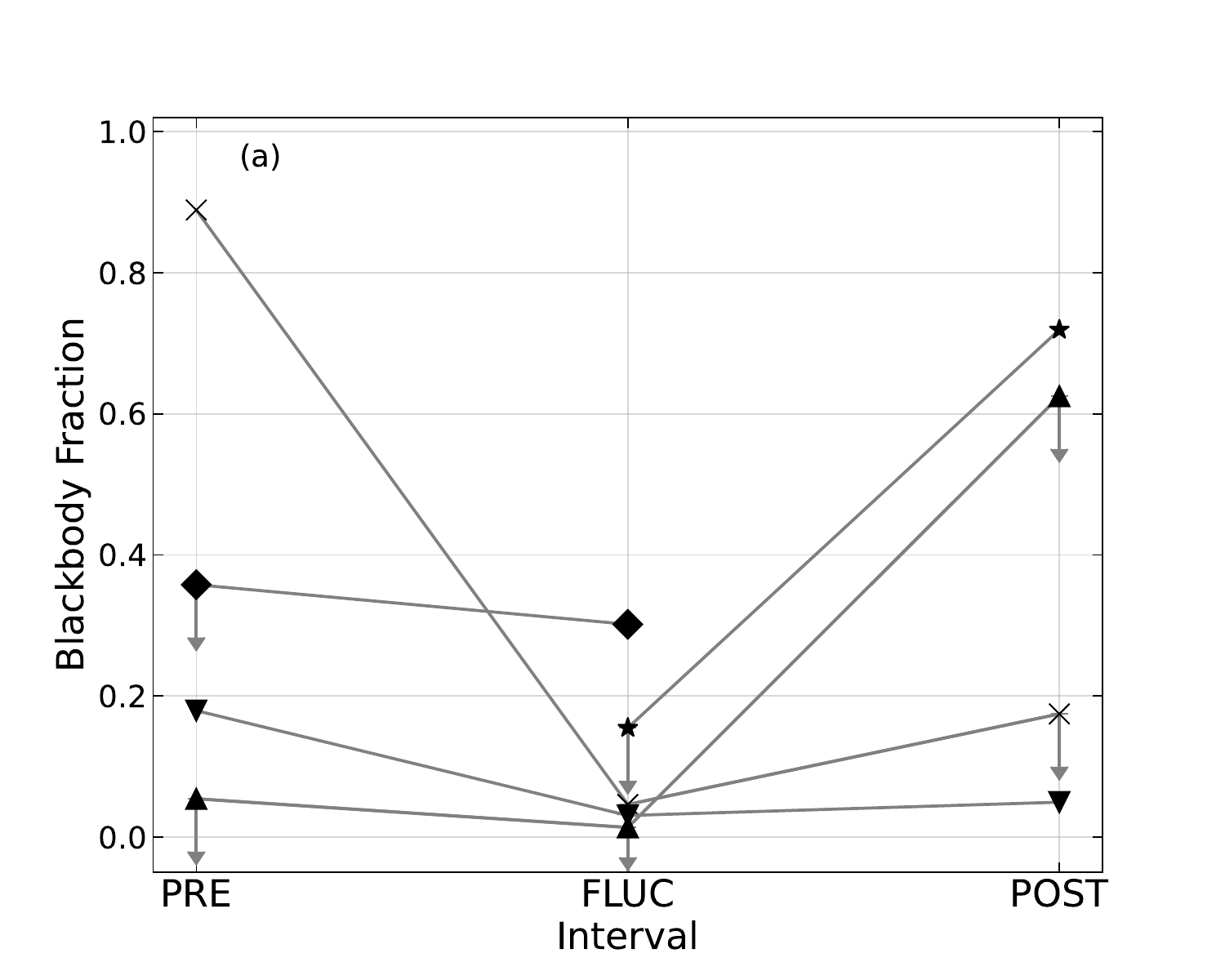}
\includegraphics[width=0.49\textwidth]{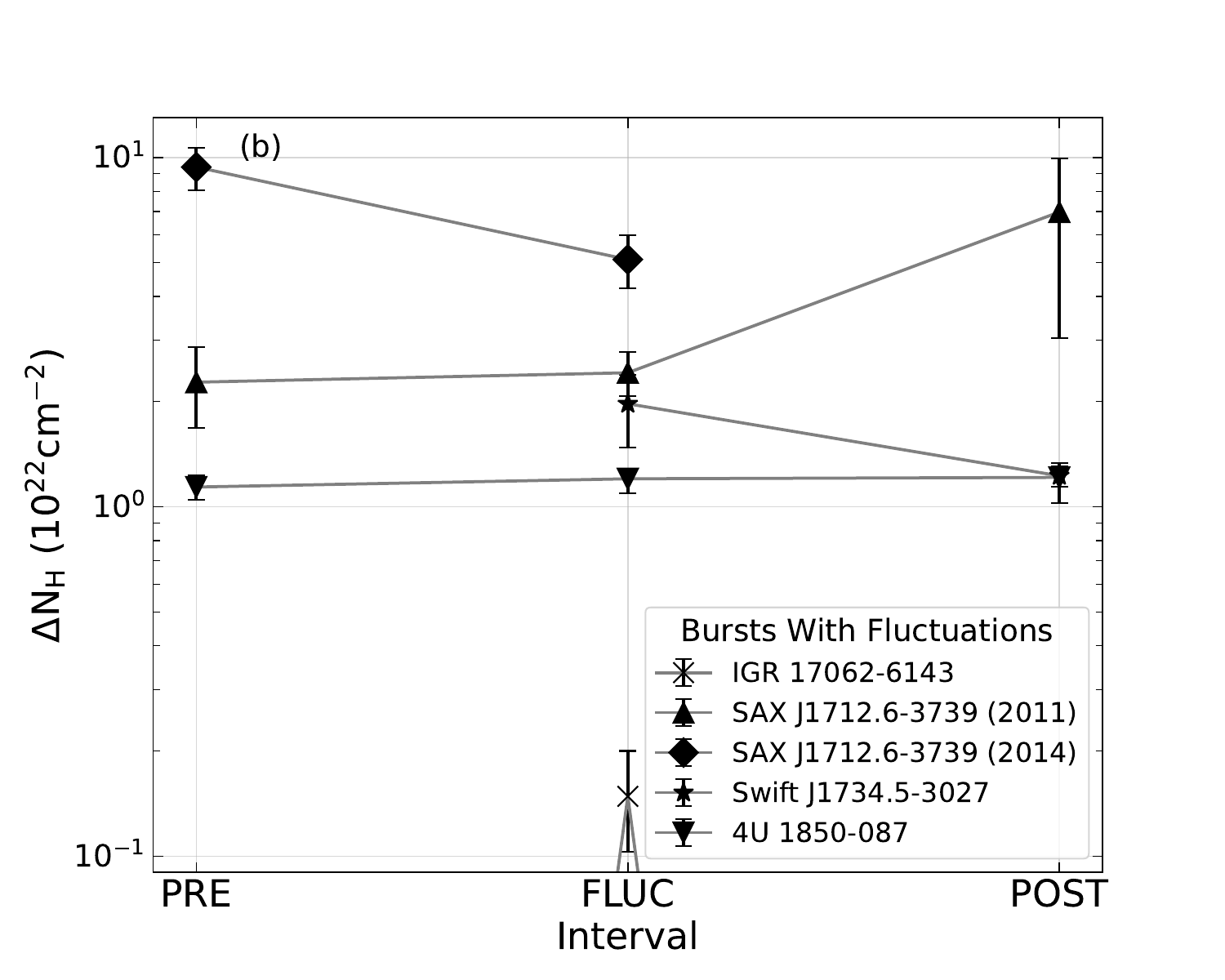}
\includegraphics[width=0.49\textwidth]{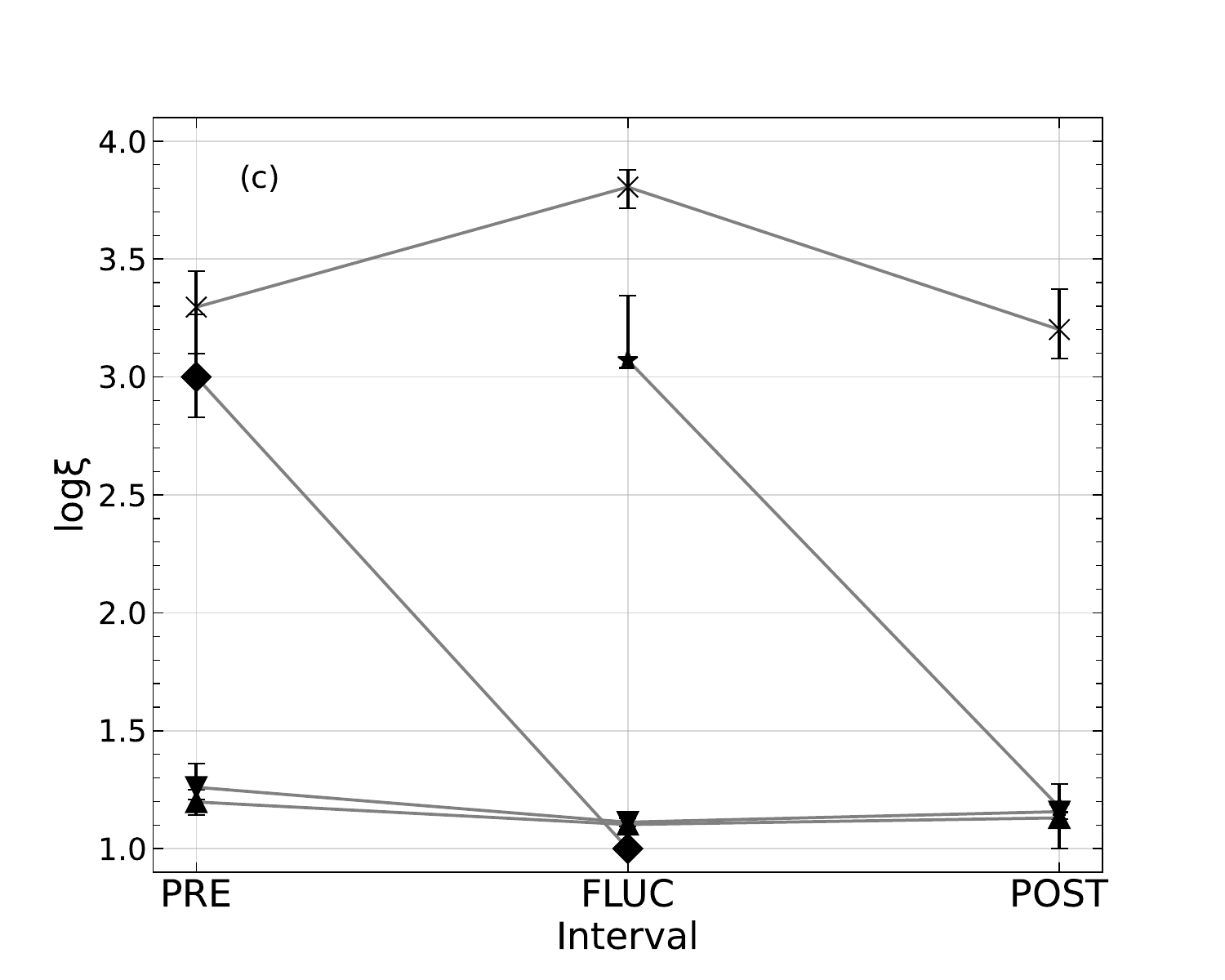}
\includegraphics[width=0.49\textwidth]{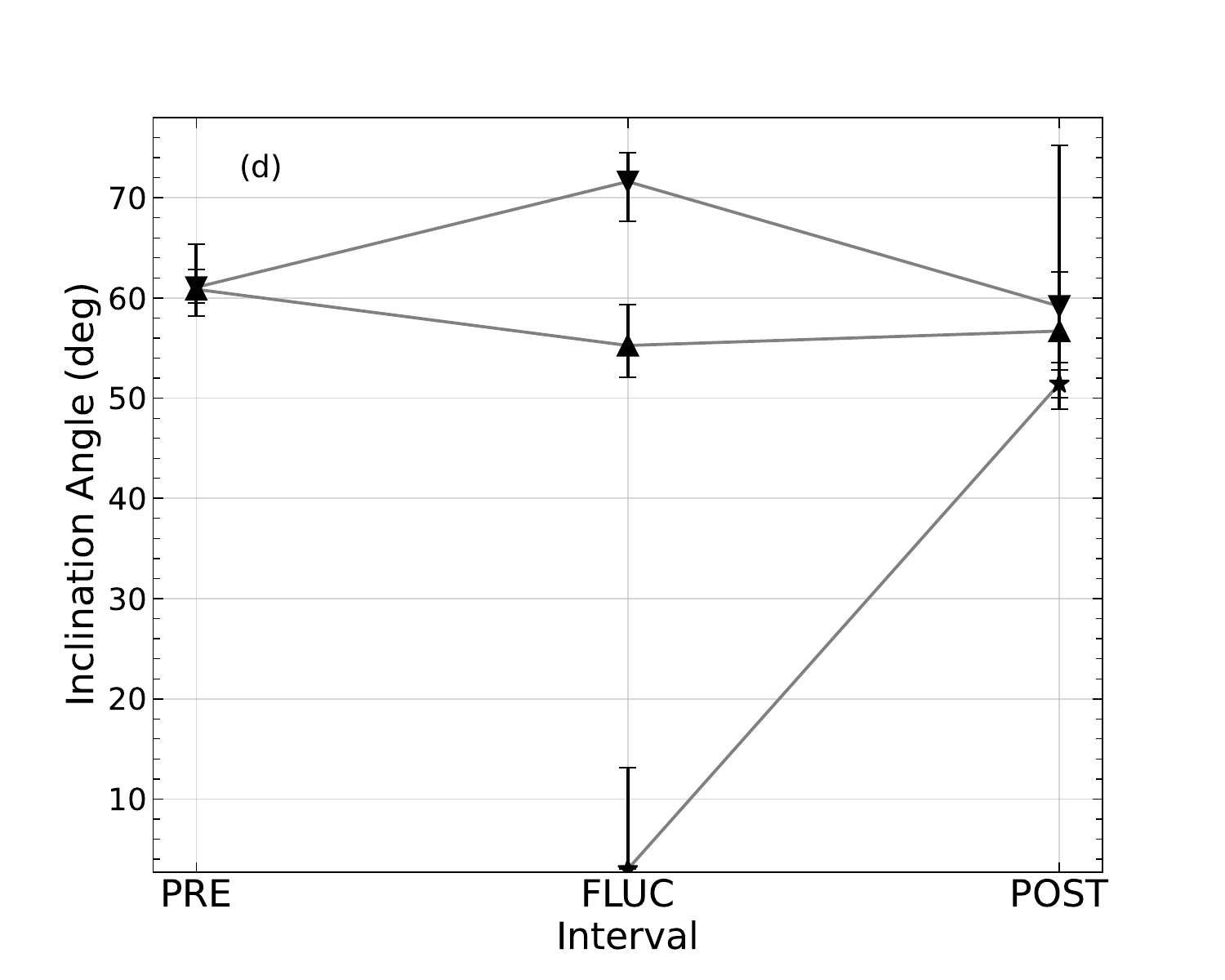}
\end{center}
\caption{Key results from fitting the PRE, FLUC and POST spectra from the 5 intermediate duration bursts that show rapid fluctuations in their light curves: (a) the fraction of the total flux in the blackbody component (including upper-limits), (b) the increase in absorbing column density, (c) the ionization parameter of the reflection component, and (d) the inclination angle of the accretion disk as measured by the reflection signal. Details of the fits are found in Sect.~\ref{sect:igr} (for \igr) and in Appendix~\ref{app:fluct} (for the other 4 sources). In panel (a) two of the upper-limit arrows extend below $0$ due to the plotting style, but the minimum possible blackbody fraction is zero. Panel (d) omits intervals where the inclination angle was frozen at 30 degrees.}
\label{fig:fluctfits}
\end{figure*}

To investigate if there are common trends in the spectral fits as the bursts move into and out of the period of rapid flux modulations, we collect four key results from the fits of each source and plot their changes with time in Figure~\ref{fig:fluctfits}\footnote{While not the focus of this investigation, it is also worth noting that, in addition to \igr, the bursts from \sax\ (2011), \swift, and \fouru\ all have best fit models where at least one interval requires a super-Solar abundance of Fe. Similar to \igr, both \sax\ and \fouru\ are candidate ultracompact X-ray binaries \citep{homer96,armas23} in which enhanced metal abundances in the accretion flow would be expected. Based on this, we suspect that \swift\ is also part of an ultracompact binary.}. 
Based on the results from \igr, we focus on (a) the fraction of the total flux contributed by the blackbody from the NS, (b) the change in the absorbing column density along the line-of-sight, (c) the ionization parameter of the reflecting region, and (d) the inclination angle of the disk (as determined by the relativistic reflection component). This last parameter is not constrained in the \igr\ fits, but is able to be determined in several of the intervals from the other 4 sources.

Panel (a) of Figure~\ref{fig:fluctfits} plots the fraction of the $0.1$--$10$~keV flux from the burst blackbody in each interval. Broadly, this property measures how much of the NS surface that is directly observed. We find that the blackbody fraction exceeds $0.5$ (i.e., contributed the majority of the flux) only in the PRE interval of \igr\ and the POST interval of \swift. In all other observations, the blackbody is found to provide a minority ($\approx 30\%$) or is consistent with no contribution to the observed spectra, even during times without fluctuations. In addition, the blackbody fraction changes across at least one interval (i.e., from PRE to FLUC; from FLUC to POST) in all bursts except \sax\ (2011) where only upper-limits are able to be measured (Table~\ref{tab:sax17-2011}). In particular, for the 4 bursts where we have both PRE and FLUC spectra, the blackbody fraction drops in the FLUC interval in 2 bursts (\igr\ and \fouru), increases in 1 burst (\sax\ (2014)), and remains undetectably small for \sax\ (2011). Similarly, we find the blackbody fraction increases from FLUC to POST in two bursts (\swift\ and \fouru) and continues to be undetectable in \igr\ and \sax\ (2011). These results clearly suggest that the environment around the neutron star is being strongly impacted by the bursts, significantly affecting the view towards the blackbody source. Moreover, these impacts change in time, with slight evidence that the view of the NS is worse during the fluctuations.

Unlike \igr, which only showed an increase in the line-of-sight column density during the FLUC interval, an enhancement in $N_{\mathrm{H}}$ is observed in all of the intervals in the other 4 bursts (Fig.~\ref{fig:fluctfits}(b)). In three bursts, the $\Delta N_{\mathrm{H}}$ is approximately the same value in the three intervals (\sax\ (2011), \swift, and \fouru). Only \sax\ (2014) and \igr\ show significant changes in $\Delta N_{\mathrm{H}}$ from one interval to another, with both changing from the PRE to FLUC intervals, albeit in opposite directions. This mix of results indicates that, with the possible exception of \igr, changes in column density during the burst are not necessarily connected to the fluctuations observed in the light curves.

 As the burst flux declines moving from the PRE to the POST interval, we would expect that the ionization parameter $\log \xi$ should lower in response. However, if the region responsible for the reflection signal changes in an interval (e.g., it arises from a smaller radius along the disk and so sees a higher flux, or has a different density and so can be more easily ionized) then the measured ionization parameter would not necessarily follow the expected decline. Panel (c) of Fig.~\ref{fig:fluctfits} plots $\log \xi$ found in each interval of the 5 bursts. The ionization parameter from the \igr\ burst increases by a factor of $3.2$ in the FLUC interval, despite a $4\times$ drop in total flux (see Table~\ref{tab:igrfits}). The ionization parameter then decreases by a factor of $4.1$ in the POST interval although the flux falls by a factor of $9.2$. None of the other bursts show a similar increase in $\log \xi$ in the FLUC or POST intervals, although the fits to both \sax\ (2014) and \swift\ give ionization parameters that drop more quickly than the total flux (Tables~\ref{tab:sax17-2014} and~\ref{tab:swift}). Specifically, $\xi$ is lower by a factor of $100$ in the FLUC interval of the \sax\ (2014) burst, despite only a $7.3\times$ drop in total flux from the PRE interval. The ionization parameter in the POST interval of the \swift\ burst is $78\times$ lower than the FLUC interval although the total flux dropped by a factor of $8.8$. The flux changes across the intervals in the \sax\ (2011) and \fouru\ bursts are more modest: the average decrease is a factor of $1.5$ for \fouru\ and $2.5$ for the 2011 \sax\ burst (Tables~\ref{tab:sax17-2011} and~\ref{tab:fouru}). The corresponding changes in the ionization parameters in these two bursts are within a factor of 2 of these flux changes. Overall, three of the bursts that exhibit rapid modulations show evolution in $\log \xi$ that appear to be disconnected from the burst flux. Such behavior may indicate large changes in the location or structure of the reflecting region in the accretion disk.

 Finally, for three of the studied bursts (\sax\ (2011), \swift\ and \fouru) the spectral fits were significantly improved when the inclination angle of the \texttt{relxillNS} reflection component was allowed to vary. Changes in the angles across the different intervals may therefore indicate variations in the geometry of the accretion disk responsible for the reflection spectrum. Figure~\ref{fig:fluctfits}(d) shows that the inclination angle appears to be approximately constant across the three intervals of the \sax\ (2011) burst, but significant changes are found in both \swift\ and \fouru. In the latter burst, the inclination angle increases from $\approx 60$ to $\approx 70$ degrees in the FLUC interval, before returning to about $60$ degrees in the POST region (Table~\ref{tab:fouru}). This range of inclinations is within the one inferred for intermediate dippers \citep{galloway16}, but \fouru\ has never previously been observed as a dipper \citep{int19}. A more striking potential change in inclination is found in the fits of the \swift\ burst (Table~\ref{tab:swift}), increasing from $\approx 3$ to $\approx 50$~degrees between the FLUC and POST intervals. Although tentative, these detected changes in inclination angle support a scenario where the fluctuations observed in the light curve are connected to changes in the accretion disk geometry.

In summary, the results shown in Figure~\ref{fig:fluctfits} show that the spectral properties in 4 out of the 5 X-ray bursts that exhibit rapid fluctuations are affected by the presence of the flux modulations. Apart from the 2011 burst of \sax, variations are seen in the amount of the blackbody seen from the NS, the ionization parameter of the disk reflecting region, and, in 2 cases, the inclination angle of the disk. In the case of \igr, the distance of the reflecting zone also appeared to change in the FLUC interval (Sect.~\ref{sect:igr}). While enhanced neutral absorption was found in all the bursts, the change in column density is largely insensitive to the presence of the fluctuations (except for \igr). In the next section, we investigate if similar effects are seen in the spectra of bursts that do not show rapid flux modulations in their light curves.

\section{Comparison to Bursts without Fluctuations}
\label{sect:control}
The results from fitting the 15 spectra extracted from the control sample of 5 long bursts that do not show fluctuations are found in Table~\ref{tab:nofluctfits}. Appendix~\ref{app:nofluct} contains plots of the best-fitting models and fit residuals for each of the 15 spectra.
It is immediately clear from examining Table~\ref{tab:nofluctfits} that the spectra from the control sample are notably different from the bursts that underwent rapid fluctuations (Sect.~\ref{sect:other}). Four of the 5 bursts are well fit by a simple absorbed blackbody spectrum, with a blackbody fraction of 1.0 in 11 of the 15 spectra. Of these four bursts, only the third interval from 4U~1246-58 requires a reflection component which is found to contribute only $\approx22$\% of the observed flux. These results indicate the view of the radiating NS in the bursts without fluctuations is not significantly altered or blocked by major changes in the geometry of the inner accretion disk.

\begin{deluxetable*}{ccccccccc}[t!]
\tabletypesize{}
%\tablewidth{0} 
\tablecaption{Best-fit parameters found from the control sample of 5 long X-ray bursts without fluctuations \label{tab:nofluctfits}}
\tablehead{
\colhead{Source} & \colhead{Interval} & \colhead{Flux} & \colhead{BB Fraction} & \colhead{$\Delta N_H$}& \colhead{$\log \xi$} & \colhead{$kT_{\mathrm{BB}}$} & \colhead{$A_{\mathrm{Fe}}$} & \colhead{$\chi^2$/dof} \\
 & & \colhead{(erg cm$^{-2}$ s$^{-1}$)} & & \colhead{($10^{22}$~cm$^{-2}$)} & & \colhead{(keV)} & & 
 } 
\startdata
4U~1246-58 & 1 & $2.7\times 10^{-9}$ & $1.0$ & $+(0.29^{+0.11}_{-0.10})$ & - & $0.84\pm 0.02$ & - & $193/160$ \\
 & 2 & $1.6\times 10^{-9}$ & $1.0$ & $0$ & $-$ & $0.76\pm 0.02$ & - & $156/153$ \\
 & 3 & $1.0\times 10^{-9}$ & $0.78$ & $0$ & $1.48^{+0.25}_{-0.48p}$ & $0.66\pm 0.02$ & $1^f$ & $275/262$ \\
 XTE~J1701-407 & 1 & $7.7\times 10^{-9}$ & $1.0$ & $+(4.6\pm 0.5)$ & - & $1.58^{+0.07}_{-0.06}$ & - & $259/236$ \\
  & 2 & $2.7\times 10^{-9}$ & $1.0$ & $+(4.5\pm 0.5)$ & - & $1.16^{+0.05}_{-0.04}$ & - & $164/173$ \\
 & 3 & $9.6\times 10^{-10}$ & $1.0$ & $+(2.9\pm 0.3$) & - & $1.04\pm 0.03$ & - & $271/224$ \\
 SAX~J1712.6-3739 (2010) & 1 & $2.1\times 10^{-9}$ & $1.0$ & $+(0.67^{+0.33}_{-0.29})$ & - & $1.31\pm 0.08$ & - & $95/96$ \\
 & 2 & $1.0\times 10^{-9}$ & $1.0$ & $+(1.3^{+0.4}_{-0.3})$ & - & $1.06\pm 0.06$ & - & $81/94$ \\
  & 3 & $4.5\times 10^{-10}$ & $1.0$ & $+(0.93^{+0.25}_{-0.23}$) & - & $0.92\pm 0.04$ & - & $137/117$ \\
XTE~J1810-189 & 1 & $4.5\times 10^{-9}$ & $<0.60$ & $+(5.8^{+0.9}_{-0.6})$ & $3.43^{+0.28}_{-0.40}$ & $1.70^{+0.15}_{-0.24}$ & $10^{+0p}_{-5.1}$ & $171/178$ \\
 & 2 & $2.5\times 10^{-9}$ & $<0.62$ & $+(4.8^{+0.8}_{-0.5})$ & $3.68^{+1.02p}_{-0.32}$ & $1.50^{+0.16}_{-0.10}$ & $10^{+0p}_{-4.5}$ & $198/174$ \\
  & 3 & $1.2\times 10^{-9}$ & $<0.68$ & $+(6.4^{+0.6}_{-1.0})$ & $2.58^{+0.55}_{-0.29}$ & $1.02^{+0.14}_{-0.05}$ & $10^{+0p}_{-5.2}$ & $206/236$ \\
SAX J1806.5-2215 & 1 & $5.2\times 10^{-9}$ & $1.0$ & $+(5.5^{+0.9}_{-0.8})$ & - & $2.07^{+0.2}_{-0.1}$ & - & $170/161$ \\
 & 2 & $1.2\times 10^{-9}$ & $1.0$ & $+(9.7^{+4.9}_{-3.5})$ & - & $1.67\pm 0.3$ & - & $124/125$ \\
 & 3 & $9.0\times 10^{-10}$ & $1.0$ & $+(5.5\pm 0.6)$ & - & $1.29\pm 0.05$ & - & $156/178$ \\
\enddata
\tablecomments{The majority of the spectra are well fit with a simple absorbed blackbody model, \texttt{TBabs*bbodyrad}. Only XTE~J1810-189 and the third interval of 4U~1246-58 requires a contribution from a \texttt{relxillNS} reflection spectrum. Interval 3 of 4U~1246-58 also exhibits a hard excess which was modeled using a \texttt{nthcomp} model with a best fit photon-index of $\Gamma=1.5^{+0.4}_{-0.5}$. A `p' in an error-bar indicates the parameter pegged at its upper or lower bound. A 'f' superscript denotes that the parameter was frozen at that value. Plots of the best-fitting models and the fit residuals are found in Appendix~\ref{app:nofluct}.} 
\end{deluxetable*}

\begin{figure*}[t!]
\begin{center}
\includegraphics[width=0.49\textwidth]{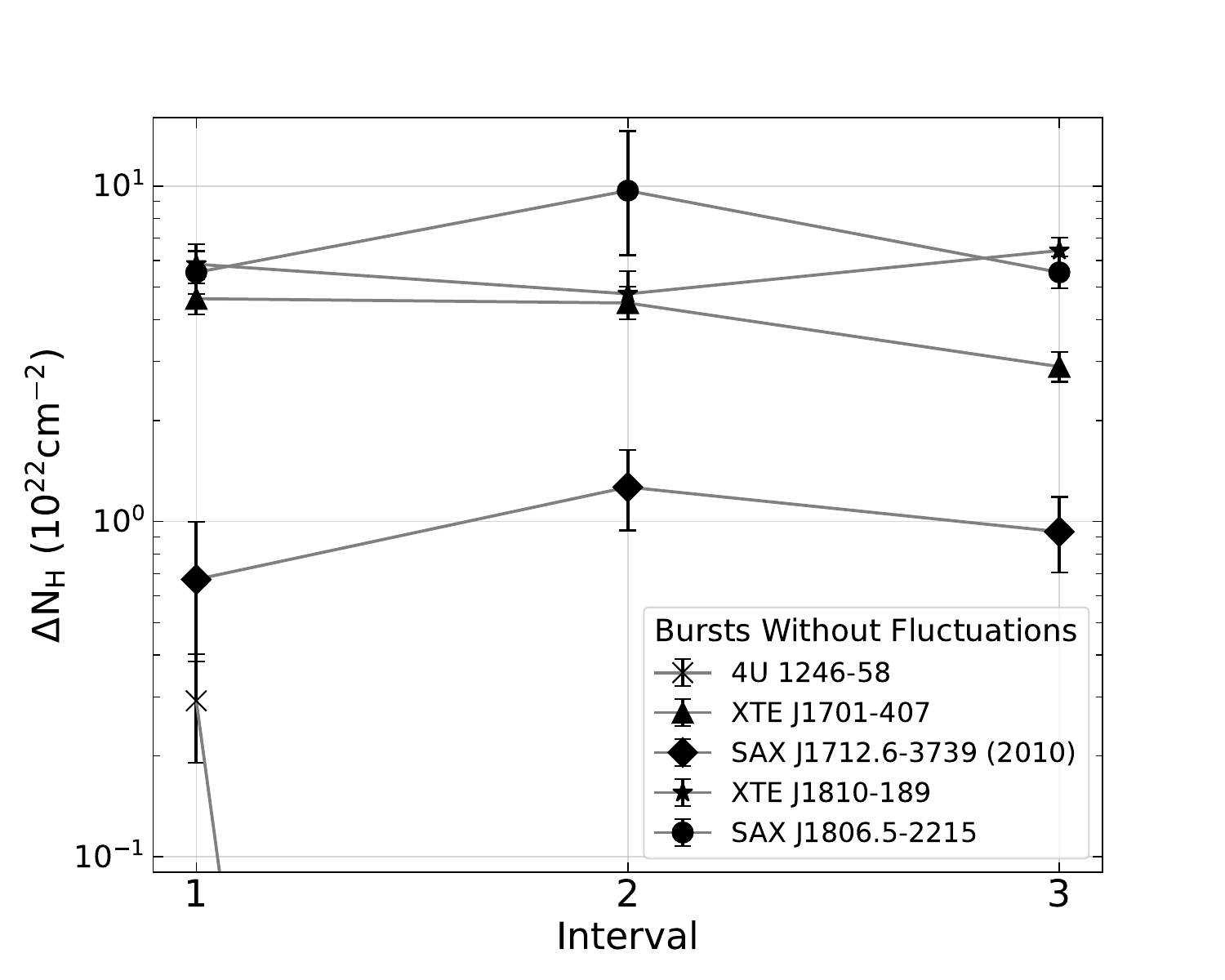}
\includegraphics[width=0.49\textwidth]{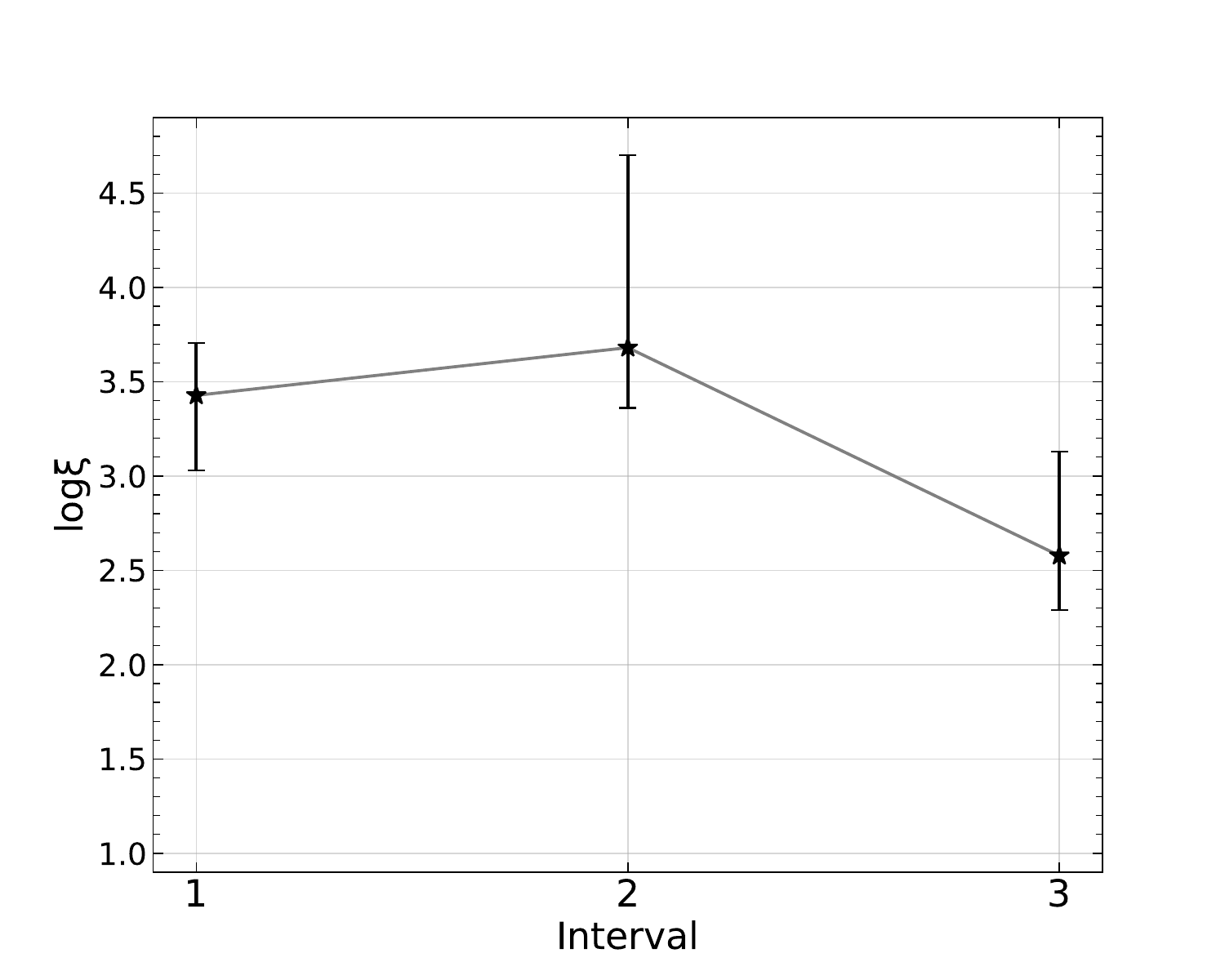}
\end{center}
\caption{As in Fig.~\ref{fig:fluctfits}, but now showing the results from fitting the spectra from the control sample of 5 bursts without fluctuations. The changes in the obscuring column density are shown in the left panel, while the right panel shows the ionization parameter from the only burst in the sample that required a strong reflection component in each interval (XTE~J1810-189). The fit parameters are listed in Table~\ref{tab:nofluctfits}, and plots showing the best models and residuals are found in Appendix~\ref{app:nofluct}. }
\label{fig:nofluctfits}
\end{figure*}

The five bursts without fluctuations all show enhanced column densities along the line-of-sight in at least one of their intervals (left panel of Fig.~\ref{fig:nofluctfits}). In 4U~1246-48, the excess $N_{\mathrm{H}}$ disappears after Interval 1, but the higher column densities in the other four sources are either approximately constant (SAX J1712.6-3739 (2010), XTE~1810-189, SAX J1806.5-2215) or decreases in the third interval (XTE~J1701-407). Other than the increase in $N_{\mathrm{H}}$ seen in the PRE interval of \igr, the variations in the column density appear to be quite similar between the bursts with and without fluctuations. To quantify this, we compute the standard deviation in $\Delta N_{\mathrm{H}}$ for each burst in the two samples. The average standard deviation for the 5 sources that show fluctuations is $1.27$, while it is $0.93$ for the control sample. However, the slightly larger value for the fluctuation sample is affected by the POST interval of \sax\ (2011) which has a particularly large error-bar, so the difference in the two standard deviations is not significant. Lastly, the fractional change in $N_{\mathrm{H}}$, $\Delta N_{\mathrm{H}}/N_{\mathrm{H,Gal}}$, is calculated to determine if there is a difference in the magnitude of the excess column densities between the two samples. The median $\Delta N_{\mathrm{H}}/N_{\mathrm{H,Gal}}$ from all 13 spectra in the fluctuations sample is $2.93$ while it is $2.96$ from the control sample. Therefore, on average, bursts from both samples show similar relative increases and variations in $N_{H}$ during the tails of the light curves. While changes in $N_{H}$ in individual bursts may be connected to fluctuations (e.g., \igr), it does not appear to be a common feature of the phenomenon.

One of the bursts without fluctuations, from XTE~J1810-189, is best fit with an ionized relativistic reflection spectrum with no contribution from a blackbody, similar to many of the bursts that showed fluctuations. The ionization parameters from the three intervals of the burst are shown in the right-panel of Figure~\ref{fig:nofluctfits}. The flux falls by a factor of 1.8 from the first to the second intervals, while the best-fit $\log \xi$ actually increases in interval 2. Similarly, the $\approx 2.1\times$ drop in flux between intervals 2 and 3 is not reflected in the best-fit ionization parameter in the last interval. Thus, the reflection signal from this burst is similar to bursts with fluctuations in the sense that changes in the ionization parameter do not track the evolution in the overall burst flux, a fact that is returned to in Sect.~\ref{sect:discuss}. 

Overall, the control sample of 5 bursts without fluctuations show significantly different spectral properties throughout the burst tails when compared to the bursts with fluctuations (see Sects.~\ref{sect:igr} and~\ref{sect:other}). Four of the five bursts in the control sample are simply described by an absorbed blackbody model with only the burst from XTE~J1810-189 dominated by ionized relativistic reflection. This striking difference in the spectra of the two samples, where the view of the NS surface is so clearly impeded in the bursts with fluctuations but is dominant in the control sample, supports the idea that a major change in the disk structure must be connected to the presence of rapid fluctuations in the light curves shown in Fig.~\ref{fig:five_lc_mod}.

\section{Discussion}
\label{sect:discuss}
Sections~\ref{sect:igr} and~\ref{sect:other} showed that the presence of rapid flux variations in the tail of long X-ray bursts is connected to spectra that are frequently dominated by ionized reflection and contains evidence for significant physical changes in the disk. The most striking example of this behavior is the burst from \igr, where the view of the NS surface disappeared once the fluctuations began, replaced by relativistically blurred reflection which then vanishes after the fluctuations cease, indicating the inner-disk is no longer in view. While none of the other four bursts with fluctuations present such a straightforward `beginning' and 'ending' to the spectral changes, they all exhibit weak to non-existent blackbodies from the NS and relativistic ionized reflection spectra. The fits also find that the spectral models frequently show large swings in best fit parameters as the source either starts and stops the flux variations (e.g., Fig.~\ref{fig:fluctfits}). In contrast, the spectra of four out of the five bursts in our control sample do not require any significant reflection components and are well described by an absorbed blackbody (Table~\ref{tab:nofluctfits}).  

\subsection{The Role of Burst Duration and Energy}
\label{sub:fluence}
The difference in spectra between the two samples suggests that there must be a physical distinction in the bursts, the accretion disks, or both, which results in rapid flux fluctuations to temporarily appear in the tails of the bursts. Interestingly, bursts from \sax\ appear in both samples, showing that different bursts from the same system may or may not exhibit flux variations. This fact again indicates that accretion disk properties do not singularly drive the fluctuations. To investigate the influence of the burst properties, we consider $t_{5\%}$ as a measure of the burst duration (Table~\ref{tab:sample}) and the observed fluence as a proxy for the energy released (Table~\ref{tab:fluences}). Three of the fluences are obtained from the literature and the remaining are estimated from our analysis using the durations and fluxes measured in each interval. These calculated fluences  underestimate the true fluences by a small factor. In the three cases where literature values are available, the estimated fluences are on average $2.3\times$ smaller than the ones from the literature. Therefore, our calculated fluences are a satisfactory estimate of the fluence from each burst. Lastly, because of the large break in the \sax\ (2014) observation, only the 2nd, FLUC interval is used to compute the fluence and we also use the duration of this interval instead of the uncertain $t_{5\%}$ value in Table~\ref{tab:sample}. 

\begin{deluxetable*}{ccc}[t!]
\tabletypesize{}
%\tablewidth{0} 
\tablecaption{Fluence of each burst in the sample. \label{tab:fluences}}
\tablehead{
\colhead{Source} & \colhead{Fluence} & \colhead{Source} \\
\colhead{} & \colhead{(erg cm$^{-2}$)} & \colhead{}
} % Note
\startdata
\multicolumn{3}{c}{Bursts with fluctuations}   \\
\hline
IGR J17062--6143 & $1.9\times 10^{-5}$ & \citet{degenaar13} \\ 
SAX J1712.6--3739 (2011) & $5.5\times 10^{-4}$ & Estimated \\ 
SAX J1712.6--3739 (2014) & $2.5\times 10^{-4}$ & Estimated \\ 
Swift J1734.5--3027 & $1.1\times 10^{-5}$$ $ & \citet{bozzo15} \\ 
4U 1850--087 & $1.1\times 10^{-3}$ & Estimated \\ 
\hline
\multicolumn{3}{c}{Bursts without fluctuations}   \\
\hline
4U 1246--58 & $1.0\times 10^{-6}$ & Estimated \\
XTE J1701--407 & $3.5\times 10^{-6}$ & \citet{linares09} \\
SAX J1712.6--3739 (2010) & $4.7\times 10^{-7}$ & Estimated \\
XTE J1810--189 & $1.3\times 10^{-6}$ & Estimated \\
SAX J1806.5--2215 & $1.1\times 10^{-6}$ & Estimated \\
 \hline
\enddata
\tablecomments{The estimated fluences are calculated using the interval times from Table~\ref{tab:sample} and the measured $0.1$-$10$~keV fluxes found from spectral fitting. Given the long break in the light curve, only the FLUC interval is used to estimate the fluence for the burst from \sax\ (2014).}
\end{deluxetable*}
\begin{figure}[t!]
\begin{center}
\includegraphics[width=0.50\textwidth]{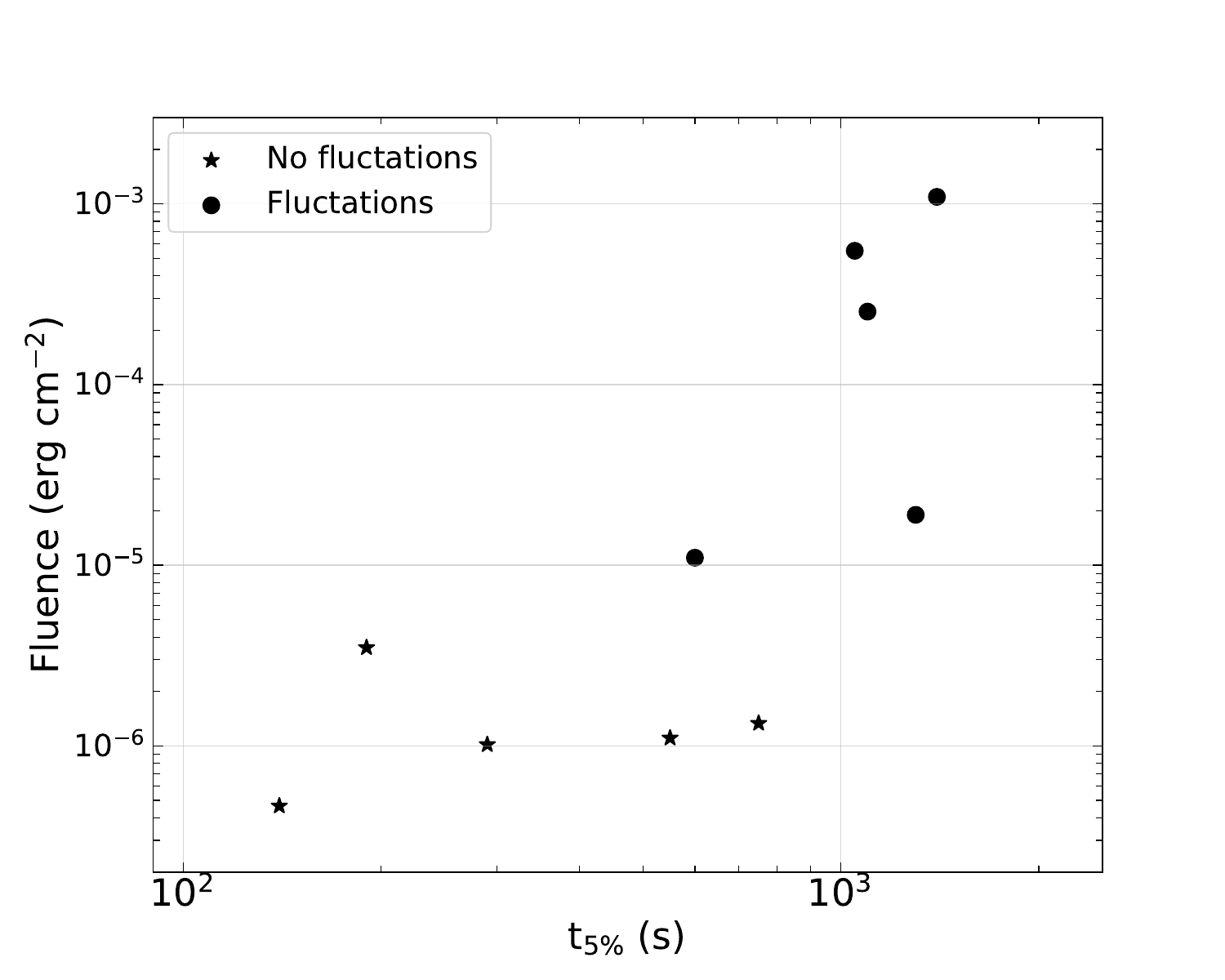}
\end{center}
\caption{The fluences of the 10 bursts in the sample (Table~\ref{tab:fluences}) plotted against their durations as measured by $t_{5\%}$ (Table~\ref{tab:sample}). The duration of the FLUC interval is used for \sax\ (2014) rather than the estimated $t_{5\%}$. The stars denote the bursts without fluctuations and the solid points are the bursts that exhibit rapid fluctuations in their light curves. The plot shows that the presence of fluctuations is connected to longer and more energetic bursts. }
\label{fig:fluence}
\end{figure}

Figure~\ref{fig:fluence} plots the fluences from Table~\ref{tab:fluences} against $t_{5\%}$ with the bursts with fluctuations shown as solid circles and the ones without fluctuations as solid stars. The figure clearly shows that the two samples reside in different regions of the plot, with the bursts that exhibit fluctuations being longer and with larger fluences than the control sample. The mean and standard deviations of $t_{5\%}$ from the fluctuation (control) sample are $1090\pm 309$~s ($384\pm 259$~s). For the fluences, the means and standard deviations are $(3.85\pm 4.53)\times 10^{-4}$~erg~cm$^{-2}$ for the bursts with fluctuations and $(1.48\pm 1.17)\times 10^{-6}$~erg~cm$^{-2}$ for the control sample. The larger fluences in the bursts with fluctuations are not simply because they are longer bursts. The average fluence of the control sample is $\approx 260\times$ smaller than the sample with fluctuations, while the average $t_{5\%}$ is only $\approx 2.8\times$ smaller. Therefore, the higher fluences in the bursts with fluctuations is reflective of a larger energy release, i.e., more powerful bursts\footnote{This conclusion is also supported by the `touchdown' times measured by \citet{int19}. This time estimates how long the burst was radiating above the Eddington limit for the NS. The average touchdown time for the bursts with fluctuations is $293$~s, while it is $89$~s for the control sample. Interestingly, the longest touchdown time in the control sample is the burst from XTE~J1810-189 (200~s), which had spectra most similar to the bursts that showed rapid fluctuations.}.

Given that the bursts with fluctuations are on average more powerful and longer than the ones without fluctuations, it is worth considering if this could influence the results of the spectral fitting. Specifically, the spectra extracted from the bursts in the control sample typically have far fewer counts than the spectra analyzed from the bursts with fluctuations (Table~\ref{tab:spectra}). To test if the strong reflection signals found in many of the bursts that show rapid variations in flux are just a result of the large number of counts we simulated a 40~s \xrt\ observation of the best-fitting model to the FLUC interval of \swift\ (Fig.~\ref{fig:xtej1810nofluct}). After grouping to a minimum of 20 counts per bin and ignoring data below $0.7$~keV and above $9.2$~keV, the faked spectrum has only 4315 counts. Fitting these data with an absorbed blackbody (i.e., \texttt{TBabs*bbodyrad}) leaves clear residuals from the reflection features ($\chi^2/$dof=$255/179$). Therefore, we conclude that the complex spectra found in the bursts with fluctuations are not a result of the larger number of counts in the data. This conclusion is supported by the clear detection of reflection in 4 spectra in the control sample (in particular, XTE~J1810-189), even with the lower number of counts.

\subsection{Missing Reflection Features in the Control Sample}
\label{sub:missing}
It is clear that the lack of strong reflection features in the spectra of four of the five bursts in the control sample (Table~\ref{tab:nofluctfits}) is not a result of sensitivity. Since there must be accretion disks around the NSs in these systems, the missing reflection features suggest that one or more physical effects are occurring which is suppressing disk reflection in these bursts. We note that enhanced soft X-ray absorption is not an explanation for the missing reflection signal since reflection is clearly identified in XTE~J1810-189, so the underlying effects are likely associated with the accretion disk itself. Reflection features from disks can be weak or non-existent if the disk is not present, very highly ionized, or not Compton thick \citep{ball04,be05,garcia22,speicher22}. Simulations of bursts impacting accretion disks show that the disks may be disrupted, but are not destroyed \citep[e.g.,][]{fbb19}. Similarly, bursts are expected to highly ionize the inner disks, weakening the reflection features, when the burst is at its peak brightness \citep{speicher22}. Neither of these effects seem to explain the lack of reflection features in the tail of four bursts in the control sample. The remaining possibility is that the accretion disk is no longer Compton thick and able to produce a reflection spectrum. This could happen because the strong X-ray heating from the burst inflates the disk, decreasing its surface density and optical depth \citep{fbb19}. Such a transient change in accretion structure is consistent with reflection emerging in the third interval of the burst from 4U~1246-58 (Fig.~\ref{fig:fourunonfluct}). Accretion disks with lower surface densities will be more easily affected by heating from bursts, and therefore more likely to not show reflection features \citep{fbb19}. According to standard \citet{ss73} accretion theory, lower surface densities will result from smaller accretion rates and/or larger values of the $\alpha$ viscosity parameter. Intermediate duration bursts are expected to arise from systems with accretion rates of $\approx 0.1$--$1$\% of the Eddington rate \citep{gk21}. These considerations further support an interpretation that there are distinct physical differences in the accretion disks of systems with bursts that exhibit fluctuations and those that do not.  

\subsection{Radiation-Driven Warp Interpretation}
\label{sub:warp}
Collecting the information from the spectral fitting and the above discussion, we find that the available evidence strongly indicates that the presence of rapid flux variations in the light curves of some X-ray bursts is connected to changes in the inner accretion disk structure. In addition, the fluctuations are more likely to be present in bursts that are both longer and more energetic. These qualities are all consistent with the predictions of a radiative warping instability impacting the accretion disk in these systems. \citet{ball23} showed that disks will more likely become unstable to radiative driven warping as burst durations and luminosities increase. Moreover, depending on the exact initial conditions of the disk, the growth of the warps can be slow and extend over large regions of the disk, or can be concentrated in the inner disk while rapidly growing and decaying. This variety of outcomes naturally explains the range of changes seen in the reflection features found in the spectral fitting of the bursts with fluctuations. For example, in the burst from \igr, the disappearance of the blackbody after the PRE interval would be due to a warp on the side of the disk closest to the observer blocking the NS from view, allowing the observer to see strong reflection from the warp on the far side of the disk. An orbiting, dynamically changing warp can explain the enhanced absorption, flux variations, higher ionization parameter and strong reflection seen in the FLUC interval (Sect.~\ref{sect:igr}). However, as found from the spectral fits from the other sources in the sample, disk warps are not simply isolated only to the times of the FLUC intervals. Rather, in most systems, disk warping would be happening both before and after this time, and the fluctuations only appear when certain conditions are met in the disk. This implies that disk warping may occur in bursts that do not exhibit rapid fluctuations. In particular, we hypothesize that the burst from XTE~J1810-189, which is in our control sample and has reflection dominated spectra and properties similar to \sax\ (2011), did actually experience radiative warping in its disk and should be part of the sample with fluctuations. This burst has both the longest $t_{5\%}$ and `touchdown' time in the control sample \citep{int19}, and therefore fits the qualities of a burst that can drive disk instabilities, although we do not observe rapid flux variability. Radiative driven warps also explains why the 2010 burst from \sax\ did not exhibit fluctuations, while the one from 2011 and 2014 did --- the 2010 burst was not sufficiently long or luminous enough to trigger the instability.

\citet{ball23} also find that the radiative warping instability strongly depends on the $\alpha$ viscosity parameter of the accretion disk, with disks with larger values more likely to become unstable. Therefore, the bursts that show rapid fluctuations are ones with disks that have larger values of $\alpha$. As mentioned above, the burst from XTE~J1810-189 is likely also one where the disk is being warped and has a larger value of $\alpha$. If the four bursts in the control sample that do not show evidence of reflection have disks with low surface densities, then this would be a result of smaller accretion rates rather than differences in the viscosity parameter. This interpretation also implies that the \sax\ (2010) burst, which did not show fluctuations or reflection features, potentially originated from a disk with a lower accretion rate than the subsequent 2011 and 2014 bursts. Careful spectral analysis of bright, long bursts can therefore probe these physical properties of the accretion disks in LMXBs.

\subsection{A Wind-Disk Interpretation}
\label{sub:wind}
Previous interpretations of the fluctuations have focused on the fact that all of the bursts undergo photospheric expansion, potentially driving a wind or nova-like shell from the NS that interacts with and potentially disrupts the surrounding accretion flow \citep[e.g.,][]{igb11,int19,degenaar13}. Our detailed spectral analysis does show that nearly all of the studied bursts have column densities significantly larger than the Galactic columns (Figs.~\ref{fig:fluctfits} and~\ref{fig:nofluctfits}), with values that slowly evolve during the burst tails (except for \igr\ and 4U~1246-58). These enhanced column densities may indicate the presence of gas expelled from the NS as a result of the burst. However, the spectral analysis clearly finds evidence for relativistic ionized reflection from the inner accretion disk in all the sources with fluctuations, showing that the inner disks have either reformed or were not significantly disrupted by this time in the burst decay. As described above, the preponderance of evidence suggests that the fluctuations are caused by changes in the accretion disk structure leading to strong variations in our view of the system. The properties of these changes appear consistent with radiation-driven warping, but there are other instabilities that could act on a fluid flow that may result in similar effects (\citealt{int19} specifically mentions the instability affecting flags in a laminar flow). Numerical modeling efforts focused on the interaction of winds from photospheric radius expansion bursts and the surrounding accretion disk will be important in determining if they could cause similar large scale changes in the disk structure. More detailed modeling of the radiation-driven warping instability in the context of bursts will also be crucial in determining the potential for observations to constrain the strength of the accretion viscosity in LMXBs.

\section{Conclusions}
\label{sect:concl}
This paper shows the results of a systematic spectral analysis of five long X-ray bursts that exhibit temporary rapid flux variations during the tail of their light curves. Where possible, spectra were extracted from intervals before, during, and after the fluctuations appeared in the cooling phase. Spectra from all five bursts had at least one interval where they were reflection dominated and the NS emission was hidden from the observer. In many cases, the ionization parameter or the inclination angle of the reflecting region changed dramatically as the burst either entered or left the time when the light curve was fluctuating. The fraction of the observed flux originating in the blackbody emitted from the NS also varied by large factors in many of the bursts. The detailed spectral analysis also identified the 1~keV emission line observed in the burst from \igr\ as arising from ionized Ne.

A similar spectral analysis was performed on a control sample of 5 long X-ray bursts that did not show variations in flux. Spectra were extracted from three intervals in the cooling tails of these bursts and we find that the spectra from 4 of the 5 bursts in the control sample are best described by a simple absorbed blackbody with nearly no contribution from reflection. Only one burst, from XTE~J1810-189, showed reflection-dominated spectra similar to the fluctuation group.

Comparing the burst durations and fluences between the two homogeneous samples shows that the bursts that have rapid fluctuations are both longer and more energetic than the ones without fluctuations. This fact, along with the spectral analysis results, are consistent with a radiative-driven warp instability affecting the accretion disks during the bursts with fluctuations. In this picture, radiatively-driven warps would be occurring throughout the cooling tail of the bursts, and this could sometimes lead to rapid flux variations in the light curve. We surmise that the burst from XTE~J1810-189 is causing radiatively-driven warps in its accretion disk despite not exhibiting fluctuations. Since the warp instability is more likely to be triggered for higher values of the disk viscosity, bursts that show signs of radiatively-driven warping, such as rapid fluctuations in the light curve, likely contain disks with high viscosity parameters. Similar considerations suggest that the bursts in the control sample without reflection signatures have low surface density disks that are inflated by X-ray heating during the bursts, lowering the optical depth below the point where X-ray reflection can efficiently occur.

Overall, this paper provides another illustration of how the interaction of X-ray bursts from neutron stars with the surrounding accretion disk can be used to probe fundamental aspects of accretion disk physics. Future high-throughput, time-resolved X-ray spectroscopy of X-ray bursts from a large-area instrument (such as the proposed \textit{STROBE-X} mission, or the upcoming \textit{eXTP} and \textit{NewAthena} missions; \citealt{ray24,extp2025,newAthena2025}) would provide data sufficient to closely follow the response of the accretion disk, providing powerful insights into the hidden mechanics underpinning these gas flows. 

\begin{acknowledgments}
D.R.B. is supported by NASA award 80NSSC24K0212, and NSF grants AST-2307278 \& AST-2407658. N.D. is supported by Vici research programme VI.C.242.101 (PI Degenaar) financed by the Dutch Research Council (NWO) under grant ID https://doi.org/10.61686/LIRQH96389. This research has made use of data and/or software provided by the High Energy Astrophysics Science Archive Research Center (HEASARC), which is a service of the Astrophysics Science Division at NASA/GSFC.
\end{acknowledgments}

%\begin{contribution}
%%This section gives authors the space to recognize author contributions. The text inside this environment is NOT counted towards the total word quanta. At a minimum, manuscripts are expected to include this text:

%Finish later.

%All authors contributed equally to the Terra Mater collaboration.

%% But authors are expected to provide more specific details, e.g. 
%%
%%SC was responsible for writing and submitting the manuscript.
%%WWM came up with the initial research concept and edited the manuscript.
%%OTS obtained the funding and edited the manuscript.
%%EBF provided the formal analysis and validation. He also edited the manuscript.
%%GEH Supervised the undergraduates, wrote the software and administers the project github and Zenodo repositories.
%%
%% Authors can use the Contributor Role Taxonomy (CRediT) at
%% for ideas on how write a good statement tailored to their needs.

%\end{contribution}

\software{HEASoft \citep{2014ascl.soft08004N}}

%\facilities{HST(STIS), Swift(XRT and UVOT), AAVSO, CTIO:1.3m, CTIO:1.5m, CXO}

%\software{astropy \citep{2013A&A...558A..33A,2018AJ....156..123A,2022ApJ...935..167A},  
%          Cloudy \citep{2013RMxAA..49..137F}, 
%          Source Extractor \citep{1996A&AS..117..393B}
%          }


\appendix
Appendix~\ref{app:fluct} contains the details of the fits to four of the bursts with fluctuations: the 2011 and 2014 bursts from SAX~J1712.6-3739, Swift J1734.5-3027 and 4U~1850-087. A description of the fits to the 2012 burst from \igr\ is found in Sect.~\ref{sect:igr}. Appendix~\ref{app:nofluct} is comprised of plots showing the best fit models and residuals from the spectral analysis of the five bursts without fluctuations that make up the control sample. 

\section{Bursts with Fluctuations}
\label{app:fluct}

\subsection{SAX J1712.6-3739 (2011)}
\label{subapp:saxJ1712-2011}
\begin{figure*}[t!]
\begin{center}
\includegraphics[width=0.32\textwidth]{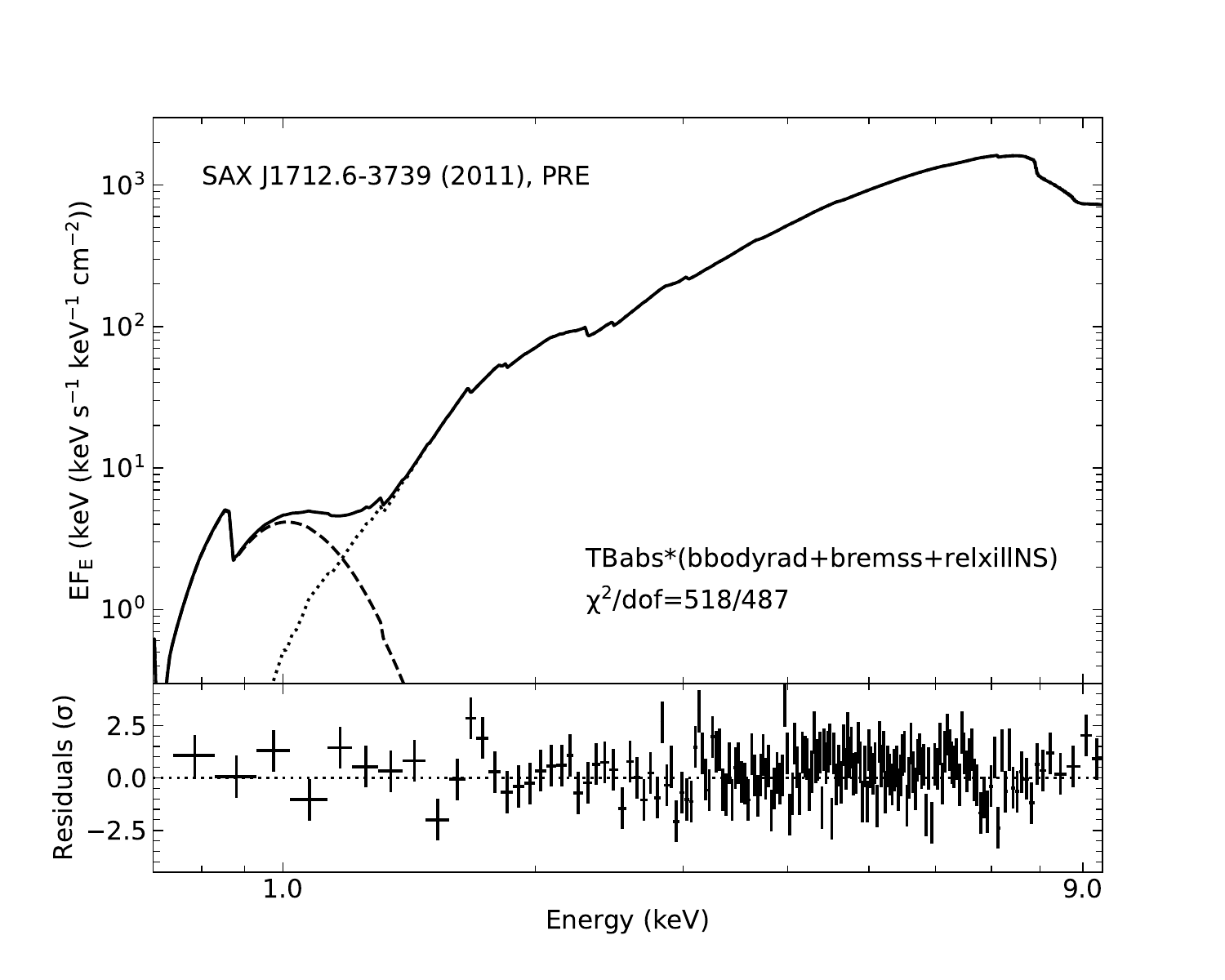}
\includegraphics[width=0.32\textwidth]{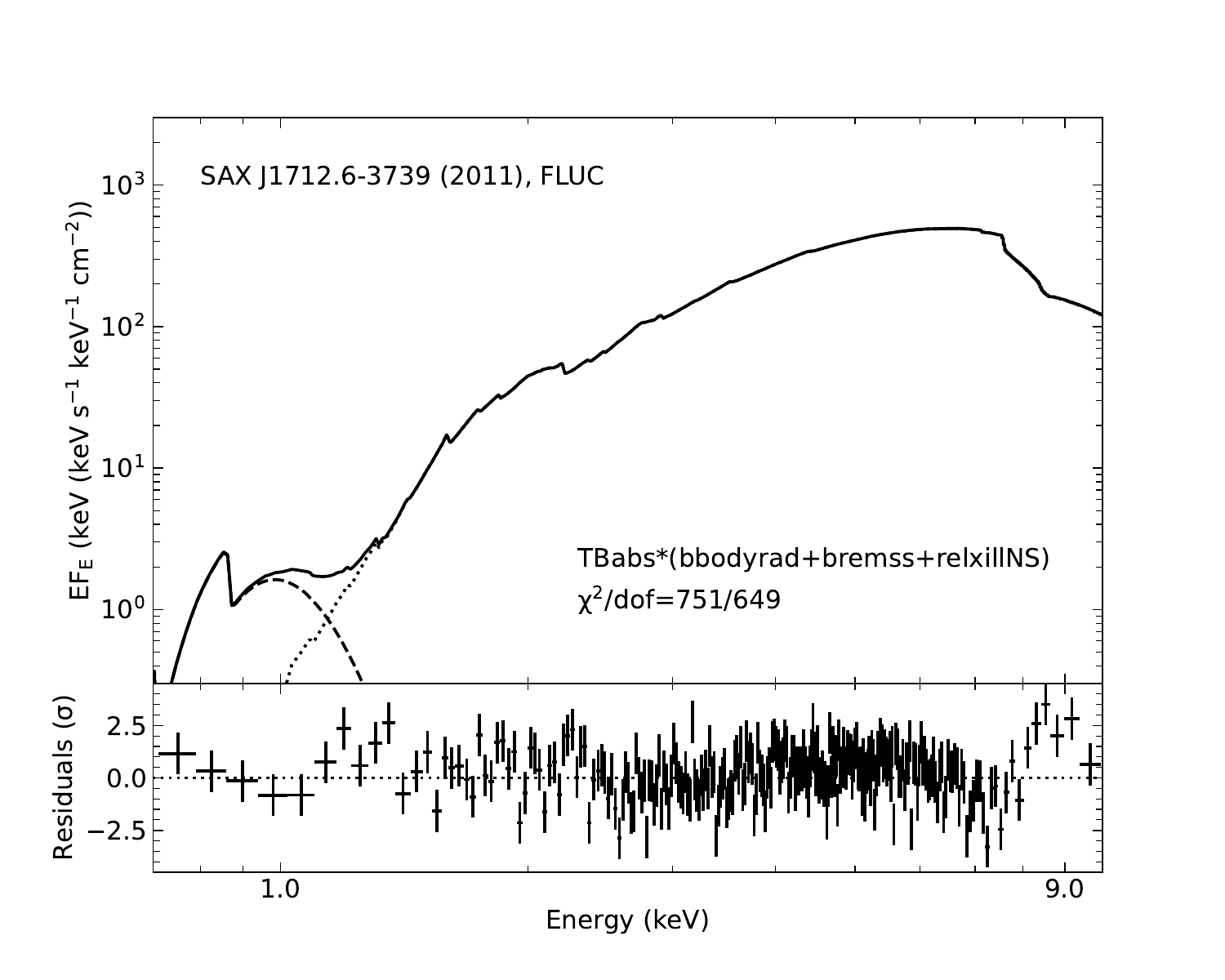}
\includegraphics[width=0.32\textwidth]{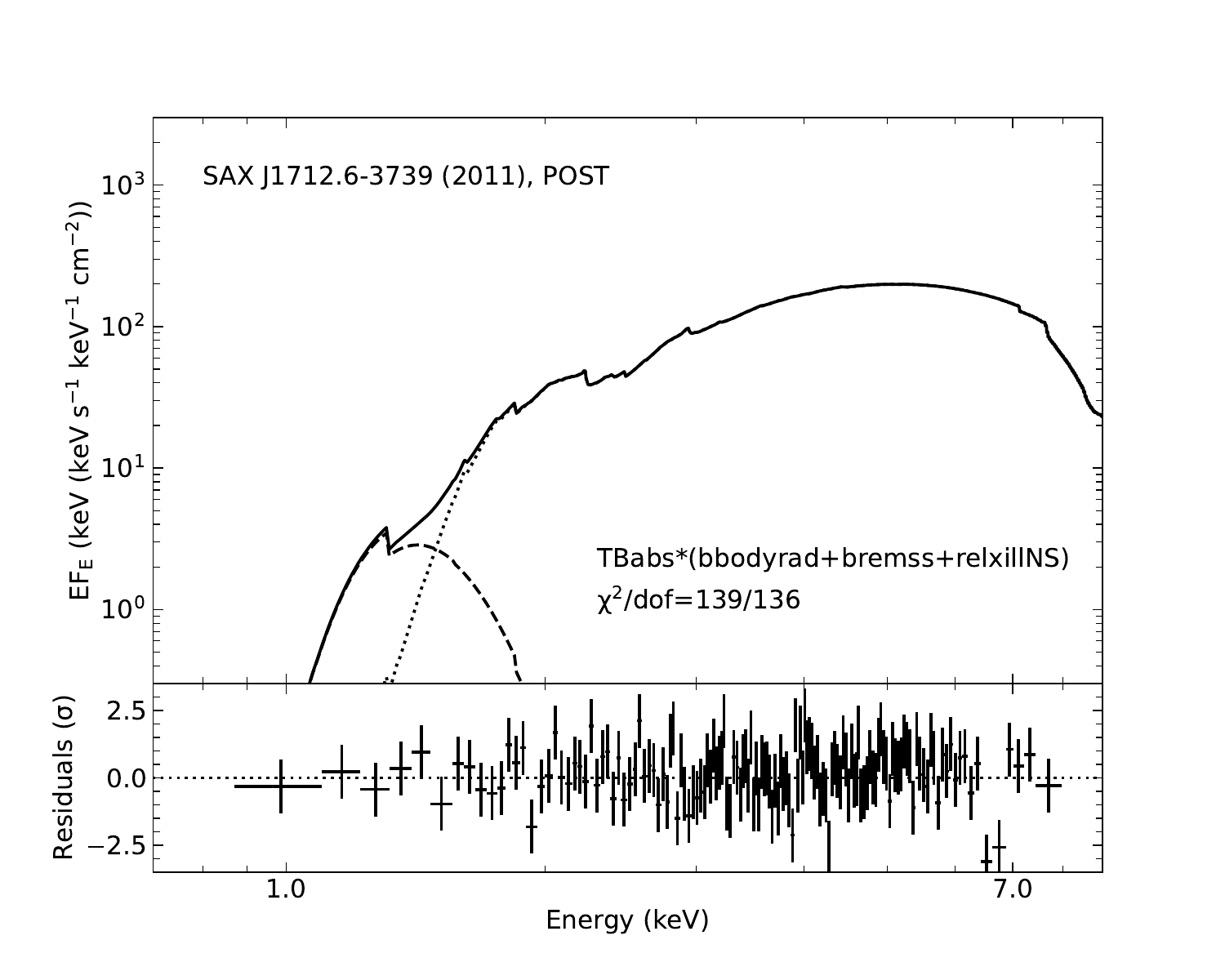}
\end{center}
\caption{Plots of the best-fit models and residuals (in units of $\sigma$) for the PRE, FLUC and POST intervals from the 2011 burst from \sax. The solid line in the upper-part of each plot shows the final model (plotted in $EF_E$ units) including the \texttt{relxill}  (dotted line) and bremsstrahlung (dashed line) components. Table~\ref{tab:sax17-2011} lists the parameters from the best-fit models.}
\label{fig:sax17-2011}
\end{figure*}

The best fits to the three intervals of the 2011 burst from \sax\ are shown in Figure~\ref{fig:sax17-2011} with the parameters listed in Table~\ref{tab:sax17-2011}. The spectrum in all 3 intervals is dominated by strongly blurred relativistic reflection with no contribution from the blackbody from the NS surface. The reflection spectrum is constrained to be from the inner accretion disk but has a relatively low ionization parameter in all cases ($\log \xi \approx 1.1$--$1.2$). The fit is improved by allowing the inclination angle to vary, yielding inclinations of $\approx 55$--$60$~degrees. Significant excess absorption is present before, during and after the time of the flux fluctuations. Without a prominent soft X-ray emission line, the soft excess is modeled with a bremsstrahlung spectrum. The fit to the FLUC spectrum leaves residuals between 7 and 8 keV that can modeled by either an absorption edge or a Gaussian absorption line. While adding these absorption features does improve the fit statistic, they do not lead to any significant qualitative changes to the parameters describing the broadband spectrum in Table~\ref{tab:sax17-2011}.

\begin{deluxetable*}{cccccccccc}[t!]
\tabletypesize{}
%\tablewidth{0} 
\tablecaption{Best-fit parameters found from the PRE, FLUC and POST intervals of the 2011 \sax\ intermediate duration burst.\label{tab:sax17-2011}}
\tablehead{
\colhead{Interval} & \colhead{Flux} & \colhead{BB Fraction} & \colhead{$\Delta N_H$}& \colhead{$\log \xi$} & \colhead{$kT_{\mathrm{BB}}$} & \colhead{$i$} & \colhead{$A_{\mathrm{Fe}}$} & \colhead{$kT_{\mathrm{bremss}}$} & \colhead{$\chi^2$/dof} \\
 & \colhead{(erg cm$^{-2}$ s$^{-1}$)} & & \colhead{($10^{22}$~cm$^{-2}$)} & & \colhead{(keV)} & \colhead{(deg)} & & \colhead{(keV)} & 
 } 
\startdata
PRE & $1.8\times 10^{-6}$ & $<0.05$ & $+(2.27_{-0.59}^{+0.60})$ & $1.20\pm 0.05$ & $1.35^{+0.04}_{-0.06}$ & $60.9^{+4.5}_{-2.7}$ & $5.61^{+3.50}_{-0.95}$ & $0.063^{+0.017}_{-0.012}$ & $518/487$\\
FLUC & $6.7\times 10^{-7}$ & $<0.01$ & $+(2.42^{+0.35}_{-0.34})$ & $1.10^{+0.03}_{-0.02}$ & $1.04^{+0.02}_{-0.03}$ & $55.3^{+4.1}_{-3.2}$ & $6.37^{+1.71}_{-1.39}$ & $0.055^{+0.011}_{-0.009}$ & $751/649$ \\
POST & $3.0\times 10^{-7}$ & $<0.63$ & $+(6.98^{+2.96}_{-3.94})$ & $1.13^{+0.14}_{-0.13p}$ & $0.64^{+0.17}_{-0.09}$ & $56.7^{+18.9}_{-6.7}$ & $8.22^{+1.78p}_{-3.28}$ & $0.085^{+0.092}_{-0.024}$ & $139/136$\\
\enddata
\tablecomments{The model \texttt{TBabs*(bbodyrad+bremss+relxillNS)} is used for all 3 intervals with the inclination angle of the reflection spectrum ($i$) free to vary. The best-fit models and the residuals to the fits are shown in Figure~\ref{fig:sax17-2011}. A bremsstrahlung spectrum with temperature $kT_{\mathrm{bremss}}$ is used to model the soft excess. Other parameters are as in Table~\ref{tab:igrfits}. A `p' in an error-bar indicates the parameter pegged at its upper or lower bound. } 
\end{deluxetable*}

\subsection{SAX J1712.6-3739 (2014)}
\label{subapp:saxJ1712-2014}
\begin{figure*}[t!]
\begin{center}
\includegraphics[width=0.49\textwidth]{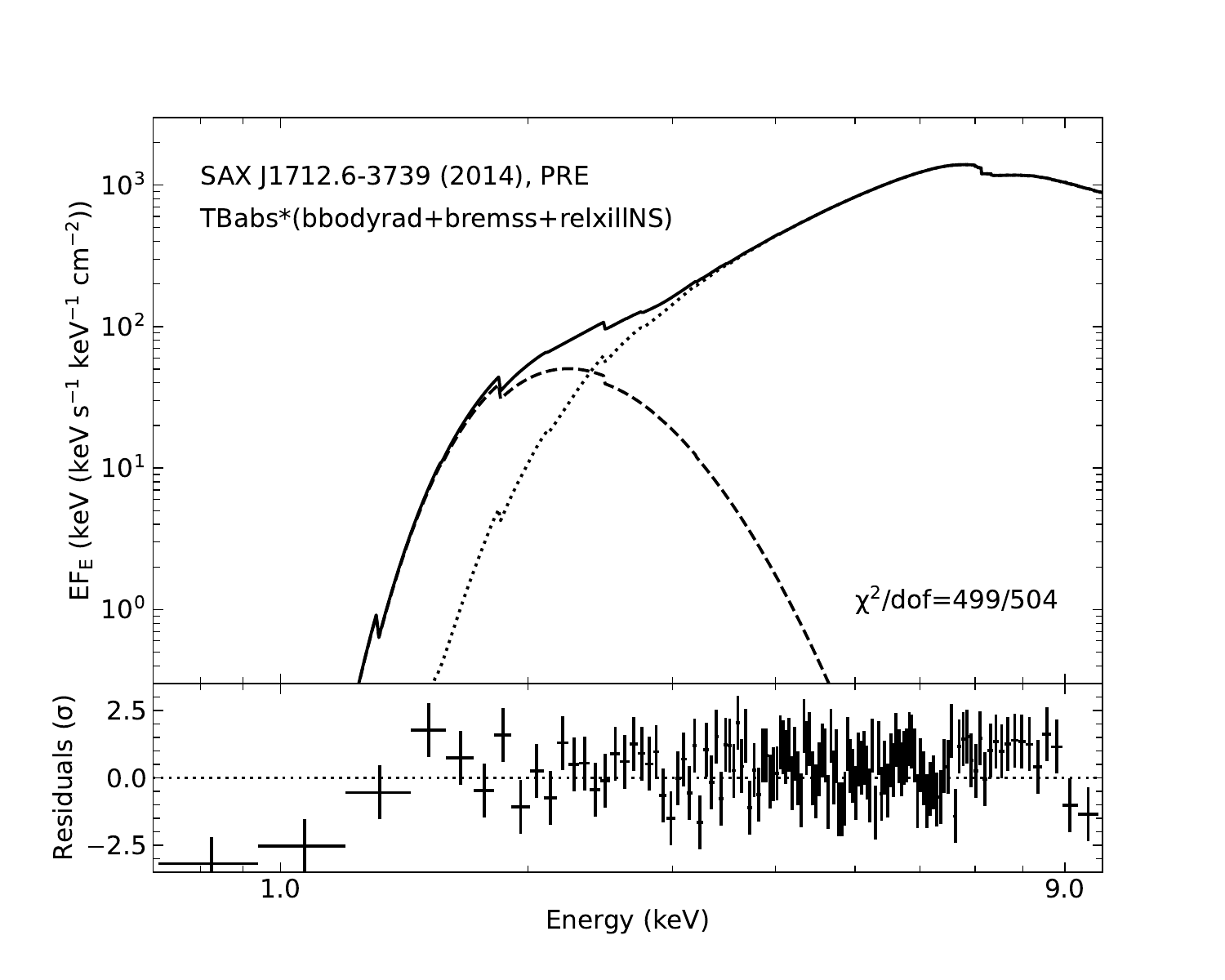}
\includegraphics[width=0.49\textwidth]{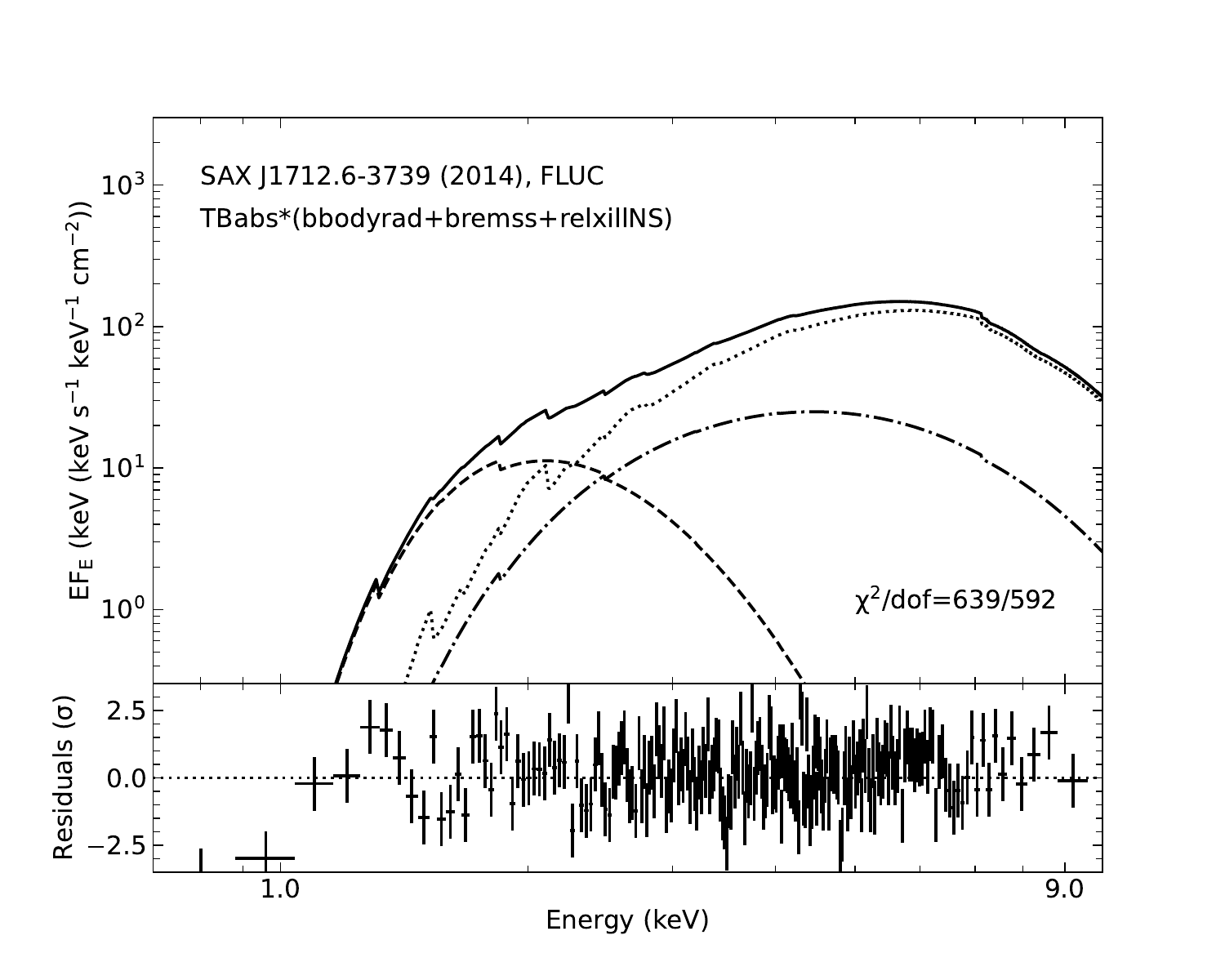}
\end{center}
\caption{Plots of the best-fit models and residuals (in units of $\sigma$) for the PRE and FLUC intervals from the 2014 burst from \sax. The solid line in the upper-part of each plot shows the final model (plotted in $EF_E$ units) including the \texttt{relxill} (dotted line), the burst blackbody (dot-dashed line in FLUC panel) and bremsstrahlung (dashed line) components. Table~\ref{tab:sax17-2014} lists the parameters from these best-fit models.}
\label{fig:sax17-2014}
\end{figure*}

Only the PRE and FLUC intervals are available from the 2014 burst from
\sax, with the PRE interval dominated by times $>3000$~s before the
fluctuations begin (Fig.~\ref{fig:five_lc_mod}). Spectra from the two
intervals are well described using the same model as the 2011 burst,
\texttt{TBabs*(bbodyrad+bremss+relxillNS)}, with the results presented
in Figure~\ref{fig:sax17-2014} and Table~\ref{tab:sax17-2014}. The PRE
spectrum is dominated by an ionized ($\log \xi=3$) relativistic
reflection model modified by a significantly enhanced absorption
column density. The primary blackbody emitted by the NS is not visible
in the spectrum. During the FLUC interval, the blackbody does provide
$26$\% of the observed flux. Relativistic reflection still dominates
the spectrum at this later time although with a lower ionization
parameter of $\log \xi = 1$. The absorbing column density is also
reduced in the FLUC interval, while still remaining much higher than
the Galactic column. The fit residuals at energies below $\approx 1$~keV seen in both intervals could be explained by a more complex absorbing structure and/or a different abundance pattern than assumed by the applied model \citep[e.g.,][]{barra25}. However, the photon flux in this energy range of the spectrum is so low that these residuals do not impact the broadband fit to the spectra.

\begin{deluxetable*}{cccccccccc}[t!]
\tabletypesize{}
%\tablewidth{0} 
\tablecaption{Best-fit parameters found from the PRE and FLUC intervals of the 2014 \sax\ intermediate duration burst.\label{tab:sax17-2014}}
\tablehead{
\colhead{Interval} & \colhead{Flux} & \colhead{BB Fraction} & \colhead{$\Delta N_H$}& \colhead{$\log \xi$} & \colhead{$kT_{\mathrm{BB}}$} & \colhead{$i$} & \colhead{$A_{\mathrm{Fe}}$} & \colhead{$kT_{\mathrm{bremss}}$} & \colhead{$\chi^2$/dof} \\
 & \colhead{(erg cm$^{-2}$ s$^{-1}$)} & & \colhead{($10^{22}$~cm$^{-2}$)} & & \colhead{(keV)} & \colhead{(deg)} & & \colhead{(keV)} & 
 } 
\startdata
PRE & $1.7\times 10^{-6}$ & $<0.36$ & $+(9.40_{-1.30}^{+1.33})$ & $3.00_{-0.17}^{+0.27}$ & $2.33^{+0.26}_{-0.12}$ & $30^f$ & $0.58^{+0.95}_{-0.08p}$ & $0.32^{+0.05}_{-0.03}$ & $499/504$\\
FLUC & $2.3\times 10^{-7}$ & $0.30$ & $+(5.10^{+0.89}_{-0.88})$ & $1.00^{+0.01}_{-0p}$ & $0.98\pm 0.05$ & $47.5^{+5.1}_{-5.3}$ & $0.50^{+0.39}_{-0p}$ & $0.41^{+0.10}_{-0.07}$ & $639/592$ \\
\enddata
\tablecomments{The model \texttt{TBabs*(bbodyrad+bremss+relxillNS)} is used for both intervals with the inclination angle of the reflection spectrum ($i$) free to vary in the FLUC interval (it is fixed at 30 degrees for the PRE spectrum). The best-fit models and the residuals to the fits are shown in Figure~\ref{fig:sax17-2014}. A bremsstrahlung spectrum with temperature $kT_{\mathrm{bremss}}$ is used to model the soft excess. Other parameters are as in Table~\ref{tab:igrfits}. A `p' in an error-bar indicates the parameter pegged at its upper or lower limit. } 
\end{deluxetable*}

\subsection{Swift J1734.5-3027}
\label{subapp:swiftj17}
\begin{figure*}[t!]
\begin{center}
\includegraphics[width=0.49\textwidth]{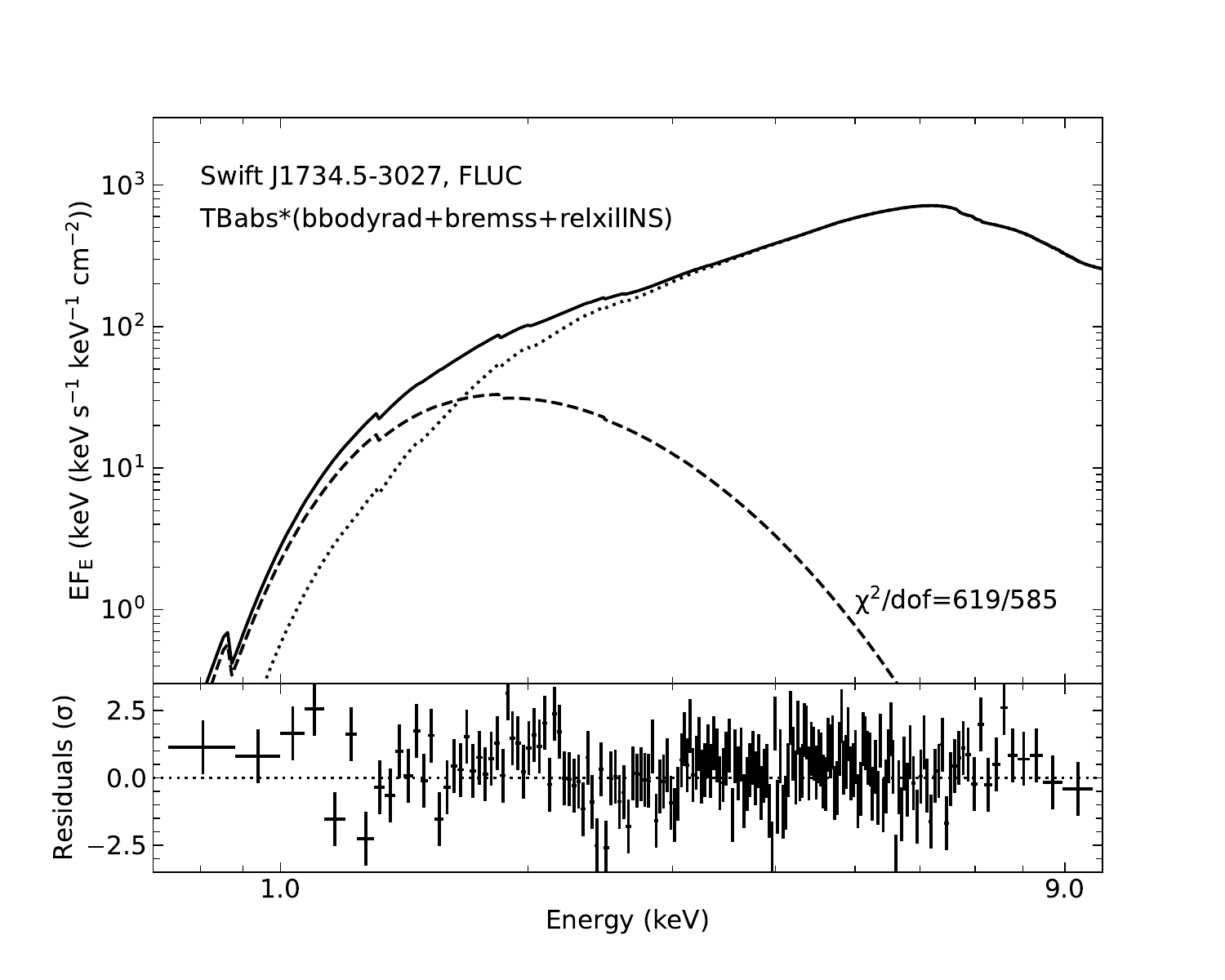}
\includegraphics[width=0.49\textwidth]{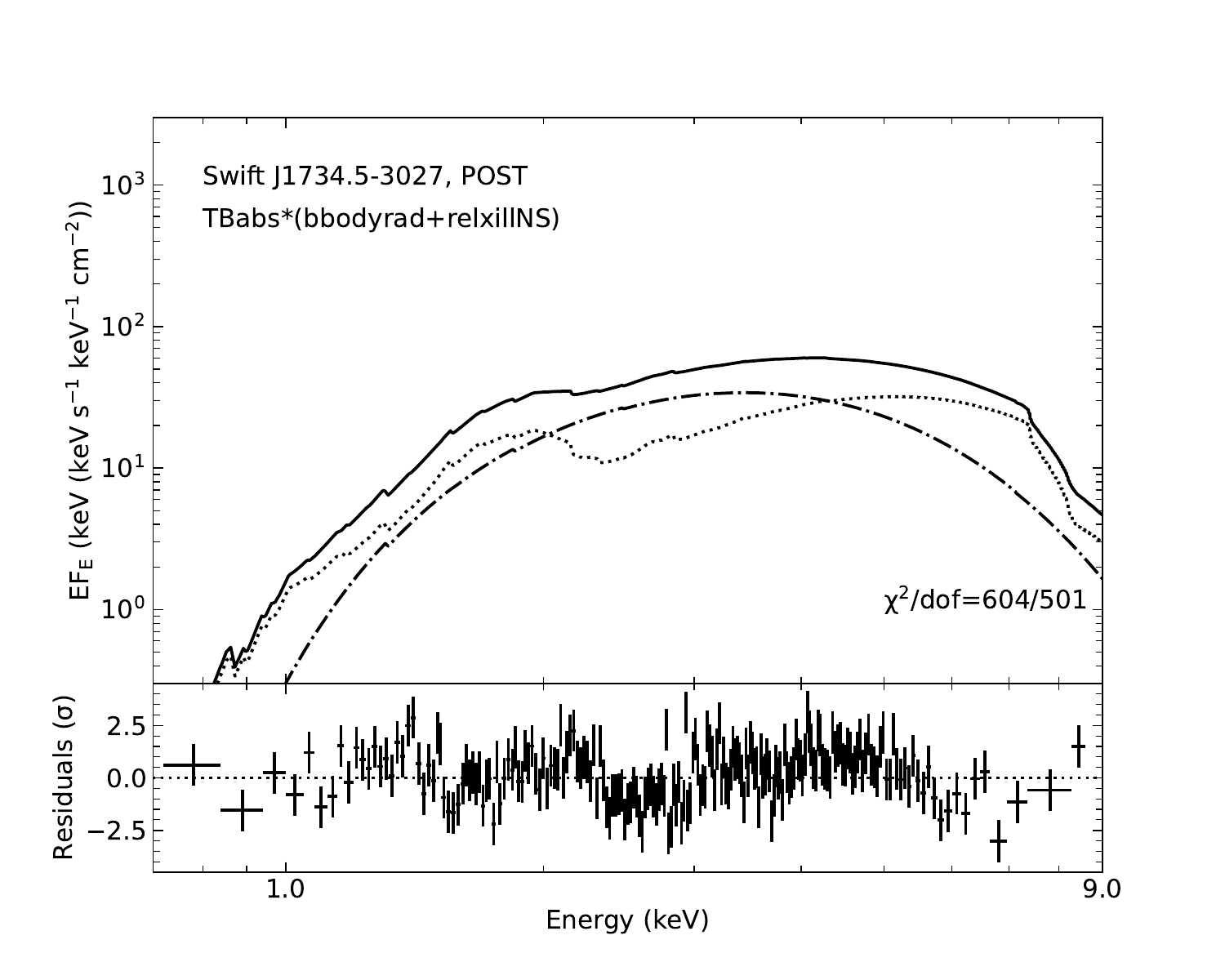}
\end{center}
\caption{Plots of the best-fit models and residuals (in units of $\sigma$) for the \swift\ FLUC and POST intervals. The solid line in the upper-part of each plot shows the final model (plotted in $EF_E$ units) including the \texttt{relxill} (dotted line), the burst blackbody (dot-dashed line in POST panel) and bremsstrahlung (dashed line) components. Table~\ref{tab:swift} lists the parameters from these best-fit models.}
\label{fig:swift}
\end{figure*}

The \swift\ FLUC interval is dominated by ionized relativistic
reflection with no flux contributed by the blackbody from the NS
(Fig.~\ref{fig:swift} and Table~\ref{tab:swift}). The POST interval
sees a large reduction in the ionization parameter of the reflector
and a return of the blackbody, which now comprises $72$\% of the total
flux. Two very different inclination angles are found from the
\texttt{relxillNS} components in the two intervals. The iron
abundances are also not consistent which may indicate that one of the
two intervals is being incorrectly modeled. An alternative fit to the
FLUC interval was found with $A_{\mathrm{Fe}}\approx 10$ with a
$\chi^2/$dof$=601/585$. However, this solution predicts a low
$kT_{\mathrm{BB}}=0.95$~keV and $\log \xi=1.1$, both of which appear
unreasonably small during this stage of the burst. The inclination
angle of this alterative solution moves to other extreme,
$i=87$~degrees. It appears that the strong fluctuations that occur
during this interval (Fig.~\ref{fig:five_lc_mod}) could make the total
spectrum a mixture of at least two qualitatively different shapes. We also note that similar residuals are also seen in the fit to the POST spectrum, indicating that the spectral complexity persists beyond the end of the fluctuations. We attempted to describe the high energy residuals in the POST spectrum with an edge or a Gaussian line, but did not find physically reasonable parameters even if the fit statistic improved.

\begin{deluxetable*}{cccccccccc}[t!]
\tabletypesize{}
%\tablewidth{0} 
\tablecaption{Best-fit parameters found from the FLUC and POST intervals of the \swift\ intermediate duration burst.\label{tab:swift}}
\tablehead{
\colhead{Interval} & \colhead{Flux} & \colhead{BB Fraction} & \colhead{$\Delta N_H$}& \colhead{$\log \xi$} & \colhead{$kT_{\mathrm{BB}}$} & \colhead{$i$} & \colhead{$A_{\mathrm{Fe}}$} & \colhead{$kT_{\mathrm{bremss}}$} & \colhead{$\chi^2$/dof} \\
 & \colhead{(erg cm$^{-2}$ s$^{-1}$)} & & \colhead{($10^{22}$~cm$^{-2}$)} & & \colhead{(keV)} & \colhead{(deg)} & & \colhead{(keV)} & 
 } 
\startdata
FLUC & $1.0\times 10^{-6}$ & $<0.15$ & $+(1.97_{-0.50}^{+0.42})$ & $3.07_{-0.03}^{+0.27}$ & $1.99^{+0.04}_{-0.07}$ & $3^{+10.1}_{-0p}$ & $1.00^{+0.47}_{-0.21}$ & $0.60\pm 0.11$ & $619/585$\\
POST & $1.2\times 10^{-7}$ & $0.72$ & $+(1.23\pm 0.08)$ & $1.18\pm 0.02$ & $0.80\pm 0.01$ & $51.4^{+2.1}_{-2.5}$ & $10^{+0p}_{-1.7}$ & $-$ & $604/501$ \\
\enddata
\tablecomments{The model \texttt{TBabs*(bbodyrad+bremss+relxillNS)} is used for the FLUC interval, but the \texttt{bremss} component is not needed for the POST spectrum. The best-fit models and the residuals to the fits are shown in Figure~\ref{fig:swift}. A `p' in an error-bar indicates the parameter pegged at its upper or lower limit. } 
\end{deluxetable*}

\subsection{4U 1850-087}
\label{subapp:fouru}
\begin{figure*}[t!]
\begin{center}
\includegraphics[width=0.32\textwidth]{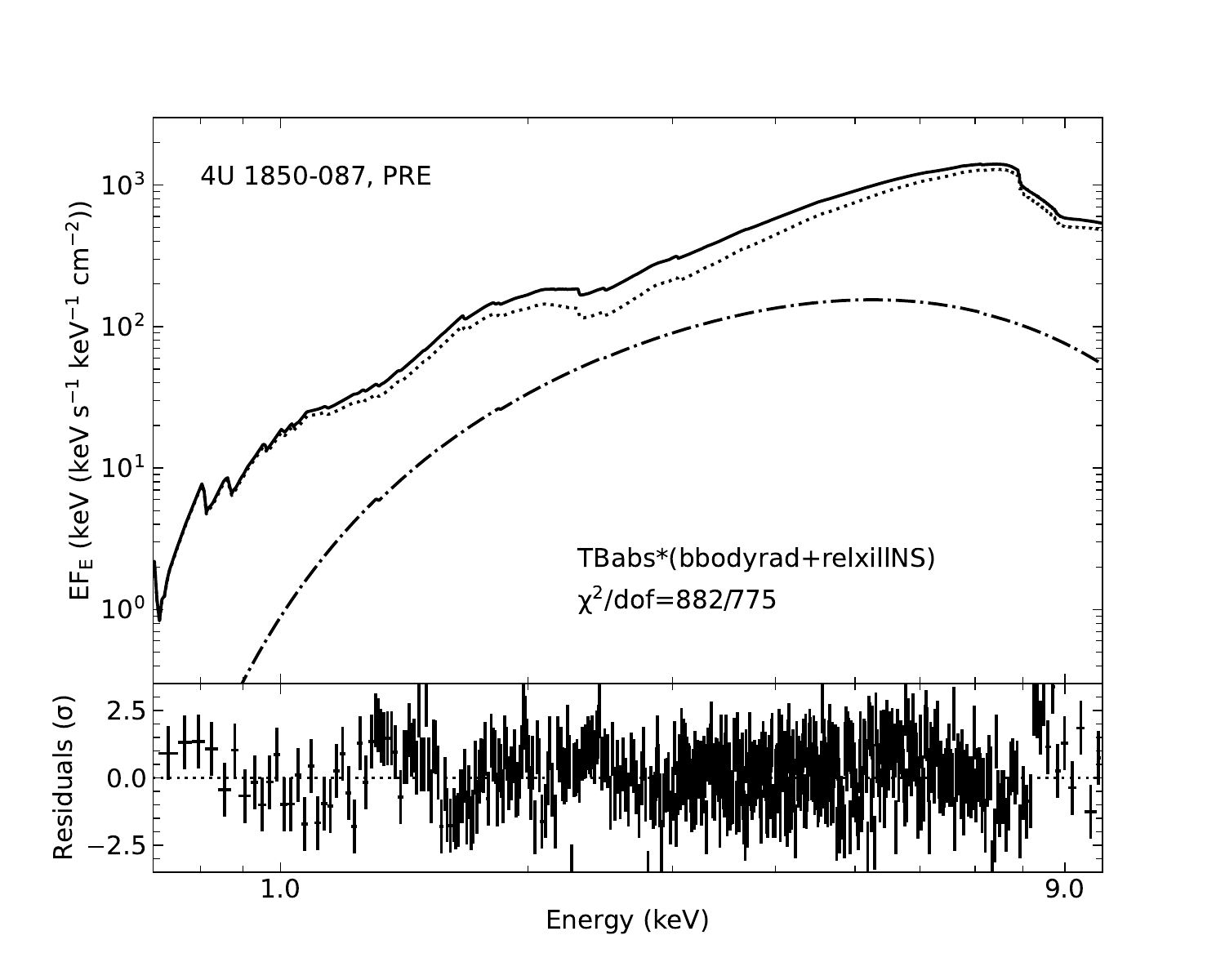}
\includegraphics[width=0.32\textwidth]{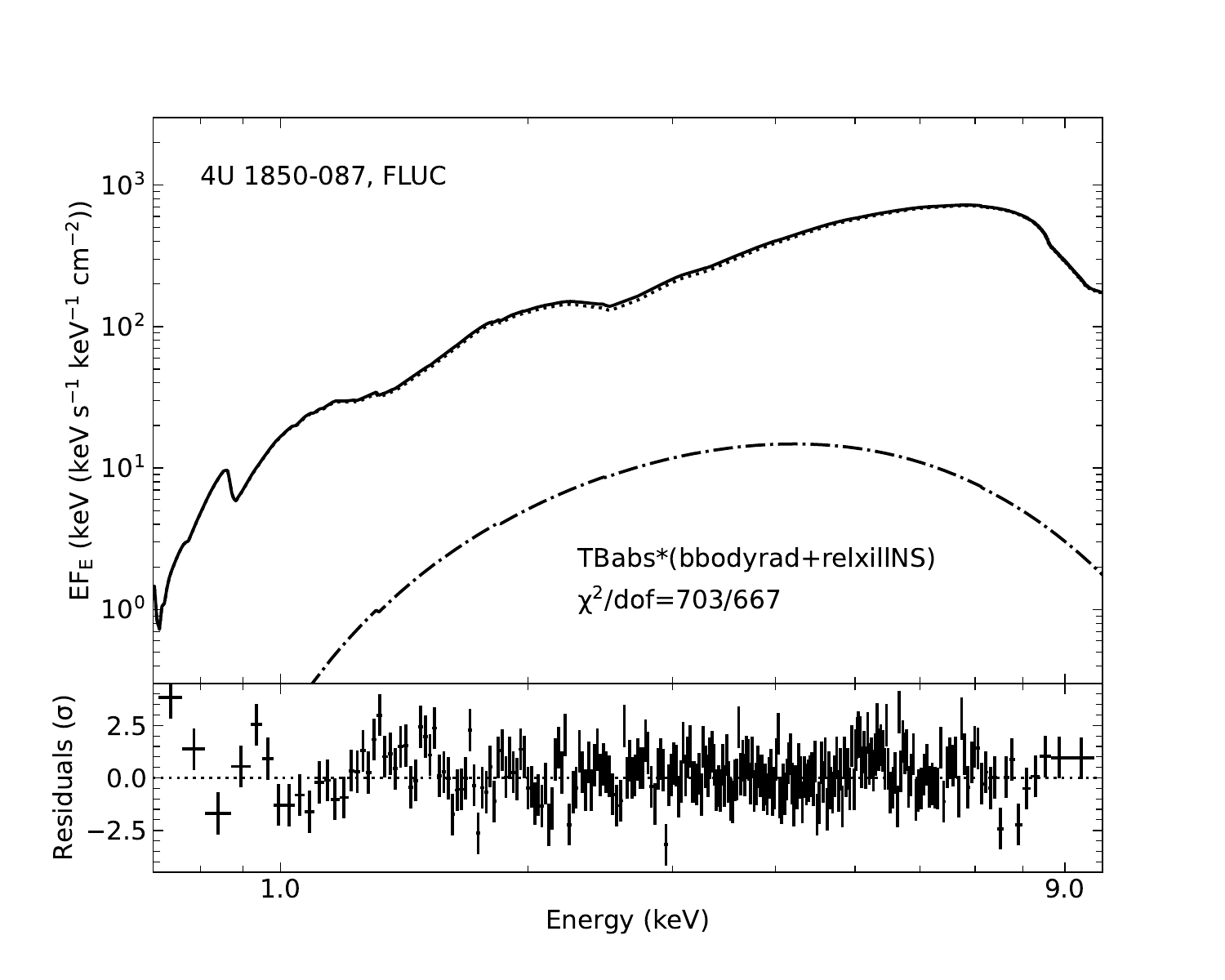}
\includegraphics[width=0.32\textwidth]{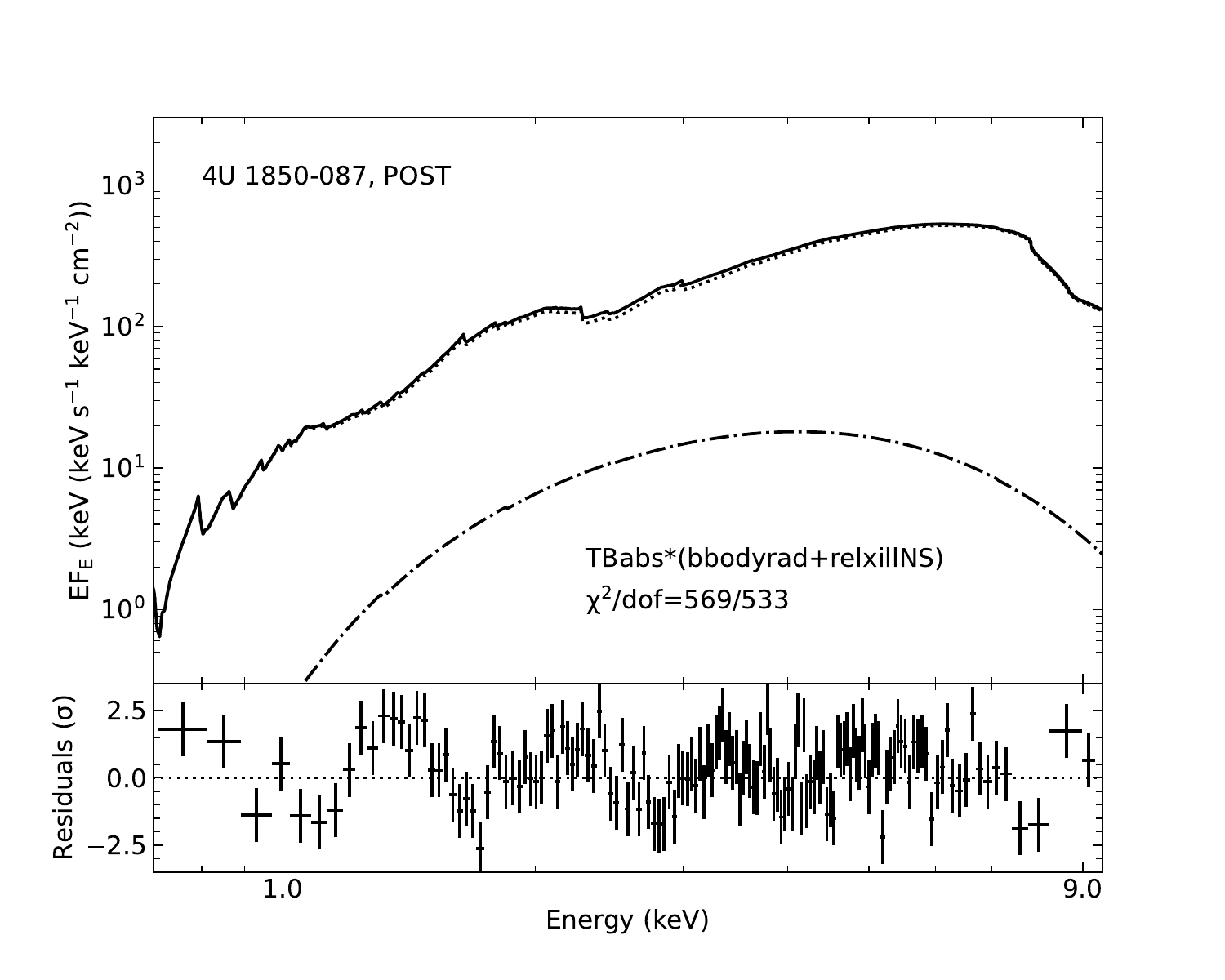}
\end{center}
\caption{Plots of the best-fit models and residuals (in units of $\sigma$) for the PRE, FLUC and POST intervals from \fouru. The solid line in the upper-part of each plot shows the final model (plotted in $EF_E$ units) including the \texttt{relxill}  (dotted line) and blackbody (dot-dashed line) components. Table~\ref{tab:fouru} lists the parameters from the best-fit models.}
\label{fig:fouru}
\end{figure*}

The spectra from the three \fouru\ intervals are all dominated by relativistic reflection from the inner accretion disk (Fig.~\ref{fig:fouru} and Table~\ref{tab:fouru}). The burst blackbody provides $\approx 20$\% of the flux in the PRE interval, dropping to $\approx 2$\% and $\approx 4$\% in the FLUC and POST intervals, respectively. The ionization parameter of the reflector is relatively low, but does show a drop in the FLUC interval, before rising again after the fluctuations stop. Allowing the inclination angle to vary improves the fits in all three intervals and does indicate a significant change in inclination during the FLUC phase. There is enhanced neutral absorption seen during the burst, but the column density is consistent with being constant in all three intervals.

\begin{deluxetable*}{ccccccccc}[t!]
\tabletypesize{}
%\tablewidth{0} 
\tablecaption{Best-fit parameters found from the PRE, FLUC and POST intervals of the \fouru\ intermediate duration burst.\label{tab:fouru}}
\tablehead{
\colhead{Interval} & \colhead{Flux} & \colhead{BB Fraction} & \colhead{$\Delta N_H$}& \colhead{$\log \xi$} & \colhead{$kT_{\mathrm{BB}}$} & \colhead{$i$} & \colhead{$A_{\mathrm{Fe}}$} & \colhead{$\chi^2$/dof} \\
 & \colhead{(erg cm$^{-2}$ s$^{-1}$)} & & \colhead{($10^{22}$~cm$^{-2}$)} & & \colhead{(keV)} & \colhead{(deg)} & & 
 } 
\startdata
PRE & $1.8\times 10^{-6}$ & $0.18 $ & $+(1.14\pm 0.09)$ & $1.26_{-0.05}^{+0.10}$ & $1.31\pm 0.06$ & $61.1^{+1.8}_{-1.6}$ & $7.07^{+2.17}_{-2.07}$ & $882/775$\\
FLUC & $1.1\times 10^{-6}$ & $0.03$ & $+(1.20^{+0.08}_{-0.11})$ & $1.11\pm 0.02$ & $1.03_{-0.04}^{+0.05}$ & $71.6^{+2.9}_{-3.9}$ & $9.02^{+0.98p}_{-2.60}$ & $703/667$ \\
POST & $8.6\times 10^{-7}$ & $0.05$ & $+(1.21^{+0.12}_{-0.19})$ & $1.16\pm 0.03$ & $1.00^{+0.09}_{-0.05}$ & $59.2^{+3.4}_{-6.3}$ & $5.0^{+4.2}_{-0.6}$ & $569/533$\\
\enddata
\tablecomments{The model \texttt{TBabs*(bbodyrad+relxillNS)} is used for all 3 intervals with the inclination angle of the reflection spectrum ($i$) free to vary. The best-fit models and the residuals to the fits are shown in Figure~\ref{fig:fouru}. This source did not require an additional component to fit a soft excess. Other parameters are as in Table~\ref{tab:igrfits}. A `p' in an error-bar indicates the parameter pegged at its upper or lower limit. } 
\end{deluxetable*}

\subsection{Comparison with Previous Results}
\label{subapp:compare}
The 5 bursts with fluctuations analyzed here were previously studied by \citet{degenaar13} (for \igr) and by \citet{int19}. Both studies employed absorbed blackbodies as the spectral model so a detailed comparison between our results and these earlier works is not possible. However, our method also measures the blackbody temperatures of hot NS surface, so it is interesting to compare the blackbody temperatures measured by the different approaches.

\citet{degenaar13} fit similar PRE, FLUC and POST spectra from \igr\ as considered here, and also found the same higher column density during the FLUC interval. For the PRE and FLUC intervals the blackbody temperatures measured by \citet{degenaar13} are $\sim 10$\% lower than the ones listed in Table~\ref{tab:igrfits}, while they are consistent for the POST interval. Given the very different spectral model used in the two analyses, this level of agreement is satisfactory.

\citet{int19} performed a time-resolved spectral analysis of the five bursts, including through the period of fluctuations. An absorbed blackbody was used in the analysis with a fixed column density. Comparing to our results we find that the blackbody temperatures derived from our fits are consistent with the \citet{int19} results for \igr\ and \swift, but are much lower for the two \sax\ bursts and the one from \fouru. In these three bursts, we find $kT_{\mathrm{BB}}$ in the PRE intervals $<2.3$~keV, while \citet{int19} reports values $\ga 2.5$~keV, and, in the case of \sax, approaching $4$~keV. As mentioned by \citet{int19}, such high temperatures are not expected for bursts $\sim 200$~s into their evolution. Indeed, the five bursts from the control sample all have $kT_{\mathrm{bb}} \la 2$~keV during their first intervals (Table~\ref{tab:nofluctfits}). The large $kT_{\mathrm{bb}}$ values found by \citet{int19} are likely impacted by the presence of strong reflection features at energies $> 6$~keV. We conclude that our analysis method, which includes the impact of reflection, is providing an accurate characterization of the spectral shape of the burst.

\section{Bursts without Fluctuations}
\label{app:nofluct}
\begin{figure*}[t!]
\begin{center}
\includegraphics[width=0.32\textwidth]{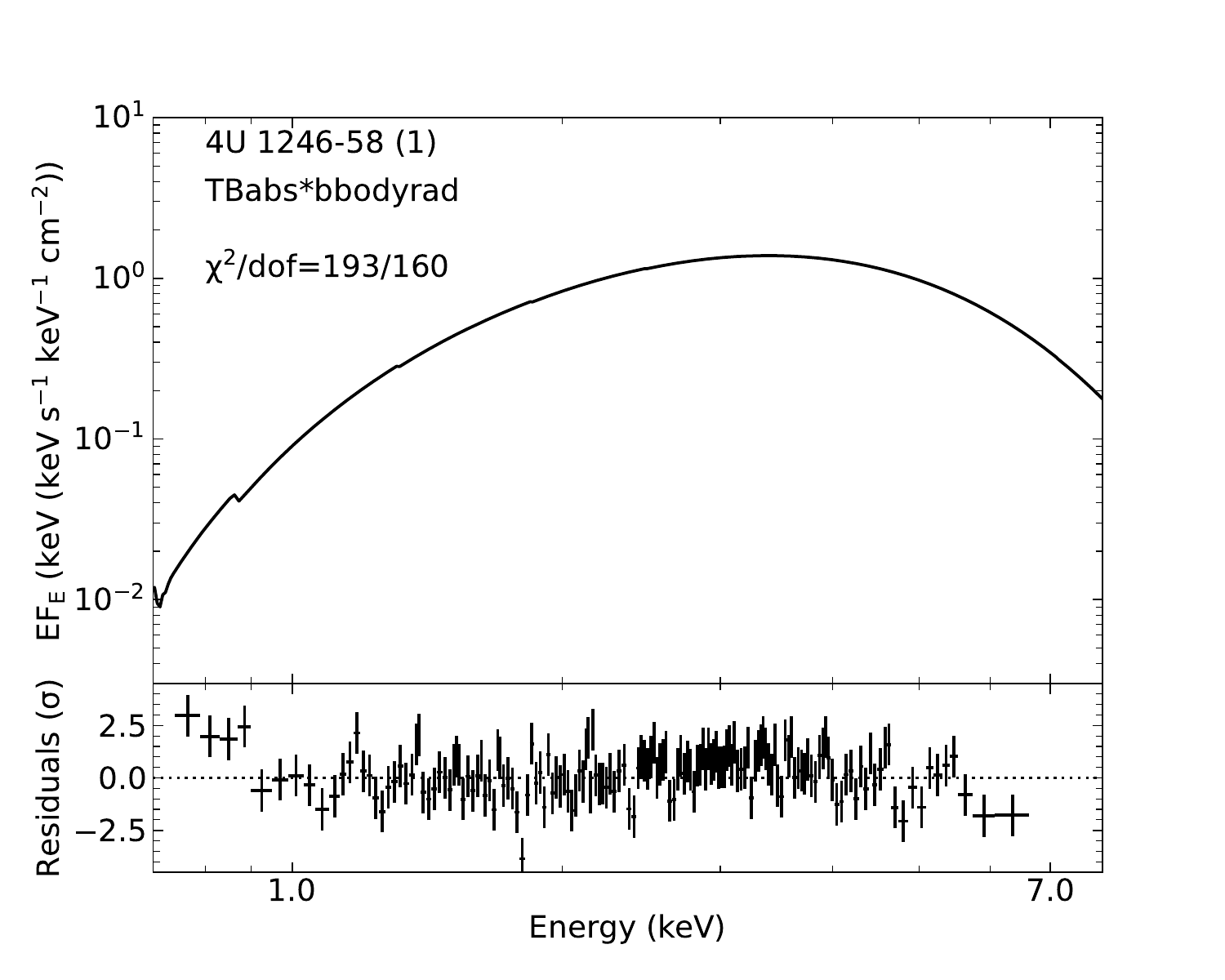}
\includegraphics[width=0.32\textwidth]{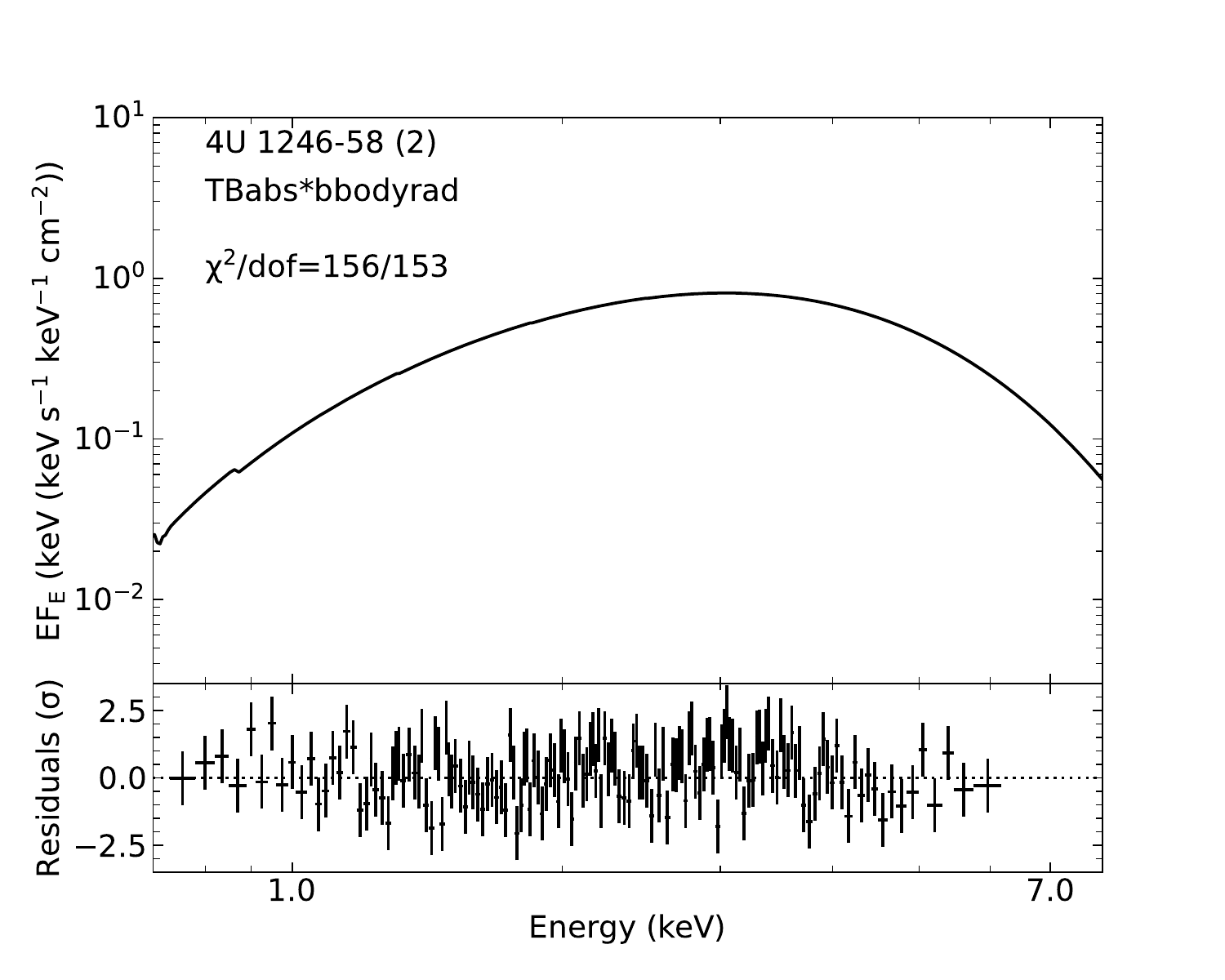}
\includegraphics[width=0.32\textwidth]{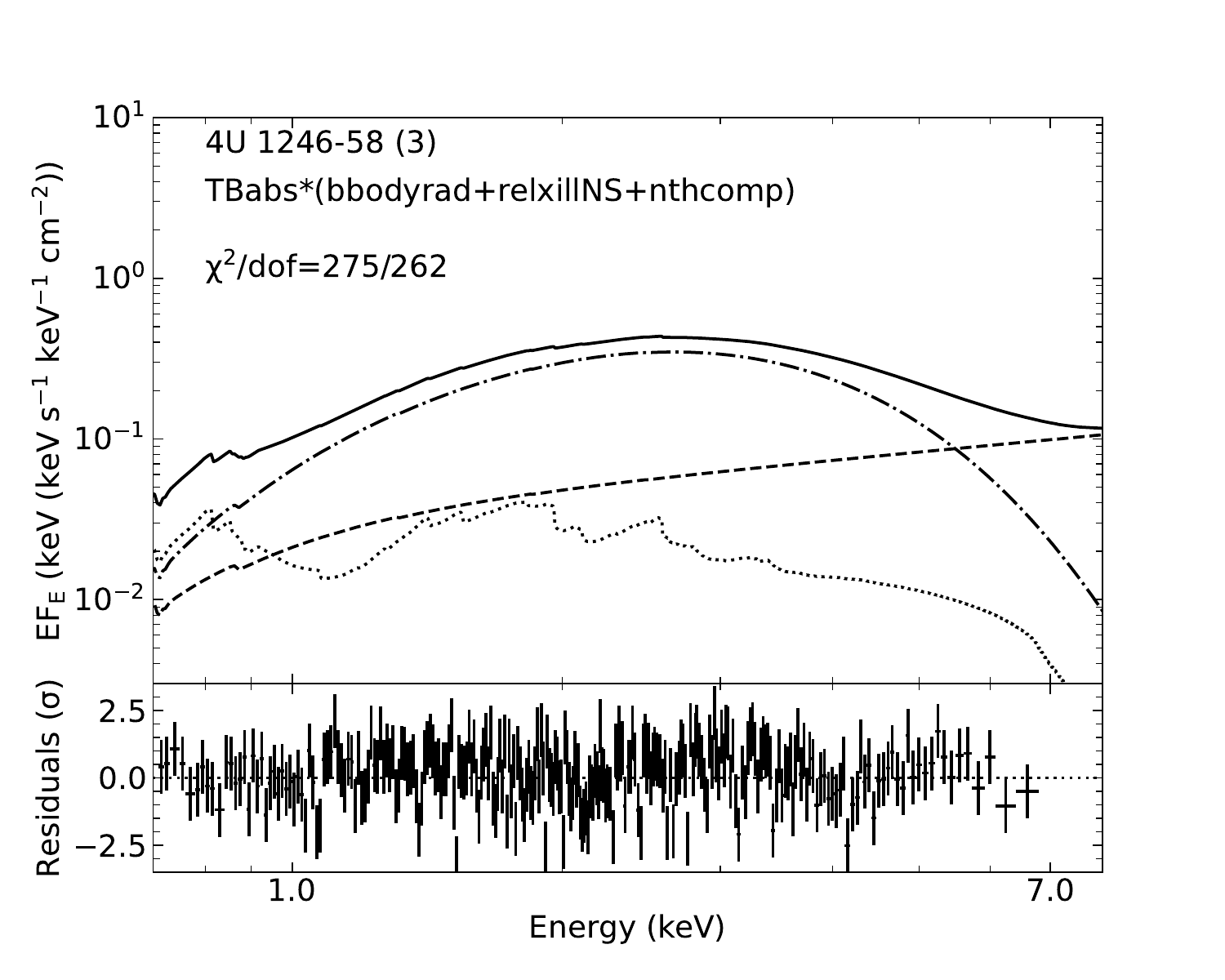}
\end{center}
\caption{Plots of the best-fit models and residuals (in units of $\sigma$) for the three intervals from 4U~1246-58. The solid line in the upper-part of each plot shows the final model (plotted in $EF_E$ units) including the \texttt{relxillNS} (dotted line), blackbody (dot-dashed line) and \texttt{nthcomp} (dashed line) components. Table~\ref{tab:nofluctfits} lists the parameters from the best-fit models.}
\label{fig:fourunonfluct}
\end{figure*}

\begin{figure*}[t!]
\begin{center}
\includegraphics[width=0.32\textwidth]{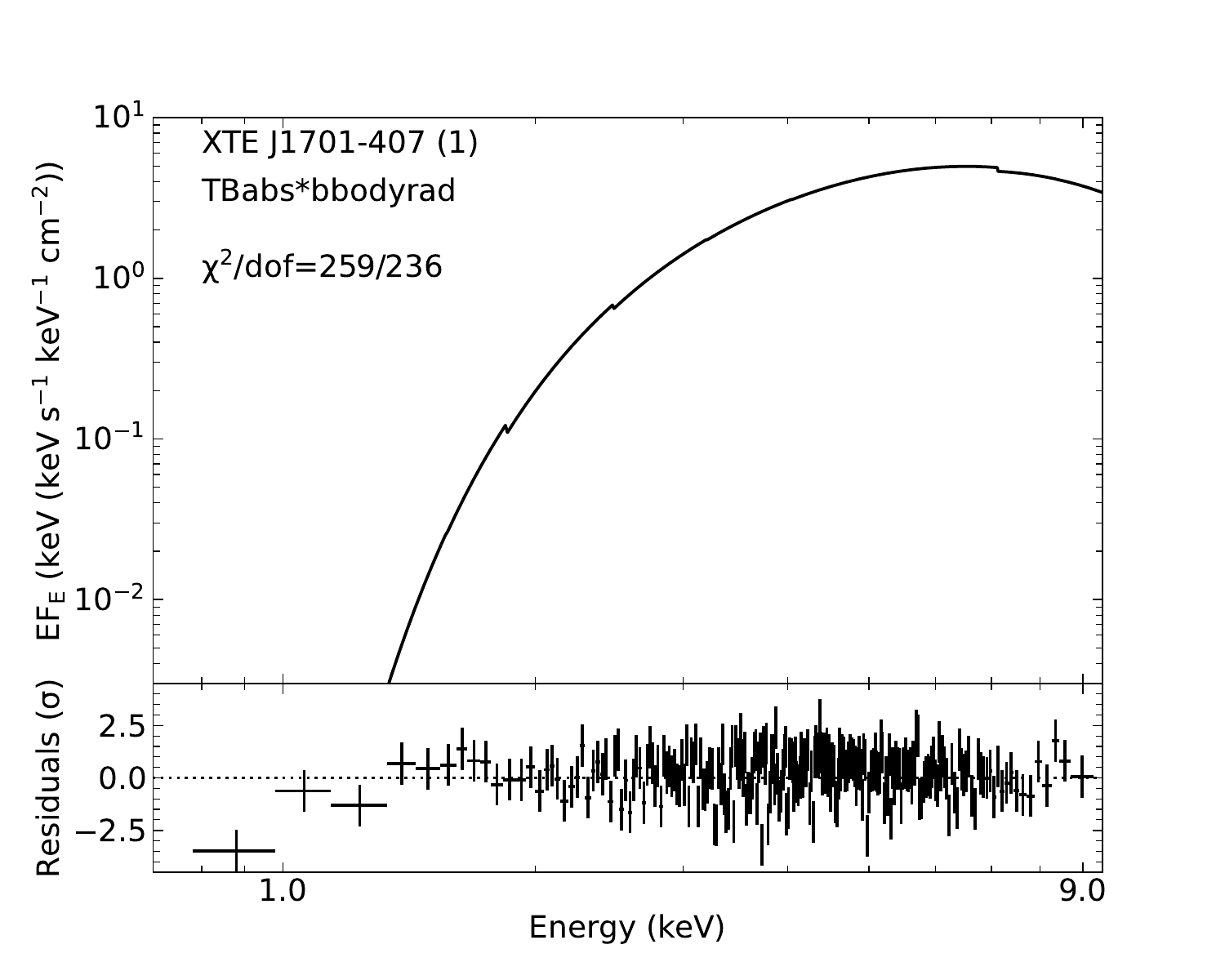}
\includegraphics[width=0.32\textwidth]{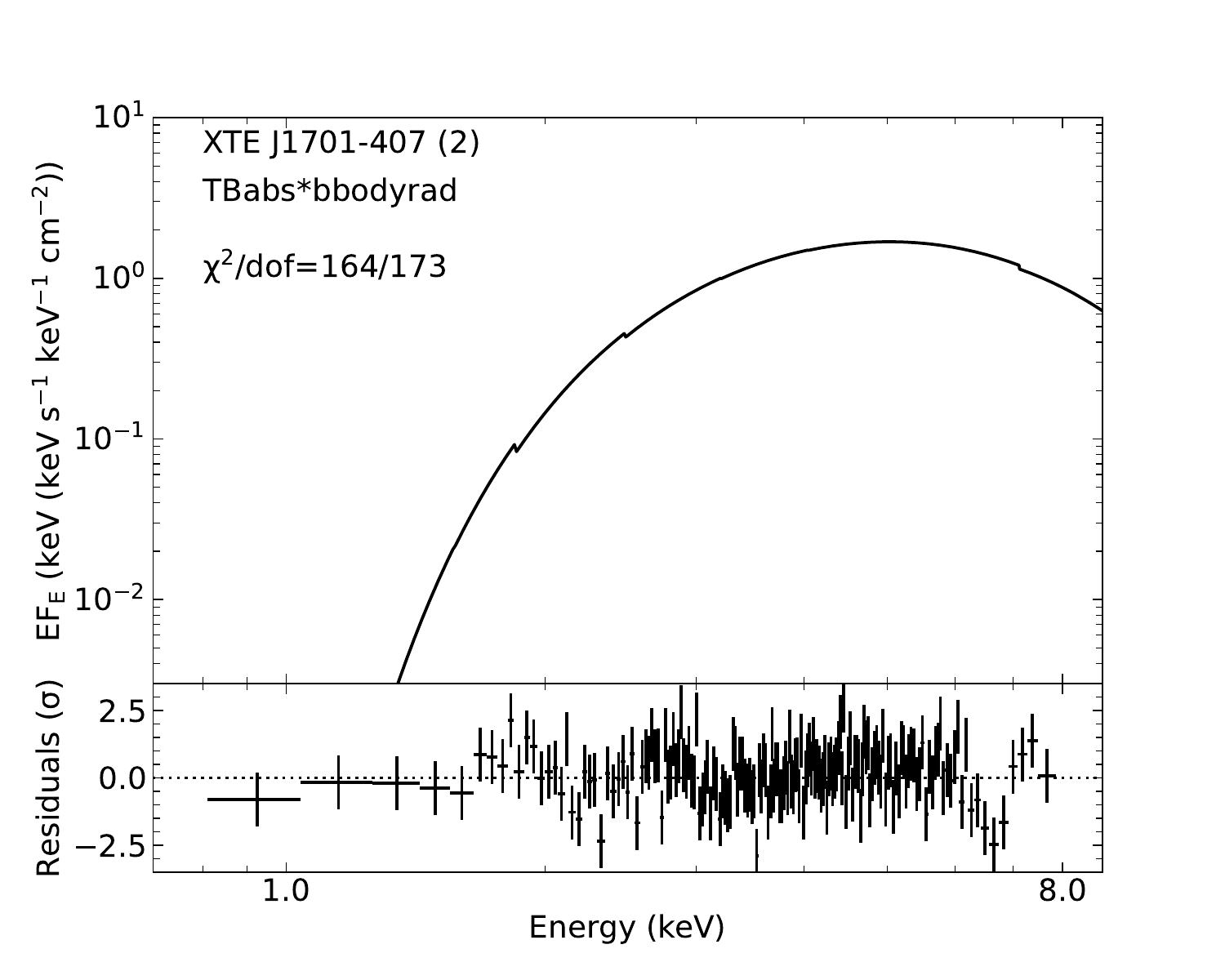}
\includegraphics[width=0.32\textwidth]{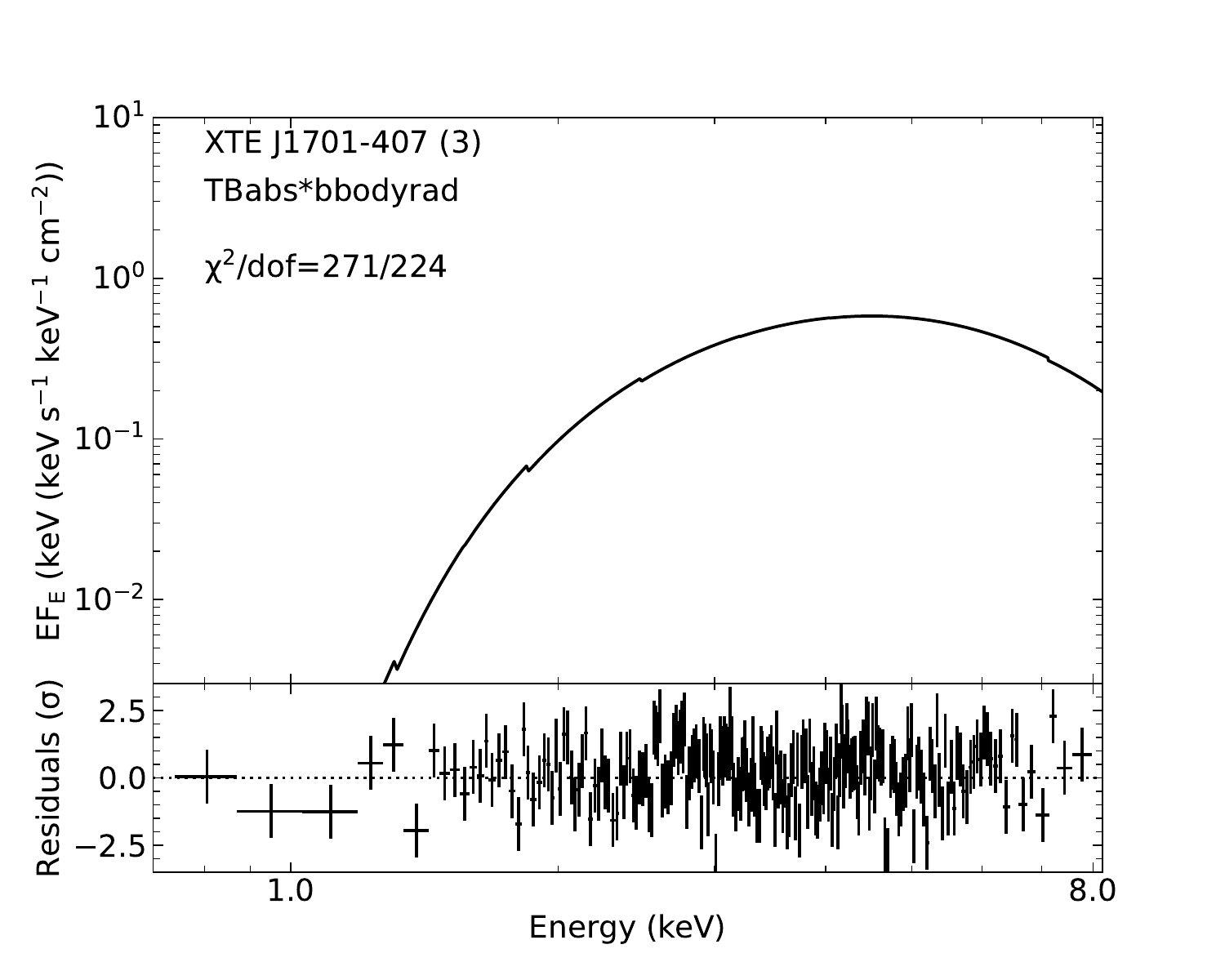}
\end{center}
\caption{Plots of the best-fit models and residuals (in units of $\sigma$) for the three intervals from XTE~J1701-407. The solid line in the upper-part of each plot shows the best-fitting model plotted in $EF_E$ units. Table~\ref{tab:nofluctfits} lists the parameters from the best-fit model.}
\label{fig:xtej1701}
\end{figure*}

\begin{figure*}[t!]
\begin{center}
\includegraphics[width=0.32\textwidth]{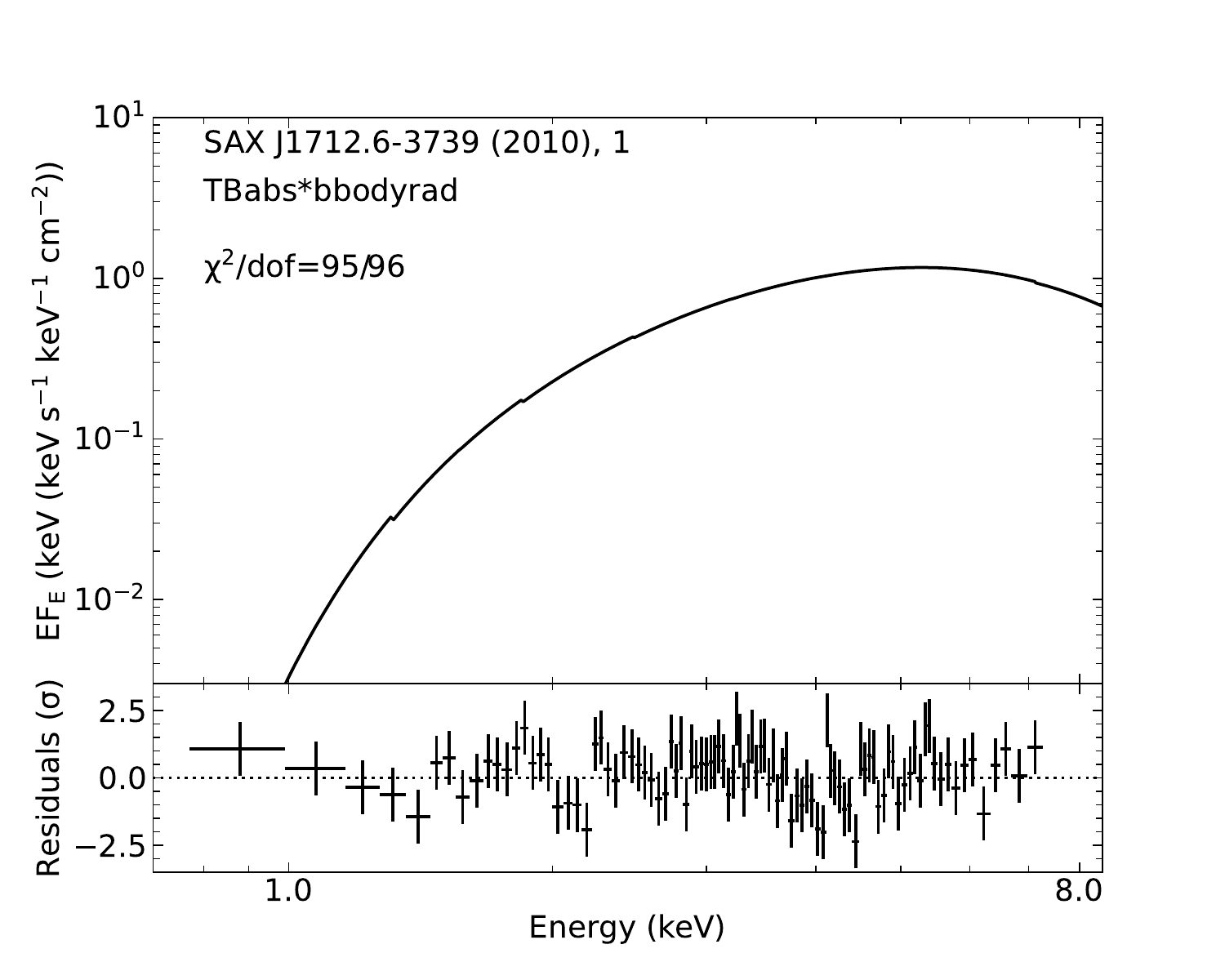}
\includegraphics[width=0.32\textwidth]{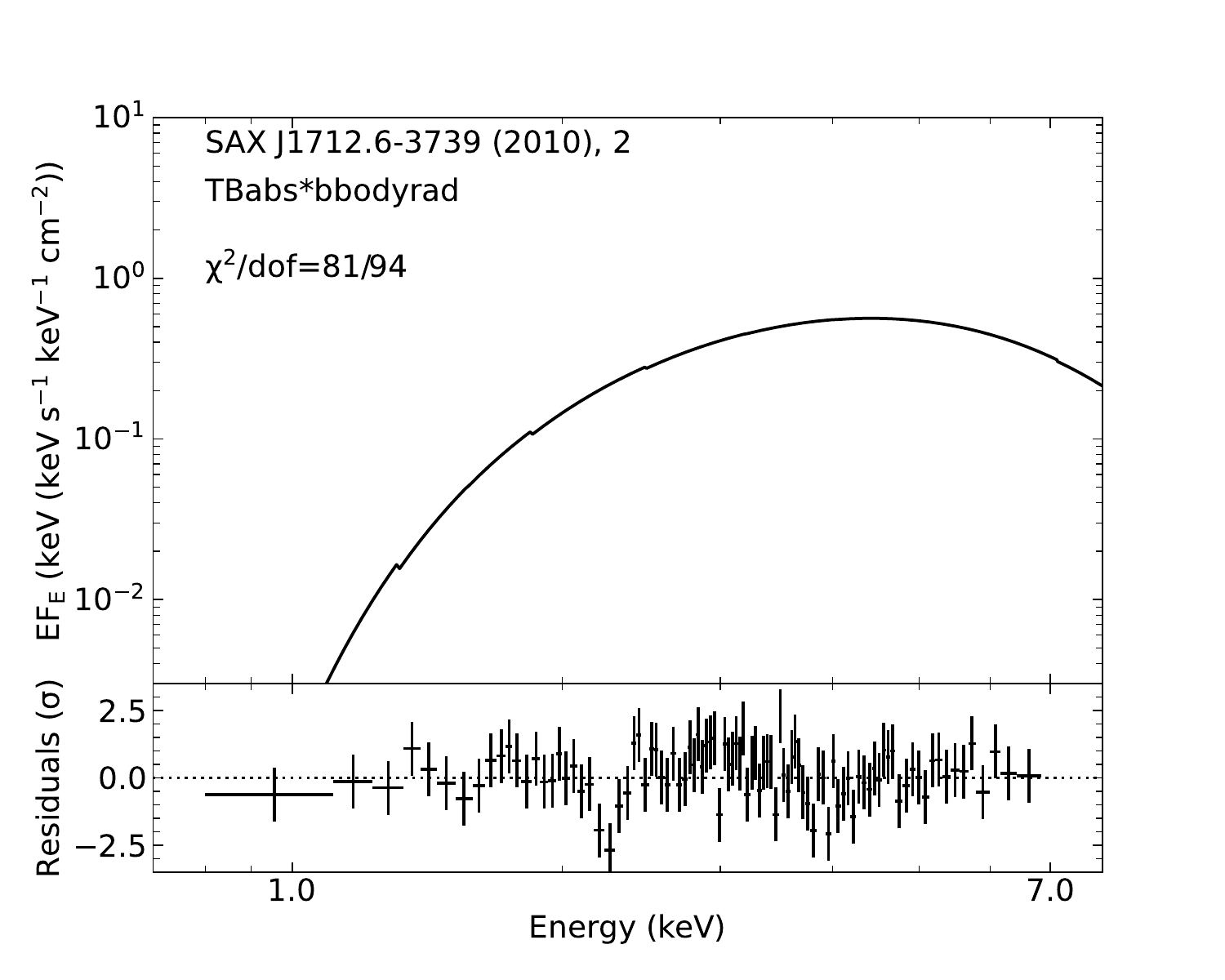}
\includegraphics[width=0.32\textwidth]{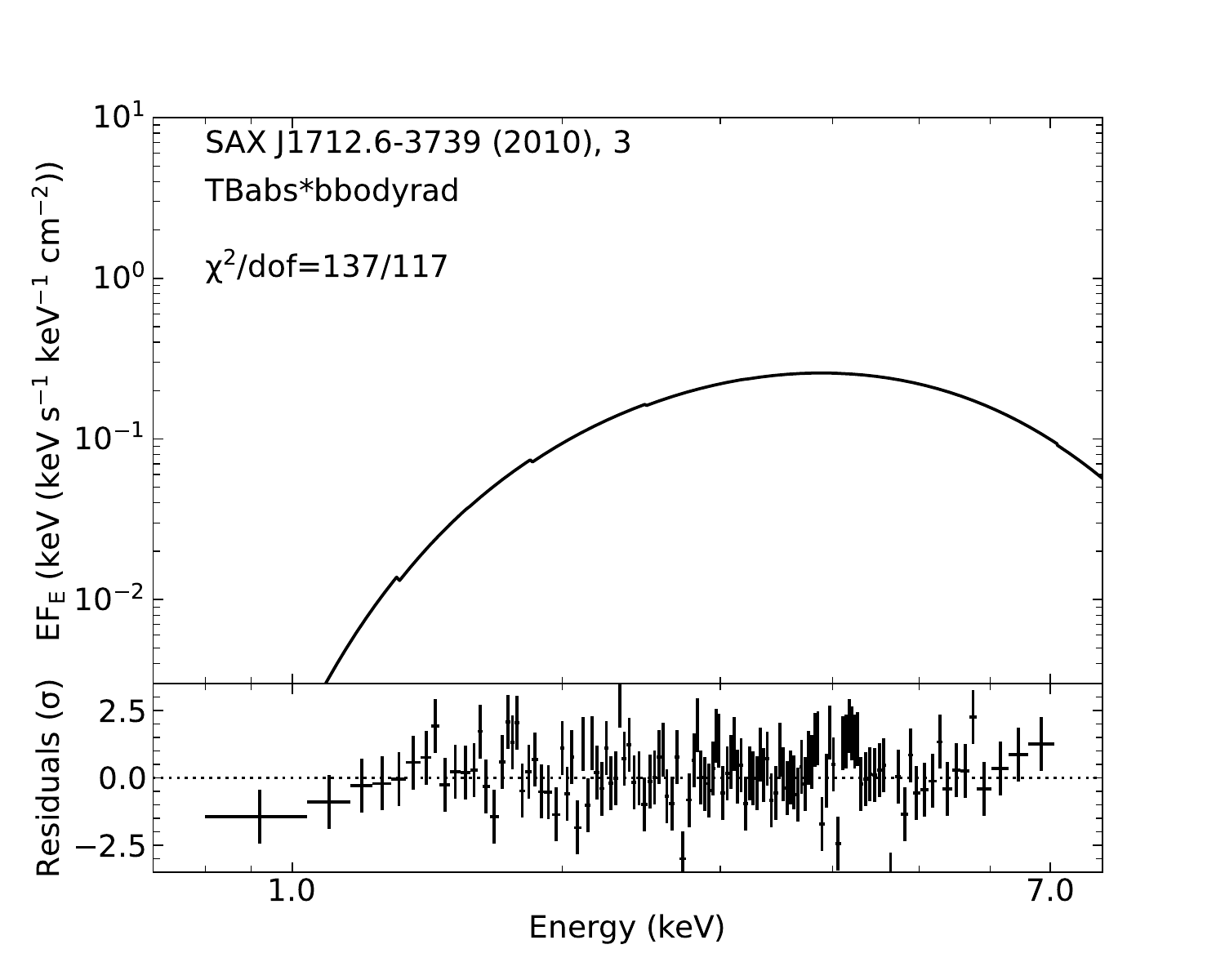}
\end{center}
\caption{Plots of the best-fit models and residuals (in units of $\sigma$) for the three intervals from the 2010 burst of \sax. The solid line in the upper-part of each plot shows the best-fitting model plotted in $EF_E$ units. Table~\ref{tab:nofluctfits} lists the parameters from the best-fit model.}
\label{fig:saxj1712nofluct}
\end{figure*}

\begin{figure*}[t!]
\begin{center}
\includegraphics[width=0.32\textwidth]{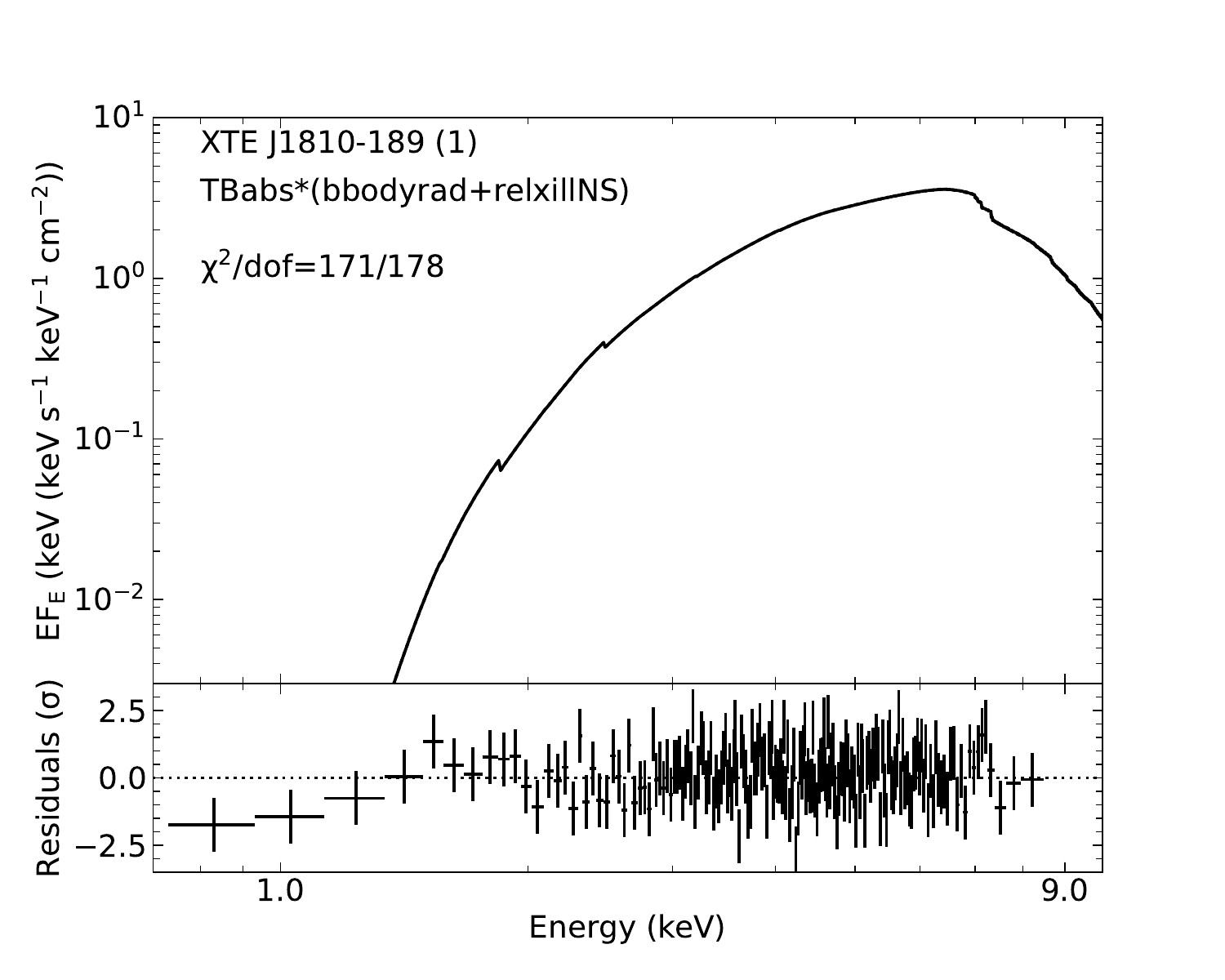}
\includegraphics[width=0.32\textwidth]{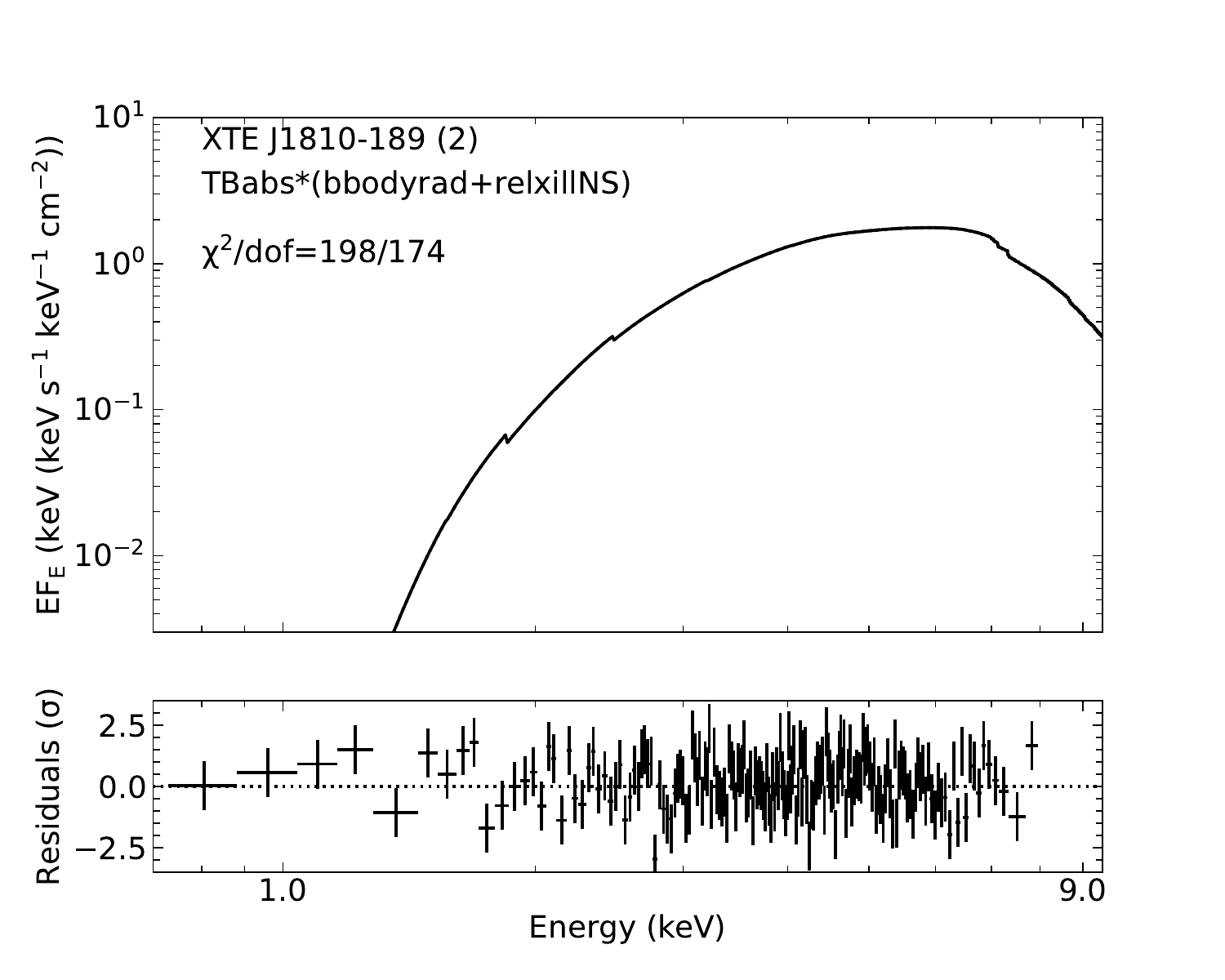}
\includegraphics[width=0.32\textwidth]{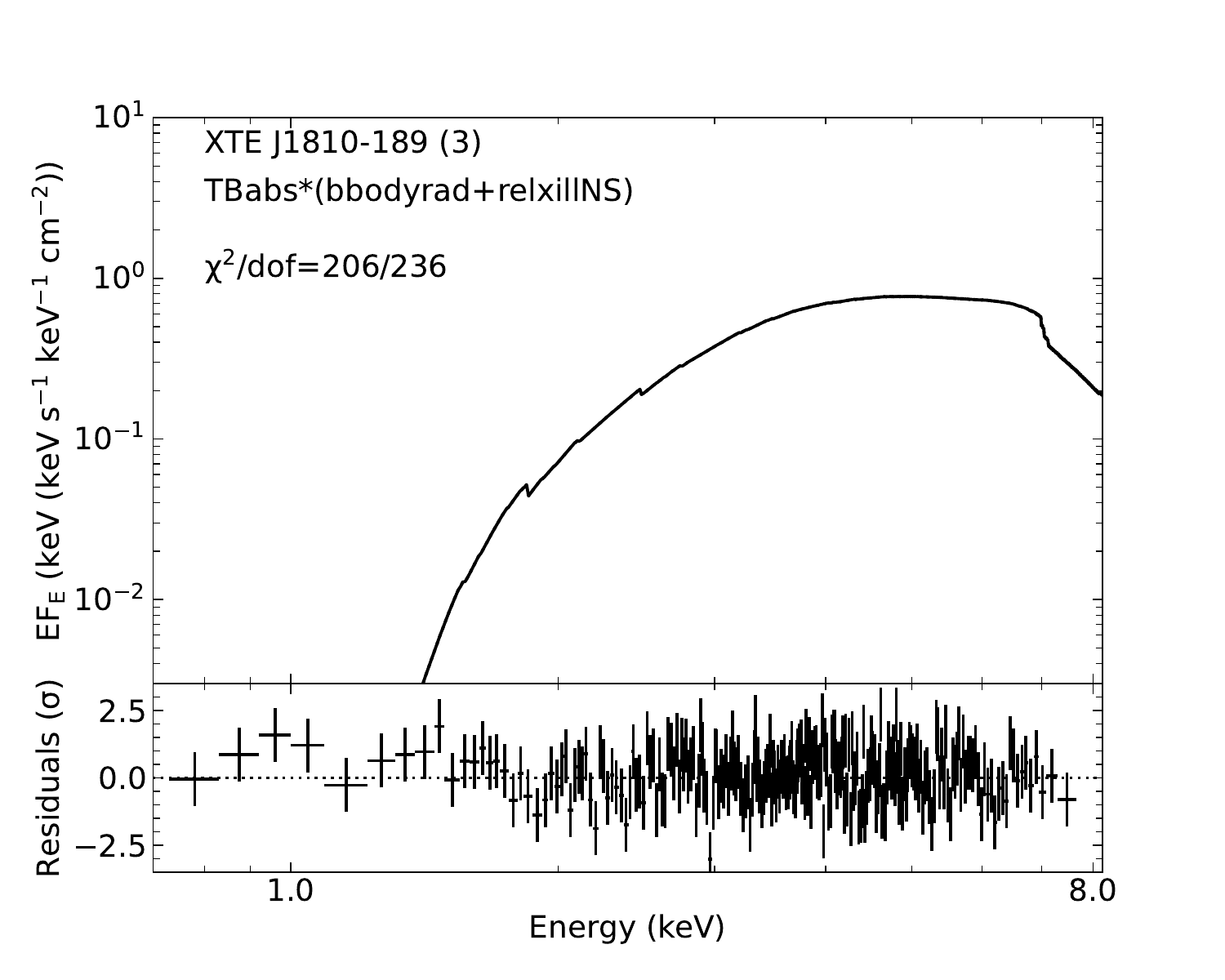}
\end{center}
\caption{Plots of the best-fit models and residuals (in units of $\sigma$) for the three intervals from XTE~J1810-189. The solid line in the upper-part of each plot shows the best-fitting model plotted in $EF_E$ units. Table~\ref{tab:nofluctfits} lists the parameters from the best-fit model.}
\label{fig:xtej1810nofluct}
\end{figure*}

\begin{figure*}[t!]
\begin{center}
\includegraphics[width=0.32\textwidth]{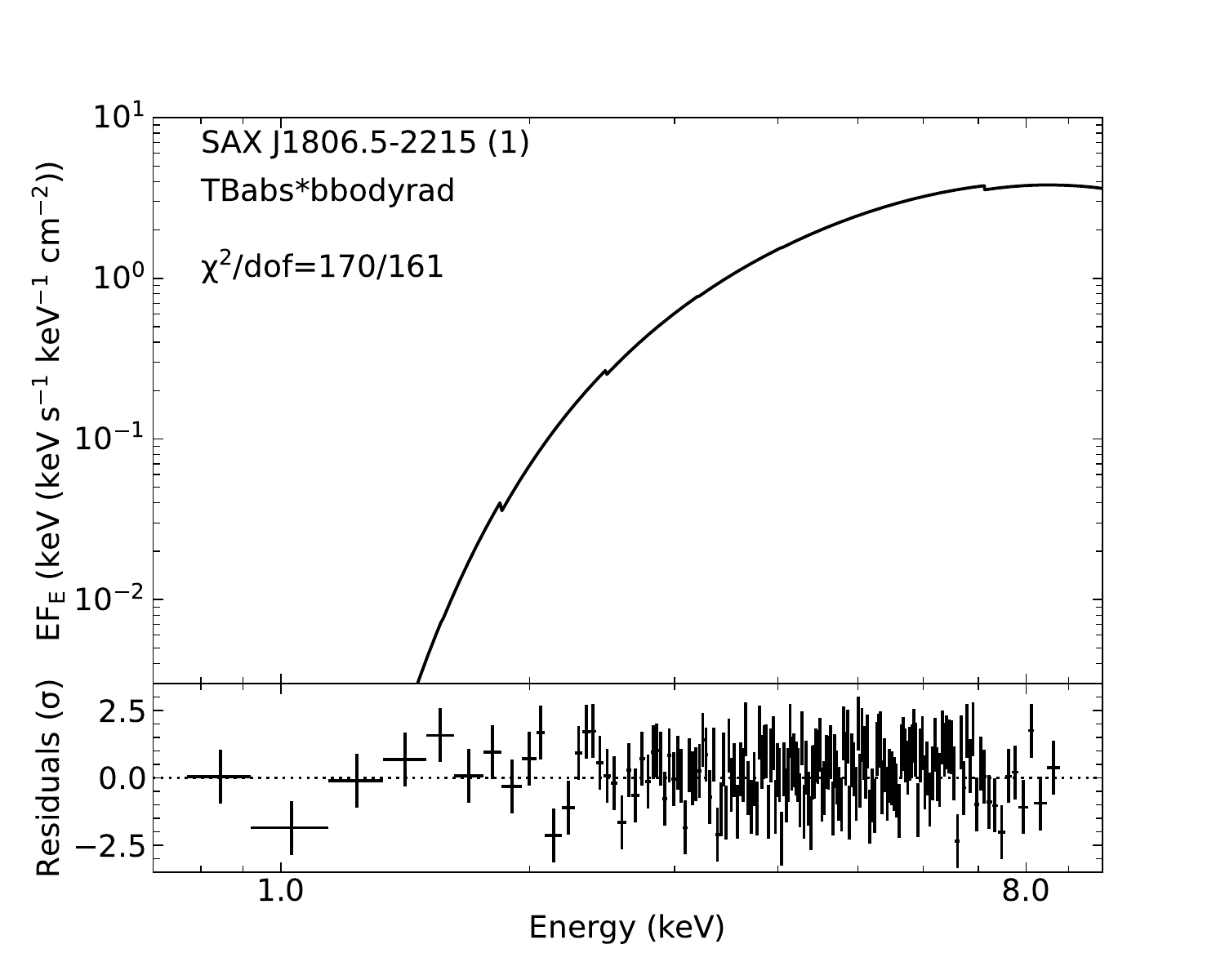}
\includegraphics[width=0.32\textwidth]{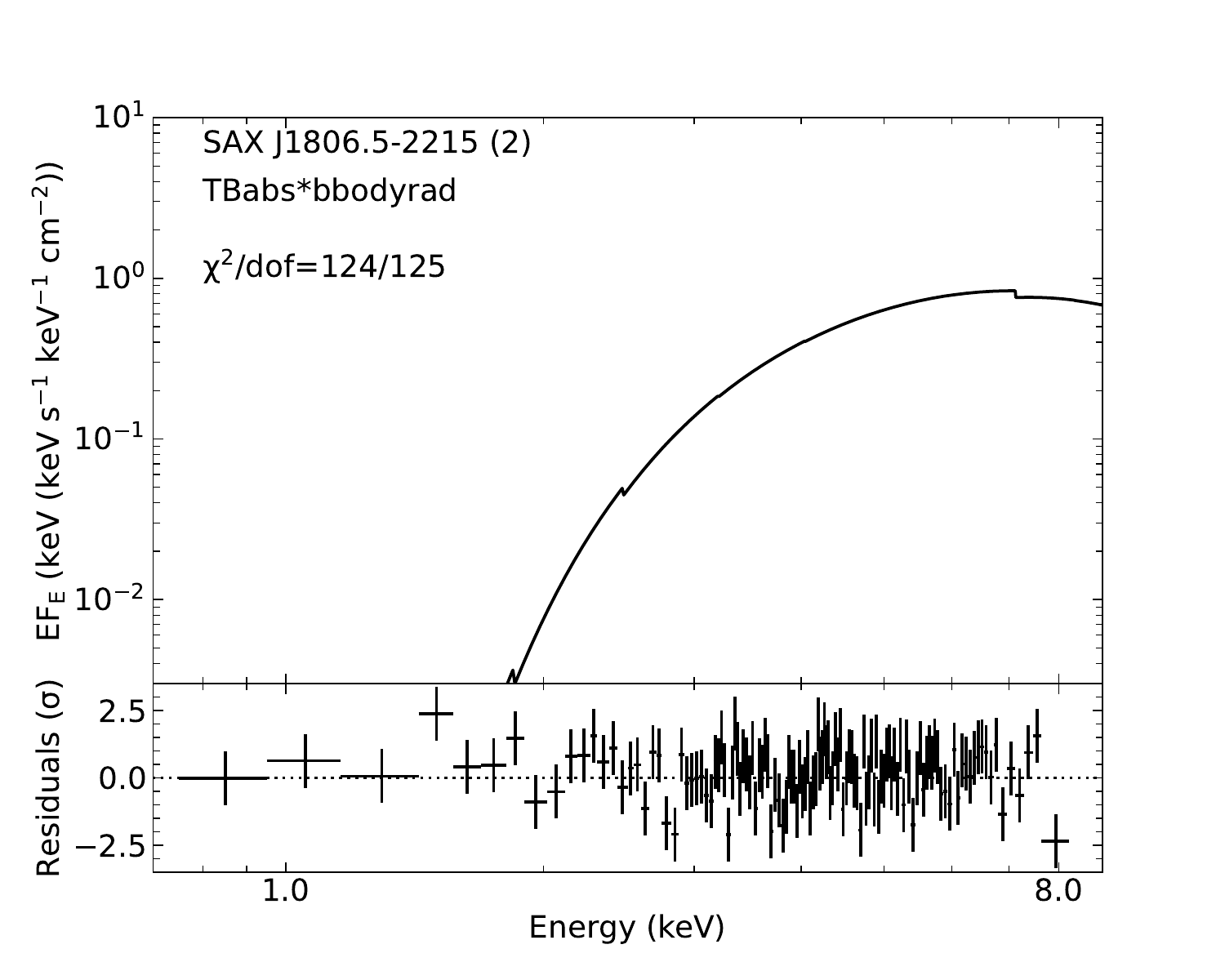}
\includegraphics[width=0.32\textwidth]{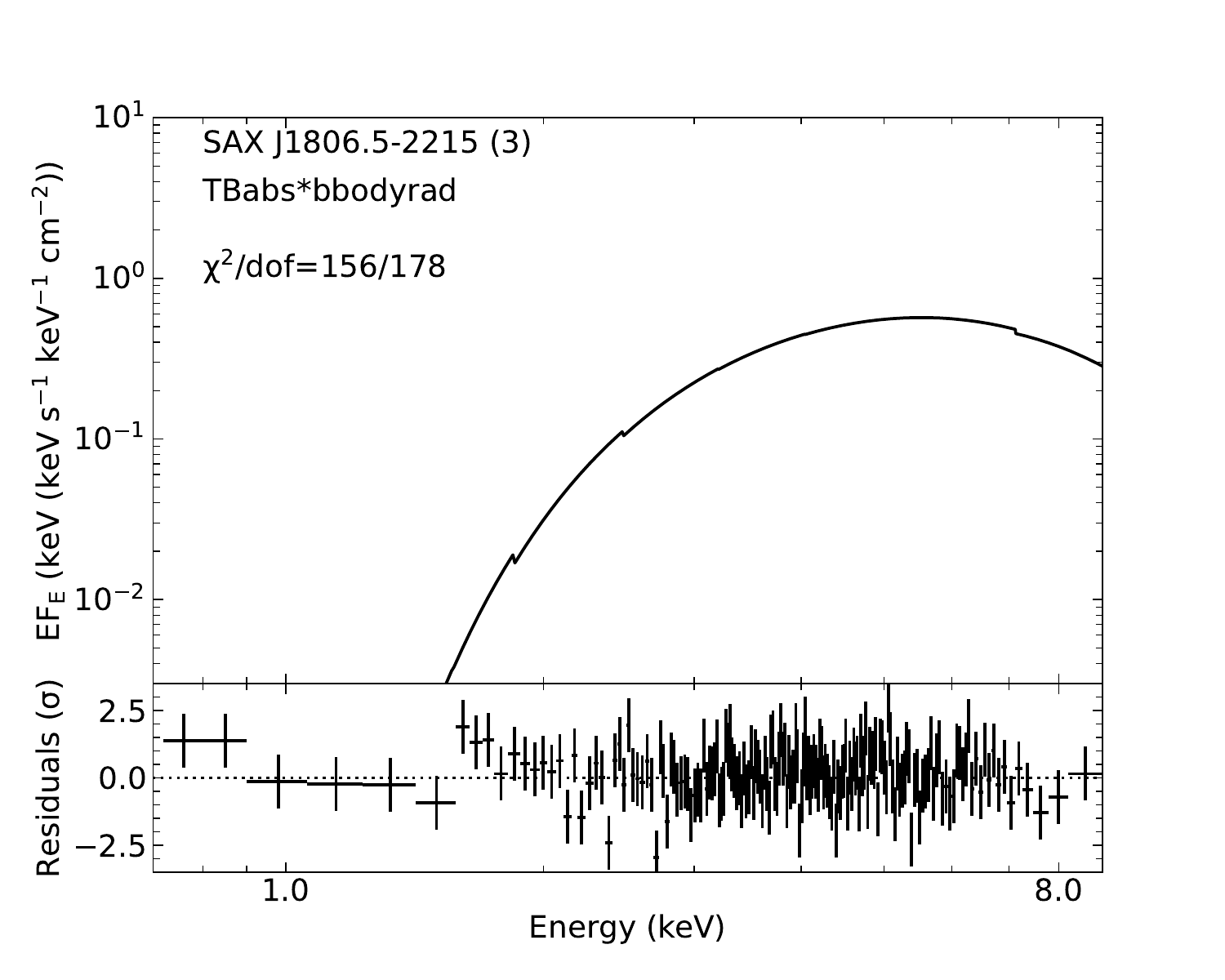}
\end{center}
\caption{Plots of the best-fit models and residuals (in units of $\sigma$) for the three intervals from SAX~J1806.5-2215. The solid line in the upper-part of each plot shows the best-fitting model plotted in $EF_E$ units. Table~\ref{tab:nofluctfits} lists the parameters from the best-fit model.}
\label{fig:saxj1806nofluct}
\end{figure*}

Figures~\ref{fig:fourunonfluct}--~\ref{fig:saxj1806nofluct} show the best fit models and the residuals from the fits to the 5 bursts that make up the control sample (e.g., Table~\ref{tab:sample}). The plots are formatted the same as those shown in Appendix~\ref{app:fluct}. The fit parameters from the analysis of these 5 bursts without fluctuations are listed in Table~\ref{tab:nofluctfits} in Sect.~\ref{sect:control}.

%

%
%\section{Author publication charges} \label{sec:pubcharge}

%% For this sample we use BibTeX plus aasjournalv7.bst to generate the

\bibliography{refs}{}
\bibliographystyle{aasjournalv7}



\end{document}